\documentclass{article}
\usepackage[T1]{fontenc}
\usepackage[utf8]{inputenc}
\usepackage[portrait]{geometry} 

\usepackage{authblk}
\usepackage{amsmath,amssymb,amsthm,amsfonts}
\usepackage{esint}
\usepackage{empheq}
\usepackage{xspace}
\usepackage{booktabs}
\usepackage{enumitem}
\usepackage[colorlinks=true,linkcolor=blue]{hyperref}
\usepackage{siunitx}

\usepackage[backend=biber,style=apa,citestyle=apa,uniquename=false]{biblatex}
\newcommand{\ww}{{WAVEWATCH III}\textsuperscript{\tiny\textregistered}}
\newcommand{\swan}{{SWAN}}
\newcommand{\mera}{{M\'ERA}}
\newcommand{\roms}{{ROMS}}
\newcommand{\nemo}{{NEMO}}
\newcommand{\cwst}{{COAWST}}
\newcommand{\wam}{{WAM}}
\newcommand{\era}{{ERA5}}
\newcommand{\eraint}{{ERA-Interim}}

\newcommand{\ecmwf}{{ECMWF}}

\newcommand{\gebco}{{GEBCO}}
\newcommand{\infomar}{{INFOMAR}}

\newcommand{\etopo}{{ETOPO1}}
\newcommand{\mars}{{MARS3D}}
\newcommand{\polcoms}{{POLCOMS}}
\newcommand{\nearoms}{{NEA\_ROMS}}

\newcommand{\punderline}[1]{\underline{\smash{#1}}}

\newcommand{\dpar}[2]{\frac{\partial {#1}} {\partial {#2}}}
\newcommand{\dcpar}[2]{\frac{\partial^2 {#1}} {\partial {#2}^2}}

\DeclareMathOperator{\pgrad}{\punderline{\nabla}}

\numberwithin{equation}{section}

\begin{document}

\title{A study of the wave effects on the current circulation in Galway Bay, using the numerical model COAWST}
\author[1,2]{Clément Calvino}
\author[2]{Tomasz Dabrowski}
\author[1]{Frederic Dias}
\affil[1]{University College Dublin, Dublin, Ireland}
\affil[2]{Marine Institute, Rinville, Ireland}
\date{}

\maketitle

\begin{abstract}
    A model coupling waves and currents is set up for Galway Bay, Ireland, using \cwst{}. The impact of the coupling on the accuracy of the model is shown to be marginal on the overall statistics. The model is in good agreement with the available in-situ observations that are used for the validation. However a closer look at the data during storms shows a deterioration of the agreement, which is barely improved by the coupling.
    A special analysis of the different processes ruling the ocean circulation and wave propagation is conducted during Storm Hector ($2018/06/14$). It shows a strong wave-induced surge by $7\,\si{\cm}$ in the back of Galway Bay, and a strong response in terms of currents along the coast of Clare of about $30\,\si{\cm\per\s}$. Most of the surge is attributed to the impact of the conservative wave-induced forcing terms, but the main process involved is not identified. The response in currents is partly attributed to the conservative wave-induced forcing terms, but mostly to the wave-enhanced surface roughness.
\end{abstract}

\section{Introduction}

Galway Bay and the Connemara region in the west of Ireland are zones of important economical activity, mostly fishery and aqua-farming. The coast is exposed to the strong Atlantic storms, causing regular flooding and damages to cities and marine infrastructures. Forecasting accurately the local ocean circulation and wave propagation is of interest for the community. The Irish Marine Institute (MI) runs several coastal ocean and wave operational models. One of them, the ocean Connemara model, focuses on the area of Galway Bay. Using these models, the MI regularly carries out specific studies for policy makers and public services. 
These studies include, for example, assistance in search and rescue missions, tracking the spread of sea lice harming salmonid farms or support to native oyster restoration efforts and oyster farming.
The quality of the services delivered and the accuracy of the forecasts depend on the capabilities of the numerical models to represent the oceanographic features and phenomena in the region.

Over the last decade several numerical studies featuring various coupled systems have shown that nearshore processes can be highly impacted by the interaction between currents, waves and winds. \textcite{bennis2020numerical} use the 3D ocean model \mars{} (\nptextcite{lazure2008external}) coupled with the third generation wave model \ww{} (\nptextcite{wavewatch2019user}), and focus on the Alderney Race (France), a region known for its strong tidal currents. \textcites{jakovljevic2018effect} also study the strong interaction between waves and tides in the Alderney Race, with an application on the tidal energy resource. They use the coupling system \cwst{} (\nptextcites{warner2008development,kumar2012implementation}), where the ocean model \roms{} (\nptextcite{shchepetkin2005regional}) interacts with the wave model \swan{} (\nptextcite{swan2020scientific}). \textcite{olabarrieta2011wave} reproduce a storm event in Willapa Bay also with the coupling system \cwst{}.

All these studies highlight a strong interaction of waves and currents. Both surf zone and tidal processes are key to correctly account for the hydrodynamics. In particular tidal currents and the sea level variation are shown to affect wave breaking, which modifies in turn the wave heights and turbulence mixing in the water column. Waves are also refracted by the tidal currents, focusing the wave rays in regions of stronger opposing currents. This last feature is also shown in \textcite{ardhuin2012numerical} where a \ww{} model coupled one-way with a \mars{} model is used for the Iroise Sea (France). A strong focusing is observed near Ouessant and Bannec islands as the waves propagate against a tidal current.

Effects are not limited to surf zone processes. Both in \textcite{olabarrieta2011wave} and  \textcite{ardhuin2012numerical}, the local generation of waves is shown to be impacted by the currents. At high frequencies the current-induced Doppler shift can be significant, and the advection of waves can be stopped when facing strong adverse currents. Currents also directly act on the wind stress itself generating the wave growth, where a relative wind is used. However it is not clear how to best account for the currents in this process. \textcite{renault2016modulation} use a coupled air-sea model to highlight a complex retro-effect of ocean currents on the wind in the surface air boundary layer. Waves are also at play in the oceanic surface layer impacting the wind stress momentum flux in the water column. \textcite{osuna2005numerical} use the coupled system \polcoms{} to investigate the interaction between waves and currents in the Irish Sea. In particular the model features a modified surface roughness, increased by the wave-induced stress, which enhances the surface momentum flux from the wind into the water column. During a storm event they show that the surface stress is significantly increased, leading to a sea level increase up to $15\,\si{\cm}$. A similar result is found in \textcite{wu2019wave}, where a similar modification of the surface roughness is done. They are using the ocean model \nemo{} (\nptextcite{nemoman}) coupled with the wave model \wam{} (\nptextcite{wamdi1988wam}).

The present paper focuses on the possible effects of wave-current interaction on the hydrodynamics of Galway Bay. A few published studies are available focusing either on the currents or waves spawning the Galway Bay area. The main results of those studies will be summarized in Section 2, along with the unpublished knowledge of the area acquired by the MI after years of monitoring and modelling in the region. A fully coupled model using \cwst{} is set up in Section 3 and described for the area. It is an upgrade of the current operational \roms{} standalone model run by the MI. In Section 4, the model is validated over the three periods $2017/01/01$-$2017/03/27$, $2017/05/04$-$2017/07/28$ and $2018/05/10$-$2018/08/03$, highlighting the statistical benefit of the coupling. A more detailed study of the processes introduced by the coupling is then conducted, highlighting the driving mechanisms for this application.

\section{Review of the Galway Bay hydrodynamics}

In this section we describe the domain that is used in the numerical model, with a focus on Galway Bay. We also offer a short review of past studies conducted in this area and summarize the main hydrodynamics features highlighted by those studies.

The Connemara region contains several aqua-farms. It features a network of small islands and bays making the hydrodynamics complex. This is highlighted in \textcite{jackson2012result} where they evaluate the cross contamination of sea lice between aqua-farms in Kilkieran Bay using the MI ocean model. However no further studies on the hydrodynamics in the Connemara region has been found.

\subsection{Irish Shelf}

\begin{figure}
  \centering \includegraphics[width=10cm]{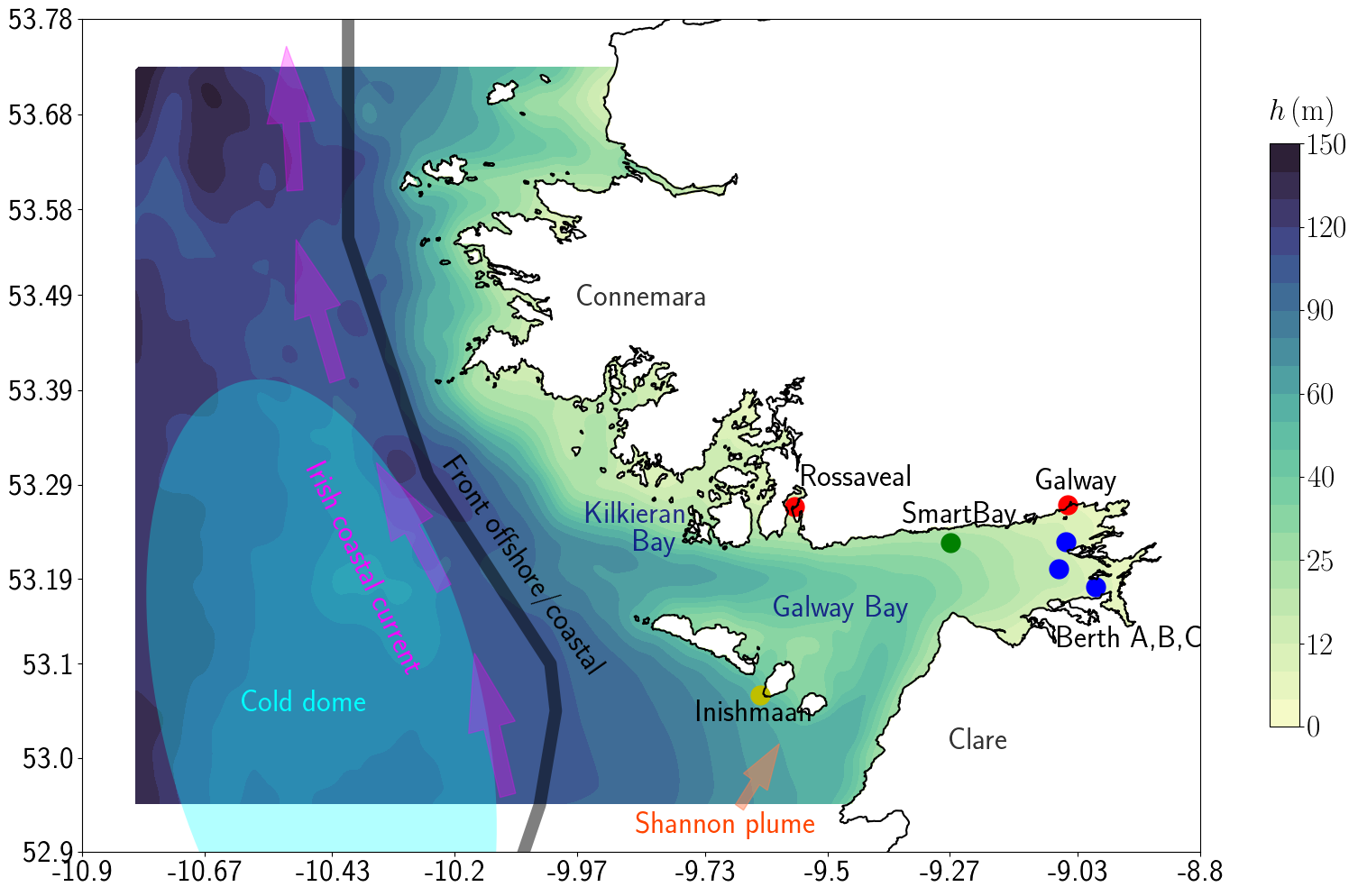}
  \caption{Summary map of the oceanographic knowledge of the western Irish shelf, approximately drawn from \textcite{nolan2004observations}.
  The red dots are tidal gauges, the green dot is the SmartBay location including one ADCP and one wave buoy, the blue dots are the Berth ADCPs (A,B and C from North to South), the yellow dot is the ADCP near Inishmaan.}
  \label{irishshelf}
\end{figure}

The Irish shelf extends far beyond the limit of the domain studied, shown in Figure \ref{irishshelf}. Overall the water depths are quite shallow, on average around $200 \,\si{\m}$ and even less (around $100 \,\si{\m}$) for the domain studied. The waters on the shelf do not interact much with the waters outside the shelf. Further to the West and not shown here is the North Atlantic Current.

A considerable work contributing to the understanding of the hydrodynamics of the western Irish coast is presented in \textcite{nolan2004observations}. A series of hydrographic sections are conducted to investigate the density structure of the water on the Irish shelf and several ADCPs are deployed. They show the existence of a front separating the inland fresh and cold waters and the shelf waters, as shown in Figure \ref{irishshelf}. The tidal stirring is reduced on the shelf. In summer the wave-induced and wind-induced vertical mixing is also weaker. The thermal stratification is therefore able to stabilise the water column and create the front. 
A jet-like flow is noticed along this front at mid-water depths, from South to North. This Irish coastal current is enhanced by the Shannon River plume. This contributes to increase the mass flux going inside Galway Bay and West of the Aran Islands. Both the location of the front and the Irish coastal current are highly seasonal and more pronounced in summer, again benefiting from the thermal stratification stabilising the water column. The residual currents are mostly wind-driven in autumn, winter and spring, but density-driven in summer, with a location and direction agreeing well with those of the front. 
The thermal stratification in summer traps cold waters at the sea bottom. A cold dome is present on the Irish shelf, adjacent to the front, as shown in Figure \ref{irishshelf}. Such a cold dome is stable because of the generation of a geostrophic gyre mid-waters, the Coriolis force balancing the density-induced pressure gradient. In theory a decrease in sea level should also be induced at the location of the dome.

In terms of waves, the west coast of Ireland is characterised by rather high significant wave heights, on average around $2.5 \,\si{\m}$ and $3 \,\si{\m}$ (\nptextcites{atan2016assessment,gallagher2014long,gallagher2016nearshore}). The coast is regularly hit by strong windstorms and hurricanes over the winter, in general moving from South-West to North-East.

\subsection{Galway Bay}

\begin{figure}
  \centering \includegraphics[width=10cm]{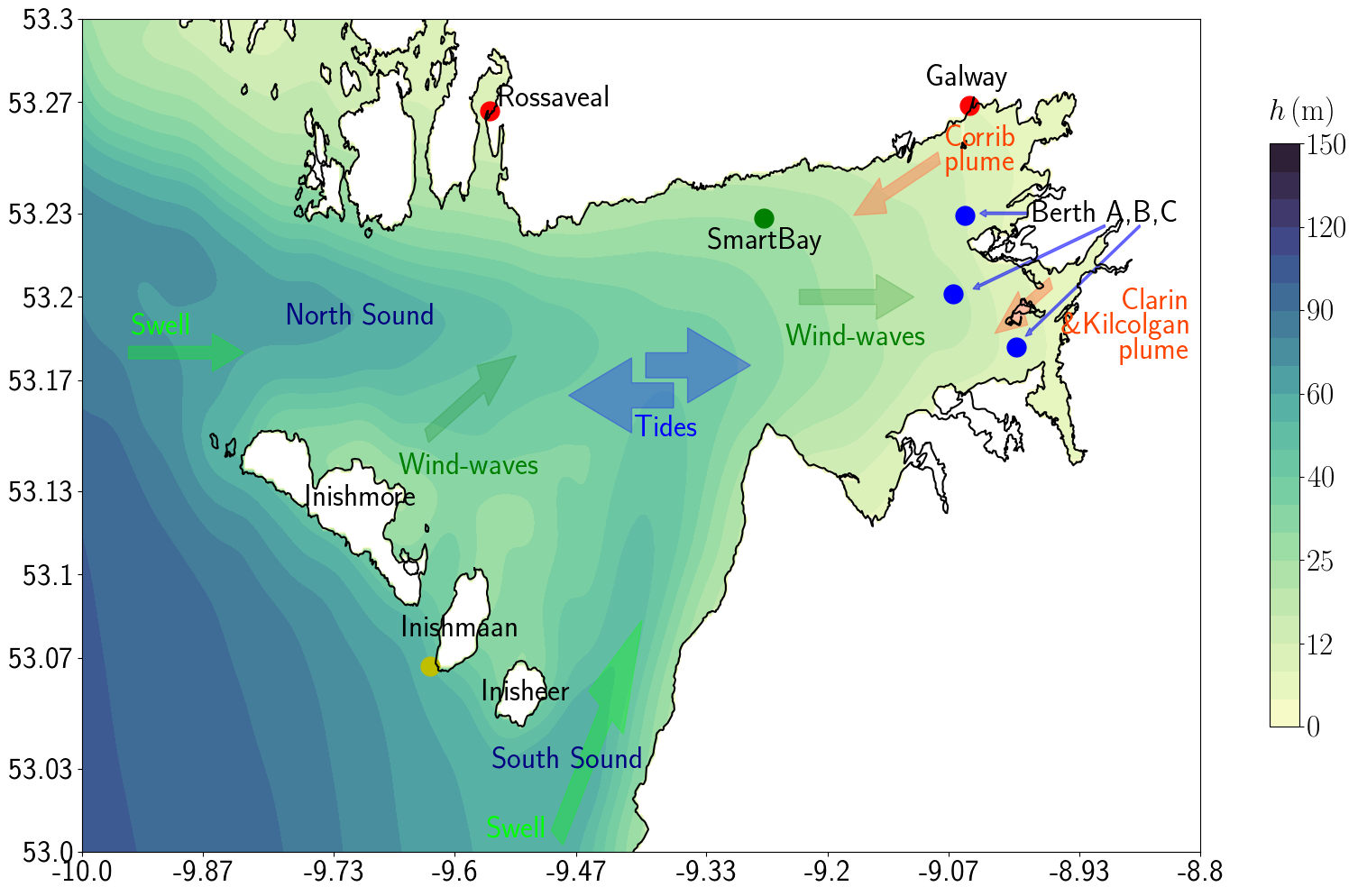}
  \caption{Summary map of the oceanographic knowledge of Galway Bay.
  The red dots are tidal gauges, the green dot is the SmartBay location including one ADCP and one wave buoy, the blue dots are the Berth ADCPs (A,B and C from North to South), the yellow dot is the ADCP near Inishmaan.
  }
  \label{galwaybay}
\end{figure}

The Galway Bay area itself has motivated more dedicated research over the years. Its sheltered location and proximity with Galway city make it an attractive area for engineering research projects. It features a marine renewable energy test site . 

The bay itself can be divided into an Inner Bay with water depths below $30\,\si{\m}$, and an Outer Bay with a maximum depth of $70 \,\si{\m}$ (see Figure \ref{galwaybay}). The bay is approximately $50 \,\si{\km}$ long and $10\,\si{\km}$ wide. It is protected from the Atlantic Ocean by the Aran islands. The main connections between the bay and the oceanic waters are the North Sound, north of Inishmore, and the South Sound, east of Inisheer. The main source of fresh water in the bay is through the Corrib River discharge located in Galway city. The bay is also highly influenced by tides, ranging from $5\,\si{\m}$ to $2\,\si{\m}$. It is common for part of the nearshore land to be flooded during strong storms or rain events.
Galway Bay is a semi-enclosed basin. As such it prevents most of the swell systems to enter the bay. A wave model consisting of nested grids is set up in \textcite{rutebento2015numerical} focusing on Galway Bay. It is shown that the dominant wave direction at the SmartBay location is from southwest, which corresponds to swell entering the bay through the South Sound. Wind waves are also shown to be of importance in the bay, with a strong occurrence of short period waves. The model also looses accuracy and underestimates higher wave heights. A poor agreement was found for all wave periods, showing the difficulty of accurately capturing both the propagation of swell inside the bay and the generation of wind waves.

A more accurate analysis of the wave characteristics inside the bay is provided by \textcite{atan2016assessment}. Their goal is to report how often optimal conditions for wave energy convectors are met. They conducted a long term analysis over $5$ years of the significant wave heights and wave periods at the SmartBay location. They find that the majority of wave systems are characterised by significant wave heights below $1 \,\si{\m}$ and periods below $4\,\si{\s}$ ($44\%$ of events), and $81\%$ of events feature significant wave heights below $2\,\si{\m}$ and periods below $5\,\si{\s}$. Those statistics strongly indicate that the bay is often dominated by locally generated wind waves. Some events with wave heights higher than $3\,\si{\m}$ are still recorded, which are unlikely to be locally generated given the maximum fetch in the bay. Those events appear on the time series as local peaks. This indicates that the strongest swell systems are still able to penetrate inside the bay. In agreement with \textcite{rutebento2015numerical}, they also find that on average most of the waves inside the bay are coming from the southwest.

Focusing on the ocean currents inside the bay, \textcite{ren2017effect} compare results from a numerical model to ADCP and radar measurements using different wind forcing conditions. The ADCP data shows that surface currents are stronger and more variable than the depth-integrated currents, sometimes twice as strong. This significant surface shear and high variability are the result of the surface wind stress. Using accurate atmospheric forcing data is therefore critical to accurately capture the currents inside the bay. They show that wind-induced currents compete against tidal-driven currents. As such the currents are more sensitive to wind during neap tides. The surface layer is in general mostly wind driven, with the mid water column being mostly driven by tides. Wind is also shown to greatly explain the weaker northward current component.
\textcite{ren2020investigation} investigate the synoptic characteristics of the surface ocean currents measured by a fixed radar and numerically modelled, using a self-organizing map analysis. Both analyses yield similar results. Tidal driven eastwards and westwards patterns are shown to be dominant but only account for $30\%$ of the data. $12$ patterns are retained, carrying significant variability. A strong rotation is noticeable in the current field. The main processes known to drive the currents include tides and winds, but also the Corrib River discharge adding a flux of fresh and cold water in the bay as mentioned in \textcite{nolan2004observations}. Such a river discharge is expected to yield a westwards flow following the coast, due to the impact of the Coriolis force.

\section{Numerical model}

The modelling system \cwst{} is used here. It couples the ocean numerical code \roms{} with the third generation wave model \swan{}. It is a widely used coupling system with a variety of modules. It includes the interaction between waves and currents, which is the main focus of this paper. The implementation presented in \textcite{kumar2012implementation} is used, following the vortex formalism derived in \textcite{mcwilliams2004asymptotic}. We will introduce all the equations solved in both the hydrodynamics and wave models, mostly referring to \textcites{uchiyama2010wave,kumar2012implementation}.

\subsection{Ocean model: \roms{}}

The hydrodynamic code \roms{} is a three-dimensional, free-surface and terrain-following numerical model. It solves the Reynolds Averaged Navier-Stokes equations with the hydrostatic and Boussinesq approximations in a finite difference framework. A split-explicit time stepping algorithm is used and detailed in \textcite{shchepetkin2005regional}.

The wave-induced conservative terms are implemented following the vortex force formalism of \textcite{mcwilliams2004asymptotic}. A multi-scale expansion is carried out introducing three components for the hydrodynamic variables, the wave component varying as the gravity waves, the long-wave component varying as the infra-gravity waves generated by non-linear wave interaction, and the current component varying as the longest temporal and spatial time scales. The conservative wave-induced forcing terms appear in the current dynamics balance. The implementation done in \cwst{} by \textcite{uchiyama2010wave,kumar2012implementation} simplifies the problem by merging the infra-gravity and current components, also ignoring several higher-order terms in the wave-induced expressions.

\subsubsection{3D equations}

For the notation, $\nabla$ is the horizontal nabla operator in the $(x,y)$ directions ($x$ is the eastward coordinate and $y$ the northward coordinate), $z$ is the vertical coordinate with $\punderline{e}_z$ the corresponding unit vector pointing upwards with $0$ the reference mean sea level. The mean sea surface elevation is $\zeta$, $-h$ is the sea bottom level and $H=h+\zeta$ the total depth. The time is $t$, $\punderline{u} = (u,v)$ is the horizontal mean Eulerian velocity, $w$ the vertical component, $(\punderline{u}',w')$ the corresponding turbulent velocities, $(\punderline{u}^{\text{St}},w)$ are the Stokes velocities, $f$ is the Coriolis parameter, $\nu$ is the molecular viscosity. The dynamic pressure $p$ corresponds to the total pressure divided by the reference water density $\rho_0$, $\rho$ is the water density and $g$ is the acceleration due to gravity.
The full model closure is not reported in this paper to save space, only the most relevant part to the present research. For a more thorough description the reader can refer to \textcite{uchiyama2010wave} and \textcite{kumar2012implementation}. The $k-\epsilon$ turbulent viscosity model with turbulent eddy viscosity $K_m$ described in \textcite{warner2005performance} is used for the turbulence closure. 

The 3D wave-averaged equations read as follows (for simplicity the terrain-following vertical and curvilinear transformation steps are not included):
\begin{empheq}{align}
 \dpar{\punderline{u}}{t} + \left(\punderline{u}\cdot\pgrad + w \dpar{}{z}\right)\punderline{u} + f \punderline{e}_z \times \punderline{u} + f\punderline{e}_z\times\punderline{u}^{\text{St}} & =  - \pgrad p - \pgrad \mathcal{K} + \punderline{J} \nonumber\\
 & \qquad + \dpar{}{z}\left( \left(\nu + K_m\right) \dpar{\punderline{u}}{z} \right) + \punderline{F}^w + \punderline{D} \,, \label{3d_momentum} \\
 \dpar{p}{z} + \frac{g \rho}{\rho_0} & = - \dpar{\mathcal{K}}{z} + K \,,\\
 \pgrad \cdot\, \punderline{u} + \dpar{w}{z} & = 0 \,.
\end{empheq} 

The wave-induced non-conservative forcing terms are contained in $\punderline{F}^w$ and $\punderline{D}$ is the horizontal mixing term.
The wave-induced conservative terms consist of the vortex force $(\punderline{J},K)$ and the Bernoulli head $\mathcal{K}$. They are evaluated as follows with $H_s$ the significant wave height, $\punderline{k}_{p}$ the peak wavenumber vector, $k_{p}$ its magnitude, and $\sigma_{p}$ the intrinsic wave peak frequency (the wave parameters are provided by the wave model, \swan{}):
\begin{empheq}{align}
& \punderline{J} = - (\pgrad\times\punderline{u})\times\punderline{u}^{\text{St}} - w^{\text{St}} \dpar{\punderline{u}}{z} \,, \\
& K = \punderline{u}^{\text{St}}\cdot\dpar{\punderline{u}}{z} \,, \\
& \mathcal{K} =  \frac{\sigma_{p} H_s^2}{32k_{p}} \frac{1}{\sinh^2 \left(k_{p} H\right)} \int_{-h}^z \dcpar{(\punderline{k}_{p}\cdot\punderline{u})}{z'}\sinh(2k_{p}(z-z'))\,\mathrm{d}z' \,.
\end{empheq}

The total mean sea level $\zeta$ (called composite sea level in \textcite{kumar2012implementation}) is decomposed into a dynamical component $\zeta^d$ and a static component $\hat{\zeta}$. The static component includes the inverse barotropic effect with $P_{\text{atm}}$ the atmospheric pressure at the surface, and the wave set-down forced by the waves labelled $\tilde{\zeta}$:
\begin{empheq}{align}
& \zeta = \zeta^d + \hat{\zeta} \,, && \hat{\zeta} = - \frac{P_{\text{atm}}}{g \rho_0} + \tilde{\zeta} \,, && \tilde{\zeta} = - \frac{H_s^2 k_{p}}{16} \frac{1}{\sinh(2H)} \,.
\end{empheq}

The Stokes velocities are given as follows, with $U^{\text{St}}$ the horizontal depth-averaged Stokes drift:
\begin{empheq}{align}
& \punderline{u}^{\text{St}} = \frac{\sigma_{p} H_s^2\punderline{k}_{p}}{16} \frac{ \cosh(2 k_{p} (h + z))}{\sinh^2 (k_{p} H)} \,, && \punderline{U}^{\text{St}} = \frac{1}{H} \int_{-h}^{\zeta} \punderline{u}^{\text{St}} \, \mathrm{d}z \,. \label{stokes_eq}
\end{empheq}



The bottom boundary condition is not impacted by the waves.
The surface boundary conditions at $z = \zeta$ read as follows, where $\mathcal{P}$ is a wave-averaged forcing term (see \nptextcite{kumar2012implementation} for its expression):
\begin{empheq}{align}
 & w\big|_{\zeta} - \dpar{\zeta^d}{t} - \left( \punderline{u}\big|_{\zeta} \cdot \pgrad \right) \zeta^d  = \pgrad\cdot \left( H\punderline{U}^{\text{St}} \right) + \dpar{\hat{\zeta}}{t} + \left( \punderline{u}\big|_{\zeta} \cdot \pgrad \right) \hat{\zeta} \,, \label{sea_level_eq1}\\
 & g\zeta^d - p\big|_{\zeta} = g\mathcal{P} \label{sea_level_eq2} \,.
\end{empheq}

Depth-integrating the conservation of mass and simplifying with the boundary conditions gives a simple expression for the evolution of the total sea surface elevation, with $\punderline{U}$ the depth-averaged horizontal velocity. The sea level is updated in \cwst{} using this expression:
\begin{empheq}{align}
 & \dpar{\zeta}{t} + \pgrad\cdot\left( H\punderline{U} \right) + \pgrad\cdot\left( H\punderline{U}^{\text{St}} \right) = 0 \,. \label{sea_level_eq3}
\end{empheq}

At the sea bottom $z=-h$, a logarithmic bottom friction model is used to evaluate the momentum bottom stress $\punderline{\tau}_b = (\tau_b^{xz},\tau_b^{yz})$, not shown here:
\begin{empheq}{align}
 & \left( (\nu + K_m) \dpar{u}{z} \right)\bigg|_{-h}= \frac{\tau_b^{xz}}{\rho_0} \,,
 && \left( (\nu + K_m) \dpar{v}{z} \right)\bigg|_{-h} = \frac{\tau_b^{yz}}{\rho_0} \,.
\end{empheq}


\subsubsection{2D equations}

The hydrodynamic code \roms{} uses a split time-stepping algorithm. The 2D depth-averaged equations read as follows:
\begin{empheq}{multline}
 \underbrace{ \vphantom{\int_{-h}^{\zeta}} \frac{1}{H} \dpar{\left(H\punderline{U}\right)}{t}}_\text{accel}
 + \underbrace{ \vphantom{\int_{-h}^{\zeta}} \frac{1}{H} \pgrad \cdot \left[ \int_{-h}^{\zeta} \left(\punderline{u}\otimes\punderline{u}\right) \,\mathrm{d}z \right] + \frac{\punderline{u}|_{\zeta}}{H} \left( \pgrad \cdot \left(H\punderline{U}^{\text{St}}\right) \right)}_\text{hadv} = \underbrace{ \vphantom{\int_{-h}^{\zeta}} \frac{1}{H} \int_{-h}^{\zeta} \left(\punderline{F}^w + \punderline{D}\right) \,\mathrm{d}z}_\text{fwav+hvisc} \\
 \underbrace{ - \vphantom{\int_{-h}^{\zeta}} f \punderline{e}_z\times\punderline{U}}_\text{cor}
 \underbrace{ - \vphantom{\int_{-h}^{\zeta}} f \punderline{e}_z\times\punderline{U}^{\text{St}}}_\text{fsco}
 + \underbrace{ \vphantom{\int_{-h}^{\zeta}} \frac{(\punderline{\tau}_w - \punderline{\tau}_b)}{H}}_\text{sstr+bstr}
 \underbrace{ - \vphantom{\int_{-h}^{\zeta}} \frac{1}{H} \int_{-h}^{\zeta} \left[ \pgrad \left(p + \mathcal{K} \right) + \pgrad \left( \int_{z}^{\zeta} K \,\mathrm{d}z \right) \right] \,\mathrm{d}z}_\text{prsgrd} \\
 + \underbrace{ \vphantom{\int_{-h}^{\zeta}} \frac{1}{H} \int_{-h}^{\zeta} \pgrad \left[ \int_{z}^{\zeta} K \,\mathrm{d}z \right]  \,\mathrm{d}z}_\text{kvrf}
 \underbrace{- \vphantom{\int_{-h}^{\zeta}} \frac{1}{H} \int_{-h}^{\zeta} \left( (\pgrad\times\punderline{u})\times\punderline{u}^{\text{St}} + w^{\text{St}} \dpar{}{z}\punderline{u} \right) \,\mathrm{d}z}_\text{hjvf} \,. \label{2d_momentum}
\end{empheq}

The terms are written so that they match the budget output given by \roms{}. They consist of the acceleration term (accel), the horizontal advection (hadv) which contains the Stokes drift contribution, the Coriolis force (cor), the wave-induced Stokes Coriolis force (fsco), the surface stress (sstr) and bottom stress (bstr), the pressure gradient (prsgrd), the $K$ vortex force (kvrf), the horizontal $J$ vortex force (hjvf), the non-conservative wave-induced acceleration (fwav) and the horizontal viscosity (hvisc).

The pressure gradient can be further decomposed as follows. The pressure solved in \roms{} is the composite pressure $p + \mathcal{K}$, so only the surface Bernoulli head has to be evaluated. Note that the inverse barotropic effect has been separated from the wave set-down effect:
\begin{empheq}{multline}
 \underbrace{ - \vphantom{\int_{-h}^{\zeta}} \pgrad \left(p + \mathcal{K} \right) - \pgrad \left( \int_{z}^{\zeta} K \,\mathrm{d}z \right)}_\text{prsgrd3D} = \\
 \underbrace{ - \vphantom{\int_{-h}^{\zeta}} \pgrad \left[g \zeta + \int_{z}^{\zeta} \frac{g \rho}{\rho_0}\,\mathrm{d}z + \frac{P_{\text{atm}}}{\rho_0} \right]}_\text{zeta3D}
 + \underbrace{ \vphantom{\int_{-h}^{\zeta}} \pgrad (g \tilde{\zeta})}_\text{zetw3D}
 + \underbrace{ \vphantom{\int_{-h}^{\zeta}} \pgrad (g\mathcal{P})}_\text{zqsp3D}
 \underbrace{- \vphantom{\int_{-h}^{\zeta}} \pgrad (\mathcal{K} |_{\zeta})}_\text{zbeh3D} \,. \label{prsgrd}
\end{empheq}

The budget terms on the right hand side are also given by \roms{}. They consist of the Eulerian sea level adjustment (zeta), the quasi-static sea level adjustment (zetw), the quasi-static pressure gradient (zqsp) and the Bernoulli head contribution (zbeh). The vertical integration is still required to obtain the corresponding 2D budget term. It is straightforward for the wave-induced terms which are only evaluated at the surface.

\subsubsection{Non-conservative wave-induced acceleration}

The wave-induced non-conservative force terms enabled for this application consist of the acceleration induced by bottom streaming  $\punderline{B}^{bf}$, by whitecapping $\punderline{B}^{wcap}$ and by depth-induced wave breaking $\punderline{B}^{b}$:
\begin{empheq}{align}
 &  \punderline{F}^w = \punderline{B}^{bf} + \punderline{B}^{sf} + \punderline{B}^{wcap} + \punderline{B}^{b} \,.
\end{empheq}

\paragraph{Bottom streaming}

The interaction of waves with the sea bed dissipates wave energy in the bottom wave boundary layer. The energy dissipated is transferred to the water column and generates bottom streaming, a wave-induced acceleration aligned with the wave propagation. The wave-induced acceleration appears as the wave orbital velocities become slightly in phase, thus generating an additional stress.

The bottom streaming formulation from \textcite{uchiyama2010wave} is used, with a body force proportional to the wave bottom dissipation $\epsilon^{bf}$ evaluated by the parameterization of \textcite{reniers2004vertical}. It decays in the upward direction following the distribution $f^{bf}$ focused at the bottom, with $k_{wd}$ a decay length related to the bottom wave orbital velocity:
\begin{empheq}{align}
 &  \punderline{B}^{bf} = \frac{\epsilon^{bf}}{\rho_0 \sigma} \punderline{k} f^{bf}(z)\,, && f^{bf}(z) = \frac{\cosh{\left[ k_{wd} \left(\zeta - z\right) \right]}}{\int_{-h}^{\zeta} \cosh{\left[ k_{wd} \left(\zeta - z\right) \right]} \,\mathrm{d}z}\,.
\end{empheq}

\paragraph{Whitecapping}

Whitecapping can occur at any water depth. This wave breaking is caused by a wave steepening of the shortest waves, mostly induced by strong winds. The wave model directly computes the dissipation term $\epsilon^{wcap}$. The induced acceleration in the water column is parameterized by the following formulation in \textcite{kumar2012implementation}. It mimics the parameterization used for the bottom friction, with $f^b$ a distribution function focused near the surface:
\begin{empheq}{align}
 & \punderline{B}^{wcap} = \frac{\epsilon^{wcap}}{\rho_0 \sigma_{p}} \punderline{k}_{p} f^{b}(z)\,, && f^{b}(z) = \frac{\cosh{\left[2\sqrt{2}\pi \left(z+h\right) / H_s \right]}}{\int_{-h}^{\zeta} \cosh{\left[2\sqrt{2}\pi \left(z+h\right) / H_s \right]} \,\mathrm{d}z}\,.
\end{empheq}

\paragraph{Depth-induced breaking}

In a similar fashion, the dissipation of energy due to depth-induced breaking $\epsilon^{b}$ is directly computed by the wave model. The associated induced acceleration in the water column is then computed as follows:
\begin{empheq}{align}
 & \punderline{B}^{b} = \frac{\epsilon^{b}}{\rho_0 \sigma_{p}} \punderline{k}_{p} f^{b}(z)\,.
\end{empheq}

\subsubsection{Wave-enhanced mixing}

Wave breaking of any sort enhances the vertical mixing in the water column (\nptextcite{agrawal1992enhanced}). This effect is implemented in \textcite{kumar2012implementation} as a surface boundary condition for the turbulent kinetic energy $k$, following \textcite{feddersen2005effect}, and the option is enabled in the model.


\subsubsection{Wave-enhanced surface roughness}



The bulk parametrization of the surface momentum and heat fluxes involves a surface roughness $z_0$ which accounts for the sea state. It is traditionally evaluated with the Charnock relation, but it has been shown that a constant Charnock coefficient fails to adequately describe several data sets. The formulation from \textcite{drennan2005parameterizing} is implemented in \cwst{}, where the surface roughness is evaluated through the wave age $c_{s,p}/u_*$, with $c_{s,p}$ the wave phase speed at the peak frequency:
\begin{empheq}{align}
 & z_0 = 3.35 H_s \left( \frac{u_*}{c_{p}}\right)^{4.5}\,.
\end{empheq}

A higher wave roughness increases the surface stress and transfer of energy from the air to the ocean. A rough sea is seen to increase the surface roughness which yields stronger wind-induced surface currents.

\subsection{Wave model: \swan{}}

The third generation wave model \swan{} (v41.20AB) is used to model the wave propagation with the \cwst{} modelling system. It solves the wave action balance accounting for current effects. The full derivation of the equations is shown in \textcite{holthuijsen2010waves}. It is worthwhile noting that a vertically uniform current is assumed in the derivation.
The wave action balance is given by Eq. (\ref{wa}) below, with $A(x,y,t, \sigma,\theta)$ the wave action, $(x,y)$ the eastward and northward horizontal coordinates, $t$ the time, $\sigma$ the intrinsic wave frequency, $\theta$ the spectral direction, $\punderline{c}_{g}=(c_{g,x},c_{g,y})$ the intrinsic wave group velocity, $\punderline{U}_{w}=(U_{w,x},U_{w,y})$ the current field passed down to the wave model, $c_{\theta}$ and $c_{\sigma}$ the rates of change of the direction $\theta$ and intrinsic wave frequency $\sigma$ respectively, $\punderline{k}=(k_{x},k_{y})$ the wave vector with $k$ its modulus, $h$ the mean water depth, $\zeta$ the sea surface elevation, and $S$ a global source term:
\begin{empheq}{align}
 & \dpar{A}{t} + \dpar{((c_{g,x} + U_{w,x}) A)}{x} + \dpar{((c_{g,y} + U_{w,y}) A)}{y} + \dpar{(c_\theta A)}{\theta} + \dpar{(c_{\sigma}A)}{\sigma} = \frac{S}{\sigma} \,, \label{wa} \\
 & c_{\theta} = \frac{D \theta}{D t} = - \frac{1}{k} \dpar{\sigma}{H} \bigg[ \left( \dpar{}{\theta} \left(\frac{{\punderline{k}}}{k}\right) \right) \cdot \pgrad (h+\zeta) \bigg] - \frac{\punderline{k}}{k} \cdot \bigg[ \pgrad \punderline{U}_{w} \cdot \left( \dpar{}{\theta}\left(\frac{{\punderline{k}}}{k}\right) \right) \bigg] \,, \label{refraction} \\
 & c_{\sigma} = \frac{D \sigma}{D t} = \dpar{\sigma}{H} \left( \dpar{(h+\zeta)}{t} + \punderline{U}_{w}\cdot\pgrad (h+\zeta) \right) - \punderline{c}_{g}\cdot(\pgrad\punderline{U}_{w}\cdot\punderline{k}) \,, \label{csigma} \\
 & \punderline{c}_{g} = \frac{\punderline{k}}{2k} \left(1 + \frac{2kH}{\sinh{2kH}}\right) \sqrt{\frac{g}{k} \tanh{kH}} \label{advection} \,.
\end{empheq}
The relative variables are all derived from the dispersion relation where the absolute wave frequency accounts for the Doppler shift due to currents:
\begin{empheq}{align}
 & \omega = \sigma + \punderline{U}_{w}\cdot\punderline{k} \,, && \sigma^2 = gk\tanh(kH) \,. \label{doppler}
 \end{empheq}
 
\subsubsection{Wave source terms}
 
The source terms used in this application are gathered in $S$ and can be divided into the following expression. $S_{nl}$ is the nonlinear wave interaction term using the standard Discrete Interaction Approximation from \textcite{hasselmann1985computations}. $S_{in}$ is the wind wave growth input term using the formulation from \textcite{yan1987improved}. $S_{ds,w}$ is the wave dissipation due to whitecapping modelled from \textcite{van2007nonlinear}, $S_{ds,b}$ is the bottom friction (\nptextcite{madsen1989spectral}) and $S_{ds,br}$ is the depth-induced wave breaking (\nptextcite{battjes1978energy}):
\begin{empheq}{align}
 & S = S_{nl} + S_{in} + S_{ds,wh} + S_{ds,bf} + S_{ds,br} \,. \label{sources}
\end{empheq}

\subsubsection{Effects of the ocean model on the wave propagation}

We briefly highlight here the different effects of the coupling for the wave propagation. Most of those effects have already been mentioned in the introduction.

\paragraph{Total sea level}

In the previous equations describing the wave propagation the total sea level is adjusted by the mean sea level computed in the hydrodynamics model. In nearshore locations, it induces a significant change, and notably modifies the depth-induced dissipative processes. During low tides more friction and more wave breaking is expected.

\paragraph{Current }

The dynamical coupling sends a current field to the wave model. The depth-integrated current of \textcite{kirby1989surface} is used, which accounts for the vertical shear in the water column. The wave action balance is derived for a vertically uniform current. With a more realistic sheared current those equations can still be used with $\punderline{U}_{w}$ an effective current field, which in reality is a higher order correction for the wave group velocity. Since this computation is done in the ocean model, an approximation is made by using the peak wavenumber $k_{p}$:
\begin{empheq}{align}
 & \punderline{U}_{w} = \frac{2 \punderline{k}_{p}}{\sinh{2k_{p}h}} \int_{-h}^\zeta \punderline{u}(z) \cosh{2k_{p}(h+z)} \,\mathrm{d}z \,. \label{eq:5}
\end{empheq}

The current field passed down to \swan{} mostly impacts the advection through Eq. (\ref{advection}). It induces additional refraction through Eq. (\ref{refraction}) and redistributes energy between the frequency bins through Eq. (\ref{csigma}). The contribution of currents to wave advection mostly impacts young waves characterised by higher frequencies and lower phase speed and group velocity. Currents can block the wave energy from propagating if they are strong enough to overcome the wave group velocity. The current-induced refraction should impact equally all waves, regardless of their age or frequency. However, the effect is spatially more coherent for narrow-banded swell systems.

\paragraph{Doppler shift}

The Doppler shift directly induced by Eq. (\ref{doppler}) can greatly modify the wave frequencies, especially for locally generated wind waves which are characterised with higher frequencies. This is because of the wave-number multiplying the current term $\punderline{U}_w$: young waves usually have high wavenumbers whereas swell systems have shorter wavenumbers.

\subsection{Summary of the runs tested}

In order to highlight the specific influence of the different processes introduced previously several run configurations are compared in this paper. The most relevant ones are gathered in Table \ref{runs}.
Three time periods are run, from $2017/01/01$ to $2017/03/27$ (labelled 1701), from $2017/05/04$ to $2017/07/28$ (labelled 1705) and from $2018/05/10$ to $2017/08/03$ (labelled 1805). The baroclinic time-step for \roms{} is set to $15 \,\mathrm{s}$ with a $1 \,\mathrm{s}$ barotropic time-step. The time-step for \swan{} is set to $3\,\si{\min}$ and the same time-step is set for the coupling between the models.

\begin{table}
\begin{center}
\resizebox{0.9\textwidth}{!}{%
\begin{tabular}{l  c  c  c  c  c  c  c  c  c c c c c c}
\toprule
Runs   & \multicolumn{5}{l}{\swan{} options}                & \multicolumn{8}{l}{\roms{} options} \\
       & \swan{} & Wind & Swell & Currents & Sea level     & \roms{} & Atmos & VF & BStr & WRough & Br & Wh & TKE \\
\midrule
ROMS-0 &         &      &       &  &                         & X       & X      &    &      &          &     &    & \\
\midrule
ROMS-T &         &      &       &  &                         & X       &        &    &      &          &     &    & \\
\midrule
SWAN-0 & X       & X    & X     &  &                         &         &        &    &      &           &     &    & \\
\midrule
CPL-0  & X       & X    & X     & X  & X                    & X       & X      & X  & X    & X          & X   & X  & X \\
\midrule
CPL-0S  & X      &      & X     & X & X                     & X       & X      & X  & X    & X          & X   & X  & X \\
\midrule
CPL-0W  & X      & X    &       & X & X                     & X       & X      & X  & X    & X          & X   & X  & X \\
\midrule
CPL-2  & X       & X    & X     & X  & X                    & X       & X      & X  & X    &            & X   & X  & X \\
\midrule
CPL-3A & X       & X    & X     & X  & X                    & X       & X      & X  & X    & X          &     & X  & X \\
\midrule
CPL-3B & X       & X    & X     & X  & X                    & X       & X      & X  & X    & X          & X   &    & X \\
\midrule
CPL-3C & X       & X    & X     & X  & X                    & X       & X      & X  & X    & X          & X   & X  &   \\
\midrule
CPL-4  & X       & X    & X     & X  & X                    & X       & X      & X  &      & X          & X   & X  & X \\
\midrule
CPL-5  & X       & X    & X     & X  & X                    & X       & X      & X  &      &            &     &    & \\
\bottomrule
\end{tabular}}

\caption{Summary of the different run configurations used in this paper. A cross means that the corresponding process is included in the model using the formulation described in the paper.
The first columns labelled \swan{} and \roms{} mean that the respective model is enabled in the run.
VF stands for vortex force, Bstr for bottom streaming, WRough for wave-induced roughness, Br for depth-induced breaking, Wh for whitecapping and TKE for turbulent kinetic energy.}
\label{runs}
\end{center}
\end{table}

\subsection{Computational grid}

The same computational grid is used for both \roms{} and \swan{}. The specifications are reported in Table \ref{specs}. The domain is shown in Figure \ref{irishshelf}, along with the bathymetry and location of the different stations.
Compared to other studies focused on nearshore processes (\nptextcite{olabarrieta2011wave,bennis2020numerical}), the resolution here is quite coarse. However, several studies (\nptextcite{osuna2005numerical,benetazzo2013wave,wu2019wave}) used even coarser domain resolutions without issues, focusing on more offshore processes.
The dimensions of the grid are a constraint of this problem. Since there is a possibility for operational applications, the wall-clock time must be kept low. The grid used here also mimics the domain already used by the MI for their standalone \roms{} operational model of Galway Bay.

The vertical grid contains $20$ levels. A terrain-following $\sigma$ transformation is applied refining the levels at the surface in order to better capture the wave-ocean and air-ocean interaction at the surface.
A coastline file from Ordnance Survey Ireland is used to generate the mask (\url{https://data.gov.ie/dataset/}). The grid is technically curvilinear but given the squared boundary it is very close to a regular grid.

\begin{table}
\begin{center}
\begin{tabular}{l c c c c}
\toprule
  Direction & Vertices & Averaged resolution & Min value & Max value  \\
\midrule
  Longitude & $641$ & ${2.98e^{-3}}^{\circ} \approx 220 \,\si{\m}$ & $-10.80^{\circ}$ & $-8.90^{\circ}$ \\
  Latitude  & $441$ & ${1.78e^{-3}}^{\circ} \approx 220 \,\si{\m}$ & $52.95^{\circ}$  & $53.73^{\circ}$ \\
\bottomrule
\end{tabular}
\caption{Horizontal grid specifications for the Connemara model grid.}
\label{specs}
\end{center}
\end{table}

\subsection{Forcing fields}

All the input data used in our model is mentioned here for clarity, along with possible processing that was deemed necessary.

\subsubsection{Bathymetry}

The bathymetry put in the model is mostly relying on the Integrated Mapping for the Sustainable Development of Ireland’s Marine Resource (\infomar{}), a 20 year programme to map Ireland’s seabed. It includes a great number of boat and LIDAR surveys. The resolution of those surveys ranges from $1\,\si{\m}$ to $7\,\si{\m}$. Following the best practices for processing bathymetry data reviewed in \textcite{sikiric2009new}, a reduction step is first conducted by averaging the raw bathymetry data around the computational grid points of the domain.

The available surveys do not cover fully the domain. The gaps are filled with the latest version of \gebco{} (\nptextcite{tozer2019global}). It is a processed product offering a global bathymetry on a regular $1/4$ arc minute resolution grid. Different sources are gathered and merged together in \gebco{}. Most of the data for Ireland comes for the \infomar{} surveys already mentioned. They use the global $1$ arc minute resolution satellite altimetry data \etopo{} (\nptextcite{amante1arc}) to fill the gaps left by \infomar{}. Using this processed product over \etopo{} is preferred as it includes slightly more sources than only the satellite data.

Additional processing is conducted in order to merge adequately the different sources and prevent instabilities in the model. Given the tidal range observed, a minimal depth of $5\,\si{\m}$ is then imposed in the domain, to ensure that the wet cells stay wet in all circumstances. Local and global smoothing is then applied to reduce the pressure gradient errors.

%

\subsubsection{Atmospheric forcing}

The atmospheric forcing is crucial either for ocean or wave forecasting. The most refined product available for Ireland is the Met \'Eireann reanalysis \mera{} (\nptextcite{whelan2018evaluation}). It covers the period $1981-2018$ and continues in real time with a one year delay. The model runs an atmospheric model and assimilates data from different sources. It is forced at the boundary by \eraint{}.

The resolution of \mera{} is $2.5 \,\si{\km}$ by $2.5 \,\si{\km}$, using a curvilinear rotated grid aligned with the main wind direction. Hourly output is available for all the necessary variables needed to force the wave model and the ocean model. Those include the wind at $10\,\si{\m}$, the surface air pressure, the air temperature and relative humidity at $2\,\si{\m}$, the net short-wave solar radiation and net long-wave radiation, the total precipitation.

\subsubsection{Lateral boundary conditions}

For \swan{}, the lateral boundary conditions come from a multi-grid large-scale \ww{} model described and validated in \textcite{calvino2021current}. Nested grids refine the model on the Irish Shelf. The \ww{} is forced by the \era{} winds (\nptextcite{hersbach2020era5}), and no boundary conditions are necessary as the coarser grid encompasses all of the Atlantic Ocean. The full spectrum is provided every $10\,\si{\min}$ to capture as best as possible the directional variability in the swell systems entering the domain. 

The boundary conditions for \roms{} consist of the surface elevation and horizontal velocities. They come from a larger scale North East Atlantic \roms{} model (\nearoms{}) run by the MI covering the Irish Shelf (\nptextcite{hazem2020regional}). It is itself forced by the GLOBAL\_ANALYSIS\_\-FORECAST\_PHY\_\-001\_024 model (\nptextcite{PHY_001_024}) for the velocities, the OSU TPX08 model (\nptextcite{egbert2002efficient}) for the tides, and the hourly \ecmwf{} High Resolution Forecasts (\url{https://www.ecmwf.int/en/forecasts/datasets/set-i}) for the atmospheric forcing. Output is extracted every $10\,\si{\min}$ at the exact boundary of the Connemara model.

\subsubsection{River discharges}

Three rivers are included in our application: the River Corrib discharging at Galway City and the Rivers Clarin and Kilkieran discharging in the east end of Galway Bay (see Figure \ref{galwaybay}). Daily climatology data computed from the records held by the Office of Public Works in Ireland is used for the discharge.

\subsection{In situ data}

For the purpose of validating the model different in-situ observations are used. Their location is shown in Figure \ref{galwaybay} by the color dots. They consist of two tidal gauges (red dots) installed in Galway Port and Rossaveal Pier, the three ADCPs Berth A, Berth B and Berth C, all deployed in the East part of the bay (blue dots), one ADCP deployed near Inishmaan (yellow dot), one ADCP at the SmartBay location together with a wave-buoy (green dot). The time periods available for each station, along with their exact location and water depth, are gathered in Table \ref{stations}. All the data is freely available either directly from the MI servers, or upon request.

\begin{table}
\begin{center}
\resizebox{0.9\textwidth}{!}{%
\begin{tabular}{l l l l l l}
\toprule
 Station                 & Lon. $(^{\circ})$ & Lat. $(^{\circ})$ & Sampling $(\si{\min})$ & Depth $(\si{\m})$ & Record window  \\
\midrule
 1701 Galway port        & $-9.048$ & $53.269$ & $6$  & NA                  & $2017/01/01$ to $2017/03/27$ \\
 1705 Galway port        & $-9.048$ & $53.269$ & $6$  & NA                  & $2017/05/04$ to $2017/08/01$ \\
 1805 Rossaveal pier     & $-9.562$ & $53.267$ & $5$  & NA                  & $2018/05/10$ to $2018/08/04$ \\
 1805 Berth A      & $-9.052$ & $53.229$ & $12$ & $13.64$ & $2018/05/15$ to $2018/07/08$  \\
 1805 Berth B      & $-9.065$ & $53.201$ & $12$ & $18.17$ & $2018/05/15$ to $2017/07/17$  \\
 1805 Berth C      & $-8.997$ & $53.182$ & $12$ & $11.78$ & $2018/05/15$ to $2017/07/17$  \\
 1705 SmartBay ADCP      & $-9.266$ & $53.227$ & $15$ & $24.52$ & $2017/05/27$ to $2017/07/12$ \\
 1701 SmartBay Wave buoy & $-9.268$ & $53.228$ & $30$ & $24.52$ & $2017/01/01$ to $2017/04/01$ \\
 1705 SmartBay Wave buoy & $-9.268$ & $53.228$ & $30$ & $24.52$ & $2017/04/18$ to $2017/07/27$ \\
 1701 Inishmaan ADCP     & $-9.627$ & $53.067$ & $12$ & $45.48$ & $2017/01/05$ to $2017/03/26$ \\
\bottomrule
\end{tabular}
}
\caption{List of the stations where in-situ data is available.}
\label{stations}
\end{center}
\end{table}

The tide gauges only record the surface elevation variation from a mean level. The datum used for the tidal gauges in Ireland is the Malin Head vertical datum, which is the mean sea level of the tide gauge at Malin Head in County Donegal. In order to avoid any bias issue linked to the reference level taken when comparing model output and observations the mean value of each time series is subtracted, only keeping the surface anomaly. This is relevant for the tidal gauges but also for the ADCPs, where the surface elevation can also be inferred.

The data available for the three Berth ADCPs and SmartBay ADCP is already processed and consists of the sea surface elevation and velocities for the part of the water column in the range of the ADCP. Additional smoothing is achieved by averaging the signal over $30\,\si{\min}$ intervals for the SmartBay ADCP and $48\,\si{\min}$ intervals for the Berth ADCPs, keeping the same sampling rate. This is done to reduce the turbulence captured by the instrument, which does not appear in the ocean model.
The ADCP was deployed near Inishmaan in early $2017$ with a $2 \,\mathrm{Hz}$ sampling rate. The original motivation for the deployment was to record and analyse high frequency time series for the surface elevation (\nptextcite{fedele2019large}). Both wave measurements and current measurements are available at this location. Raw data is available for this ADCP. It is first processed into $12 \,\si{\min}$ time-averaged spectrum time series, and then into the desired mean wave parameters.
The SmartBay wave buoy measures a wide range of mean wave parameters such as the significant wave height, peak wave direction and peak period.

The instruments were not all recording constantly. Model output data is truncated to match accordingly the observation. The few missing data in the time series are filled by linear interpolation. This is necessary for the tidal analysis to ensure a constant sampling rate throughout the whole series. Spectral output for \swan{} and three dimensional hydrodynamic variables for \roms{} are requested at the location of the stations every $10 \,\si{\min}$ in the model. The mean wave parameters of interest are then computed from the spectral output to match the exact quantity recorded by the station.

\section{Validation of the model}

The validation of the model is conducted by comparing the solution from the standalone runs ROMS-0 and SWAN-0 and the coupled model CPL-0 against available stations. The goal is also to observe the expected improvement coming with the coupling.

The statistical definitions are given in Appendix \ref{statistics}. All the directions refer to the nautical convention, and are expressed in degrees. It means that they are reported as the direction coming from, with a clockwise rotation and the origin is coming from the North. The time is always reported in Coordinated Universal Time (UTC).

\subsection{Sea surface elevation}

A constant bias in terms of surface elevation is observed in the model. It reduces the sea level by $20 \mathrm{-} 35\,\si{\cm}$ everywhere in the domain. This a legacy from the boundary conditions used. It was decided not to fix the boundary conditions as it would also break the consistency between the barotropic velocities and the surface elevation at the boundaries. Instead the mean values are removed from both the observation and the model data. As a result, static or quasi-static effects over duration longer than a few months will not be captured, but it doesn't exclude wave effects characterised by short time scales. However it forces the validation in terms of spectral analysis using a tidal analysis to evaluate how well the model propagates the different tidal components.

\subsubsection{Tidal analysis}

A tidal analysis covering all the stations recording the surface elevation is conducted. The model is forced at the boundary by the MM, MF, Q1, O1, P1, K1, N2, M2, S2, K2 constituents. However, because of the Rayleigh criteria, the P1 and K2 constituents are removed. The Matlab tool described in \textcite{pawlowicz2002classical} is used. The observation data sets are linearly filled for missing values when needed.
For clarity Table \ref{ta_obs} first includes the output of the tidal analysis on the observed data. The errors between the models, either standalone or coupled, and the observations are then shown in Table \ref{ta_errors} for both the tidal amplitude and Greenwich phase.

The M2 constituent dominates for all the stations, followed by the S2 and N2 constituents. The remaining constituents are of second importance with amplitudes one order of magnitude smaller.
It can be noticed that the two tidal analyses done for Galway Port during different time periods give slightly different outputs with a difference of $4 \,\si{\cm}$ for the M2 constituent, and $14 \,\si{\cm}$ for the S2 constituent. This is expected from conducting a tidal analysis over only three months.
The two tidal analyses for the Galway Port tidal gauge and the ADCP deployed near Inishmaan during the $1701$ period show very similar tidal constituents, both in amplitude and phase. The same goes for the Galway Port and SmartBay tidal analyses during the $1705$ period. The three outputs for the Berth ADCPs show excellent consistency, which is coherent since those three ADCPs were deployed around the same area and during the same period of time.

Looking now at Table \ref{ta_errors}, there is a clear indication that the wave coupling has a marginal impact on the tidal propagation in the model. The errors are similar for both the standalone and the coupled runs. There is also a good agreement between the model and the observations, especially against the three tidal gauge records with a maximum error of $6 \,\si{\cm}$ for the principal M2 component. Slightly higher errors are observed for the five ADCPs, up to $8 \,\si{\cm}$, but they remain reasonable.

\begin{table}
\begin{center}
\resizebox{0.9\textwidth}{!}{%
\begin{tabular}{l l l  l  l  l  l  l l l}
\toprule
                   &                    & MM     & MF     & Q1     & O1     & K1     & N2     & M2     & S2    \\
\midrule
1701 Galway Port   & Amplitude $(\si{\m})$      & $0.11$ & $0.04$ & $0.02$ & $0.07$ & $0.10$ & $0.27$ & $1.53$ & $0.60$ \\
                   & Phase $(^\circ)$     & $295$  & $146$  & $226$  & $322$  & $ 87$  & $113$  & $138$  & $181$  \\
1701 Inishmaan     & Amplitude $(\si{\m})$      & $0.10$ & $0.01$ & $0.01$ & $0.07$ & $0.09$ & $0.25$ & $1.44$ & $0.56$ \\
                   & Phase $(^\circ)$     & $284$  & $300$  & $217$  & $324$  & $ 89$  & $113$  & $138$  & $181$  \\
1705 Galway Port   & Amplitude $(\si{\m})$      & $0.07$ & $0.04$ & $0.02$ & $0.07$ & $0.13$ & $0.36$ & $1.51$ & $0.45$ \\
                   & Phase $(^\circ)$     & $196$  & $215$  & $277$  & $319$  & $ 63$  & $117$  & $140$  & $171$  \\
1705 SmartBay      & Amplitude $(\si{\m})$      & $0.03$ & $0.02$ & $0.02$ & $0.07$ & $0.14$ & $0.35$ & $1.49$ & $0.41$ \\
                   & Phase $(^\circ)$     & $224$  & $233$  & $258$  & $315$  & $ 64$  & $111$  & $137$  & $169$  \\
1805 Rossaveal     & Amplitude $(\si{\m})$      & $0.06$ & $0.05$ & $0.02$ & $0.07$ & $0.14$ & $0.27$ & $1.49$ & $0.42$ \\
                   & Phase $(^\circ)$     & $207$  & $252$  & $284$  & $322$  & $ 68$  & $137$  & $146$  & $179$  \\
1805 Berth A & Amplitude $(\si{\m})$      & $0.01$ & $0.04$ & $0.02$ & $0.07$ & $0.14$ & $0.34$ & $1.54$ & $0.44$ \\
                   & Phase $(^\circ)$     & $234$  & $204$  & $276$  & $316$  & $ 59$  & $124$  & $138$  & $164$  \\
1805 Berth B & Amplitude $(\si{\m})$      & $0.03$ & $0.02$ & $0.02$ & $0.07$ & $0.14$ & $0.34$ & $1.51$ & $0.43$ \\
                   & Phase $(^\circ)$     & $200$  & $207$  & $283$  & $316$  & $ 60$  & $124$  & $137$  & $166$  \\
1805 Berth C & Amplitude $(\si{\m})$      & $0.02$ & $0.02$ & $0.02$ & $0.07$ & $0.14$ & $0.34$ & $1.52$ & $0.43$ \\
                   & Phase $(^\circ)$     & $204$  & $211$  & $279$  & $315$  & $ 60$  & $124$  & $138$  & $167$  \\
\bottomrule
\end{tabular}}
\caption{Tidal constituents for the different records used in the validation, including tidal gauges and ADCP measurements.}
\label{ta_obs}
\end{center}
\end{table}

\begin{table}
\begin{center}
\resizebox{0.9\textwidth}{!}{%
\begin{tabular}{l l l l l l l l l l}
\toprule
\multicolumn{1}{l}{Amplitude error $(\si{\m})$} & MM     & MF     & Q1     & O1     & K1     & N2     & M2     & S2     \\
\midrule
1701 Galway Port   & ROMS-0  & $0.00$ & $0.02$ & $0.01$ & $-0.01$ & $0.01$ & $0.06$ & $0.06$ & $0.01$ \\
                   & CPL-0   & $0.00$ & $0.02$ & $0.01$ & $-0.01$ & $0.01$ & $0.06$ & $0.06$ & $0.01$ \\
1701 Inishmaan     & ROMS-0  & $0.00$ & $0.03$ & $0.02$ & $-0.01$ & $0.01$ & $0.05$ & $0.03$ & $-0.01$ \\
                   & CPL-0   & $0.00$ & $0.03$ & $0.02$ & $-0.02$ & $0.01$ & $0.05$ & $0.03$ & $-0.01$ \\
1705 Galway Port   & ROMS-0  & $-0.02$ & $0.01$ & $0.00$ & $-0.02$ & $0.01$ & $-0.02$ & $0.08$ & $0.05$ \\
                   & CPL-0   & $-0.02$ & $0.01$ & $0.00$ & $-0.02$ & $0.01$ & $-0.02$ & $0.08$ & $0.05$ \\
1705 SmartBay      & ROMS-0  & $0.02$ & $0.03$ & $-0.01$ & $-0.02$ & $0.00$ & $-0.03$ & $0.06$ & $0.07$ \\
                   & CPL-0   & $0.02$ & $0.03$ & $-0.01$ & $-0.02$ & $0.00$ & $-0.03$ & $0.06$ & $0.07$ \\
1805 Rossaveal     & ROMS-0  & $-0.03$ & $0.00$ & $0.00$ & $-0.02$ & $0.00$ & $0.04$ & $0.03$ & $0.04$ \\
                   & CPL-0   & $-0.03$ & $0.00$ & $0.00$ & $-0.02$ & $0.00$ & $0.04$ & $0.03$ & $0.04$ \\
1805 Berth A & ROMS-0  & $0.02$ & $0.01$ & $0.00$ & $-0.02$ & $0.01$ & $-0.01$ & $0.05$ & $0.04$ \\
                   & CPL-0   & $0.02$ & $0.01$ & $0.00$ & $-0.02$ & $0.01$ & $-0.01$ & $0.05$ & $0.04$ \\
1805 Berth B & ROMS-0  & $-0.01$ & $0.02$ & $0.00$ & $-0.01$ & $0.01$ & $-0.01$ & $0.08$ & $0.05$ \\
                   & CPL-0   & $-0.01$ & $0.02$ & $0.00$ & $-0.01$ & $0.01$ & $-0.01$ & $0.07$ & $0.05$ \\
1805 Berth C & ROMS-0  & $0.00$ & $0.03$ & $0.00$ & $-0.02$ & $0.01$ & $-0.01$ & $0.08$ & $0.06$ \\
                   & CPL-0   & $0.00$ & $0.03$ & $0.00$ & $-0.02$ & $0.01$ & $-0.02$ & $0.08$ & $0.05$ \\
\midrule
\multicolumn{1}{l}{Phase error $(^\circ)$} & MM     & MF     & Q1     & O1     & K1     & N2     & M2     & S2     \\
\midrule
1701 Galway Port   & ROMS-0  & $-14$ & $  2$ & $-2$ & $-2$ & $  2$ & $  4$ & $-1$ & $-2$ \\
                   & CPL-0   & $-15$ & $  3$ & $-3$ & $-1$ & $  2$ & $  4$ & $-1$ & $-2$ \\
1701 Inishmaan     & ROMS-0  & $-1$ & $-152$ & $  9$ & $-5$ & $  1$ & $  2$ & $-2$ & $-4$ \\
                   & CPL-0   & $-1$ & $-150$ & $  8$ & $-5$ & $  1$ & $  2$ & $-2$ & $-4$ \\
1705 Galway Port   & ROMS-0  & $-2$ & $-16$ & $-13$ & $ 0$ & $  6$ & $ 0$ & $-3$ & $-4$ \\
                   & CPL-0   & $-1$ & $-14$ & $-14$ & $ 0$ & $  6$ & $ 0$ & $-3$ & $-4$ \\
1705 SmartBay      & ROMS-0  & $-36$ & $-39$ & $  5$ & $  4$ & $  6$ & $  4$ & $-1$ & $-2$ \\
                   & CPL-0   & $-36$ & $-38$ & $  4$ & $  4$ & $  6$ & $  4$ & $-1$ & $-2$ \\
1805 Rossaveal     & ROMS-0  & $  4$ & $-7$ & $-32$ & $-4$ & $  4$ & $-19$ & $-8$ & $-10$ \\
                   & CPL-0   & $  5$ & $-8$ & $-33$ & $-4$ & $  4$ & $-19$ & $-8$ & $-10$ \\
1805 Berth A & ROMS-0  & $-15$ & $ 45$ & $-24$ & $  1$ & $ 13$ & $-7$ & $-1$ & $  3$ \\
                   & CPL-0   & $-14$ & $ 44$ & $-25$ & $  1$ & $ 13$ & $-7$ & $-1$ & $  3$ \\
1805 Berth B & ROMS-0  & $ 18$ & $ 42$ & $-30$ & $  1$ & $ 12$ & $-7$ & $-1$ & $  1$ \\
                   & CPL-0   & $ 19$ & $ 41$ & $-30$ & $  1$ & $ 12$ & $-7$ & $-1$ & $  1$ \\
1805 Berth C & ROMS-0  & $ 14$ & $ 38$ & $-26$ & $  2$ & $ 12$ & $-7$ & $-1$ & $  1$  \\
                   & CPL-0   & $ 16$ & $ 37$ & $-26$ & $  2$ & $ 12$ & $-7$ & $-1$ & $  1$ \\
\bottomrule
\end{tabular}}
\caption{Tidal constituent errors between the \roms{} standalone ROMS-0 run or \roms{} and \swan{} coupled CPL-0 run and the observations, for the different records used in the validation (computed as model minus observation).}
\label{ta_errors}
\end{center}
\end{table}

\subsubsection{Storm surges}

The sea surface elevation in coastal areas is globally explained by the tidal signal, but non-tidal processes can occasionally greatly impact the sea surface. Severe coastal flooding is often caused by a combination of wind, tide and atmospheric pressure. Accurately capturing those events depends greatly on the accuracy and resolution of the atmospheric forcing used.

\begin{figure}
  \centering \includegraphics[width=10cm]{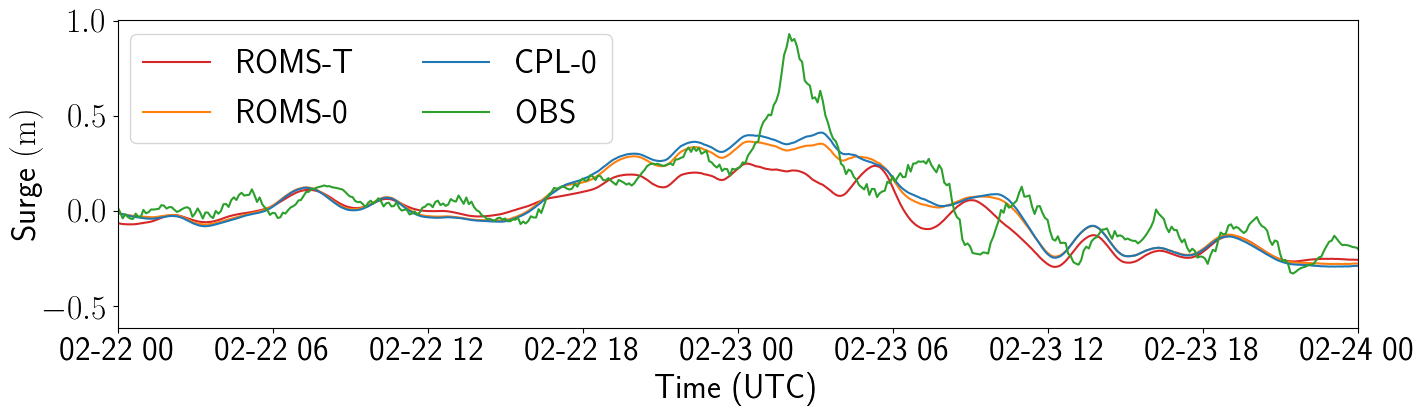}
  \centering \includegraphics[width=10cm]{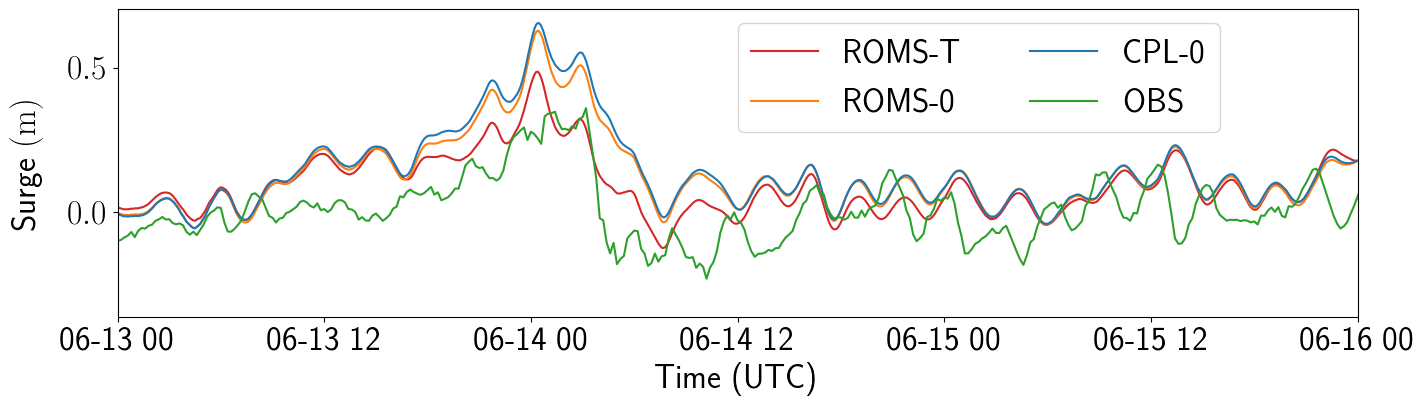}
  \caption{Detided signals at the Galway Port tidal gauge location during Storm Doris ($2017/02/23$) on the top and at the Berth A ADCP location during Storm Hector ($2018/06/14$) on the bottom. Model runs include \roms{} without atmospheric forcing (ROMS-T), \roms{} with atmospheric forcing (ROMS-0) and \roms{} coupled with \swan{} (CPL-0). The tidal only run ROMS-T still captures some part of the surge. Local atmospheric effects and current effects both contribute to increase the surge.}
  \label{surge}
\end{figure}

The surge signal is extracted by computing the detided signal. Figure \ref{surge} shows the surge in Galway Port during Storm Doris ($2017/02/23$), and the surge at the Berth A location during Storm Hector ($2018/06/14$). 
For reference, the output from a \roms{} standalone run without any atmospheric forcing is included as well in those figures, only capturing the tidal signal.
Despite the detidal processing, residual errors are still left in the signals. This is partly due to the refined temporal resolution of the records ($6\,\si{\min}$) and the relatively short observation duration (around $3$ months). It was still decided not to filter or average any further the signal as it would result in damping the surge in the process.

It is interesting to note that in both cases the run with only tidal forcing already reasonably picks up the surge. We believe that the forcing boundary conditions already capture a good portion of the atmospheric impact on the sea level. It also explains why sea surface anomalies are still observed with run ROMS-T despite not modelling anything else than the tidal propagation.
The capability of the model to capture accurately surges seems to depend significantly on the capabilities of the parent model used to generate the boundary conditions.
For Storm Doris the model fails to capture accurately the strong surge which reaches almost $1 \,\si{\m}$, whereas for Storm Hector the model over-predicts the surge at the Berth A ADCP location.

In the case of Storm Doris, ROMS-T predicts a $20 \,\si{\cm}$ surge. This is increased by $9 \,\si{\cm}$ ($45\%$ increase) when locally taking into account the atmospheric forcing, and increased by another $3\,\si{\cm}$ ($10\%$) with the action of waves in CPL-0. For Storm Hector, ROMS-T predicts a $49 \,\si{\cm}$ surge, increased by $10 \,\si{\cm}$ ($20\%$ increase) for run ROMS-0, and increased by another $3\,\si{\cm}$ ($5\%$) with CPL-0.
Although the overall performance of the model in each case is different, the local atmospheric forcing and wave coupling consistently increase the amplitude of the surge. The atmospheric forcing in the Connemara model has a higher resolution than in the parent model, which explains a better capture of the local low pressure less spatially averaged, giving a stronger inverse barometric effect.
The positive impact of currents will be studied below. It still corresponds to a significant increase in sea level, which is of the same order as that observed in \textcite{wu2019wave} for a different application in the Baltic Sea.

\subsection{Velocities}

The depth-integrated barotropic velocities are compared between the models and the ADCP measurements. A careful manual integration is conducted for the model data, in order to integrate the same portion of the water column captured by the measurements. Indeed, sea bed deployed ADCPs usually cannot measure the first few meters at the bottom. At every time-step the actual portion within the ADCP measurement range is used to integrate the velocities in the model. In practice it contributes to slightly increase the magnitude of the barotropic velocities computed since the smaller near-bed velocities are truncated.

\subsubsection{Overall statistics}

\begin{figure}
  \centering \includegraphics[width=10cm]{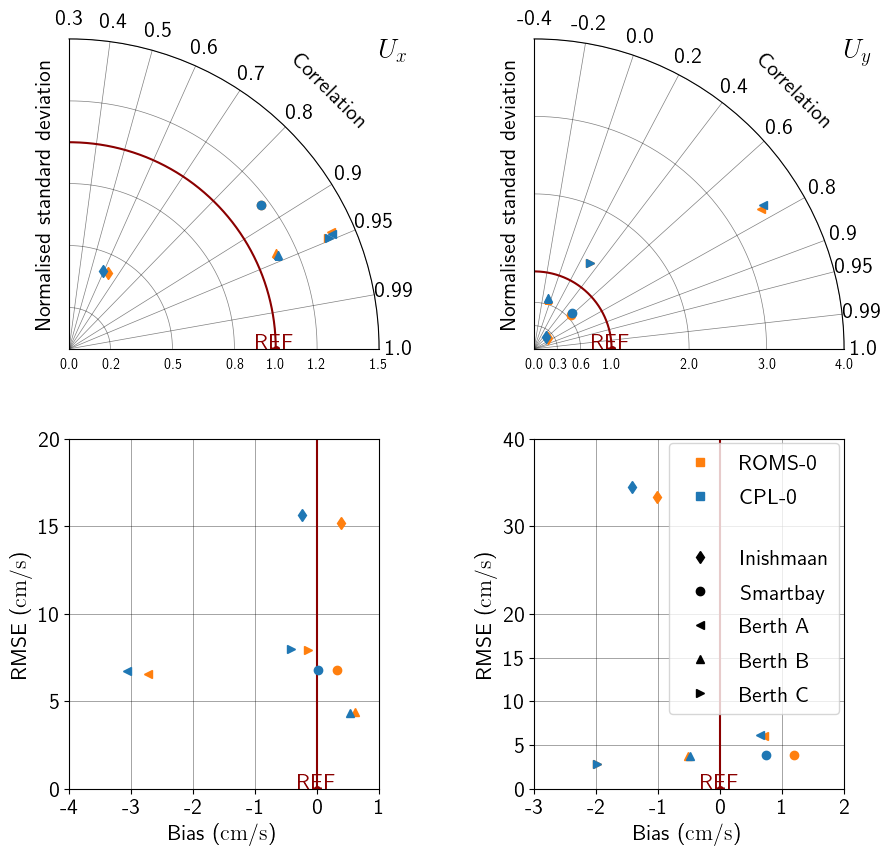}
  \caption{Taylor and RMSE versus bias diagrams for the barotropic eastward velocities (left, UBAR) and northward velocities (right, VBAR), comparing \roms{} standalone and \roms{} coupled with \swan{} (Run 0) against measurements from five ADCPs.}
  \label{taylor_adcp}
\end{figure}

The statistical agreement is shown in Figure \ref{taylor_adcp}. The model clearly fails to capture correctly the velocities at the location of the ADCP deployed near Inishmaan. For the other four ADCPs the agreement is reasonable. Looking at the correlation coefficients, the eastward components are well reproduced by the model, especially for the three Berth ADCPs. On the other hand poor agreement is found for the northward components.
The biases are around $0.5 \,\si{\cm\per\s}$ which is overall small compared to the amplitude of the signals around $20 \mathrm{-} 40 \,\si{\cm\per\s}$. The currents are mostly tide driven and mean values are centered around $0$, so the bias is not the best statistics to evaluate the accuracy of the model. 
Except for the Inishmaan ADCP, the root-mean-square errors are in-between $3 \mathrm{-} 8 \,\si{\cm\per\s}$ for both components.
The poor correlation observed in terms of northward velocities does not reflect on the corresponding biases or root-mean-square errors. Indeed it corresponds to the weaker component of the flow.
Comparing with the literature, those errors are of the same order as those obtained in \textcite{ren2017effect}, namely around $4.5 \mathrm{-} 6 \,\si{\cm\per\s}$, where a different numerical model is used for Galway Bay.

The effect of waves on the currents is overall small. Depending on the station and the component picked, the correlation is either improved or deteriorated by less than $0.5\%$.
A consistent negative bias on the eastward component still appears for all the stations, around $-0.3\,\si{\cm\per\s}$. A similar bias is also noticed for the northward component of the flow at the SmartBay location only.

\subsubsection{Time series}

\begin{figure}
  \centering \includegraphics[width=10cm]{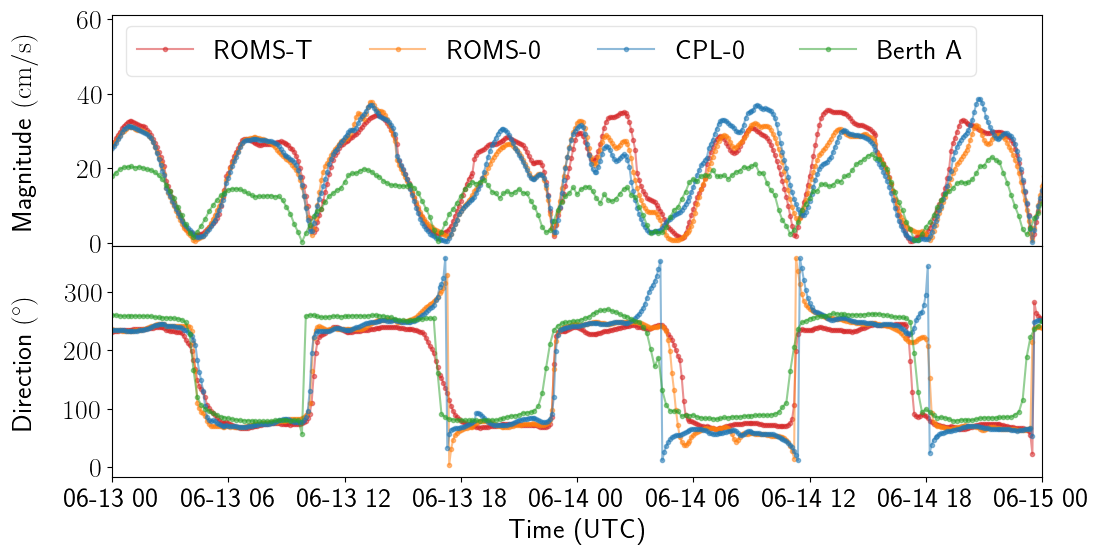}
  \centering \includegraphics[width=10cm]{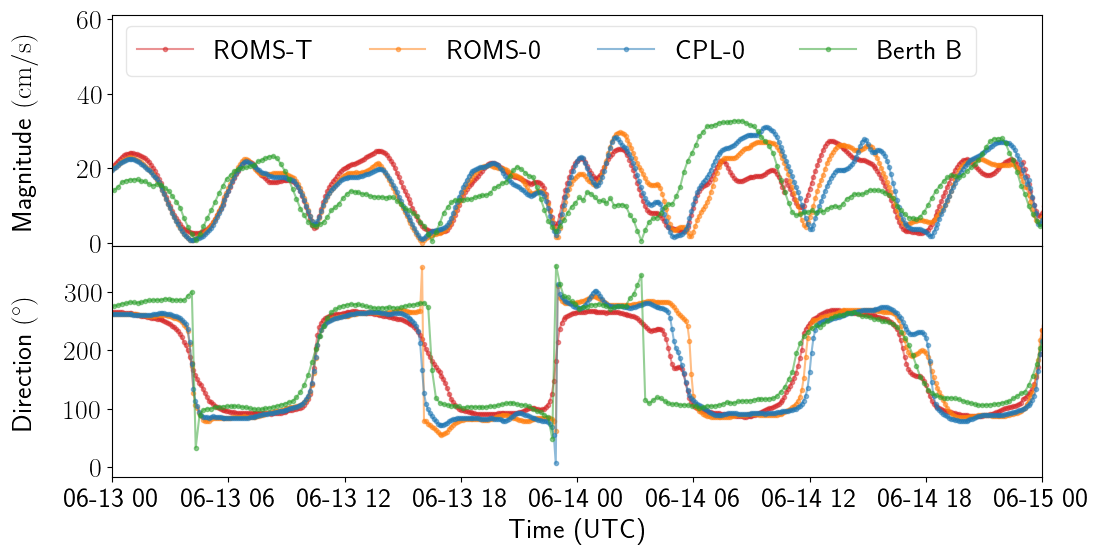}
  \centering \includegraphics[width=10cm]{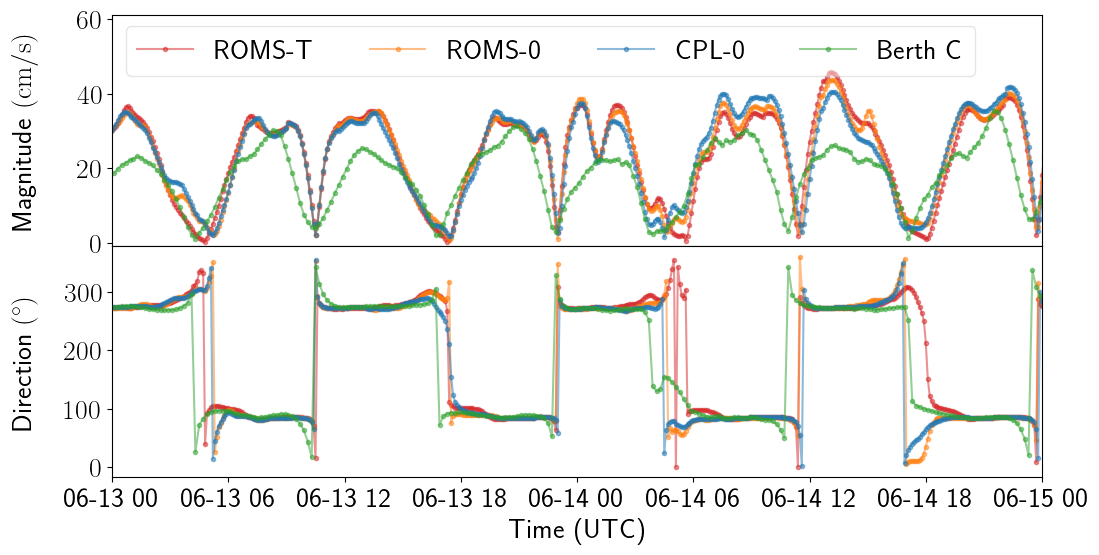}
  \caption{Time series of the depth-integrated velocities for the three  ADCPs Berth A (top), Berth B (middle) and Berth C (bottom), showing the standalone runs ROMS-T and ROMS-0, the coupled run CPL-0 and the ADCP observations. The velocities are shown in terms of magnitude and direction and the time window is centered around Storm Hector.
  The three model configurations only show differences when the storm gains in intensity starting around $2018/06/13$ at $20\mathrm{:}00\mathrm{:}00$. The impact of the storm on the record is not obvious and only strongly noticeable for Berth B with an usual increase of the tidal ebb flow.}
  \label{ts_adcp}
\end{figure}

Time series of the depth-integrated velocities around Storm Hector ($2018/06/14$) are shown in Figure \ref{ts_adcp} for the three ADCPs Berth A, Berth B and Berth C. Instead of the eastward and northward components, the velocities are shown in terms of magnitude and direction. Those two components clearly highlight the strong tidal dominance on the mass flux.

Outside the time window of the storm, the direction of the flow is well captured especially at Berth C. Small discrepancies appear during the storm, especially at the transition during the flood and ebb.
The flow rotation in-between tides shows a chaotic behavior for all stations. For instance looking at the Berth B record, on $2018/06/13$ $04\mathrm{:}00\mathrm{:}00$ the flow turns clockwise from flood to ebb (direction increases), whereas on $2018/06/13$ $16\mathrm{:}00\mathrm{:}00$ the flow turns anticlockwise from flood to ebb (direction decreases).
It is suspected that the wind driven surface currents are responsible for this behaviour. It is not always accurately captured by the models and contributes to deteriorate the statistics especially on the weaker northward component.
The velocity magnitude is rather well captured, but the model tends in general to overestimate the current velocities at its peak, by $10\,\si{\cm\per\s}$ at worst.

Only during the storm window significant differences are observed between the different runs, both in terms of magnitude and direction. Looking at the time series on $2018/06/14$ between $00\mathrm{:}00\mathrm{:}00$ and $18\mathrm{:}00\mathrm{:}00$, it is difficult to assess if the model benefits from the coupling. For instance in the case of Berth A, the slack transition is perfectly captured at $04\mathrm{:}00\mathrm{:}00$ by the coupled model CPL-0 in terms of magnitude when ROMS-0 and ROMS-T show a delay. However, CPL-0 predicts a clockwise rotation of the flow when both the record and standalone runs indicate an anticlockwise rotation.
The only observation that can be inferred is that wave effects on the currents are only meaningful during storm events, but the capabilities and accuracy of the model are not improved drastically by the coupling.

\subsection{Wave parameters}

Two observations are available when it comes to the wave parameters: the ADCP deployed near Inishmaan exposed to the Atlantic swell, and the SmartBay wave buoy inside Galway Bay. Different sea states and agreements are found for the locations.
Only the processed mean wave parameters are available for the wave buoy at SmartBay so spectral partitioning is not possible. In order to separate the contribution from the swell and the local wind wave we loop through the peak period time series. If the value is above a specified threshold then we assume that the swell dominates and all the wave parameters are attributed to the swell component. Otherwise they are attributed to wind waves. The threshold value is arbitrarily picked. A value of $6.8\,\si{\s}$ is found to work best for the SmartBay wave buoy and $8.0 \,\si{\s}$ for the ADCP near Inishmaan.

\subsubsection{Inishmaan ADCP}

\begin{table}
\begin{center}
\resizebox{0.9\textwidth}{!}{%
\begin{tabular}{l l l l l l l l}
\toprule
Inishmaan 1701 & Run & \multicolumn{2}{l}{Significant wave height} & \multicolumn{2}{l}{Peak period} & \multicolumn{2}{l}{Peak direction} \\
Sea partition &     & RMSE $(\si{\m})$ & Correlation & RMSE $(\si{\s})$ & Correlation & RMSE $(^\circ)$ & Correlation \\
\midrule
Total sea     & SWAN-0 & $0.42$ & $0.95$ & $1.68$ & $0.74$ & $16.37$ & $0.42$ \\
              & CPL-0  & $0.42$ & $0.95$ & $1.61$ & $0.75$ & $16.93$ & $0.40$ \\
Swell         & SWAN-0 & $0.41$ & $0.95$ & $1.69$ & $0.71$ & $16.31$ & $0.42$ \\
$98\%$        & CPL-0  & $0.42$ & $0.95$ & $1.62$ & $0.72$ & $16.91$ & $0.40$ \\
Wind wave     & SWAN-0 & $0.47$ & $0.67$ & $0.54$ & $0.48$ & $19.50$ & $0.34$ \\
$2\%$         & CPL-0  & $0.44$ & $0.74$ & $0.52$ & $0.49$ & $18.36$ & $0.41$ \\
\bottomrule
\end{tabular}}
\caption{Mean wave statistics for the Inishmaan ADCP, divided into swell dominated sea state and wind waves dominated sea state. An excellent agreement is found for the significant wave height, mostly attributed to the swell which dominates at this location open to the Atlantic Ocean.}
\label{inishmaan}
\end{center}
\end{table}

The relevant wave statistics for the ADCP near Inishmaan are shown in Table \ref{inishmaan}. The total sea statistics are computed without discriminating against the peak period, and as expected the sea state is dominated by swell ($98\%$).
The errors for the significant wave height are around $0.4 \mathrm{-} 0.5 \,\si{\m}$. It gives a relative error of $15\%$ with a mean value of $3.1 \,\si{\m}$ recorded by the ADCP, which is deemed reasonable. A worse agreement is found for the peak period and a poor agreement is found for the peak direction. Those values consist of higher order moments of the wave spectrum so this is not surprising. The high peak direction errors around $16 \mathrm{-} 17 ^\circ$ can also be explained by the local nearshore bathymetry, which is badly captured in the model and is strongly responsible for wave refraction.
A better agreement is found for the swell dominated sea states than for the wind wave dominated sea states. The errors for the peak period must be compared with the mean values in each case, $11.7\,\mathrm{s}$ for the swell and $7.4\,\mathrm{s}$ for the wind waves, which gives an average relative error of $6.1\%$ for swell and $6.5\%$ for wind waves. The correlation coefficients are always significantly better for the swell dominated sea states than for the wind waves. It shows that the model captures well the propagation of the swell from the boundary conditions.

Overall there is only a marginal impact of the currents on the waves at this particular location. The mean wave parameters stay unmodified for the total sea. Looking into more details, the highly occurring swell dominated sea states are almost not affected by the currents, but the scarcer wind wave states see a small impact from the currents, which increases the agreement, thus reducing the errors for all three wave parameters.

\subsubsection{SmartBay}

\begin{table}
\begin{center}
\resizebox{0.9\textwidth}{!}{%
\begin{tabular}{l l l l l l l l}
\toprule
SmartBay 1701 & Run & \multicolumn{2}{l}{Significant wave height} & \multicolumn{2}{l}{Peak period} & \multicolumn{2}{l}{Peak direction} \\
Sea partition &     & RMSE $(\si{\m})$ & Correlation & RMSE $(\si{\s})$ & Correlation & RMSE $(^\circ)$ & Correlation \\
\midrule
Total sea     & SWAN-0 & $0.22$ & $0.91$ & $3.95$ & $0.55$ & $26.00$ & $0.69$ \\
              & CPL-0  & $0.21$ & $0.91$ & $3.69$ & $0.56$ & $25.55$ & $0.71$ \\
Swell         & SWAN-0 & $0.15$ & $0.88$ & $4.68$ & $0.27$ & $28.79$ & $-0.03$ \\
$65\%$        & CPL-0  & $0.15$ & $0.89$ & $4.37$ & $0.30$ & $28.14$ & $0.08$ \\
wind wave      & SWAN-0 & $0.32$ & $0.90$ & $1.85$ & $0.41$ & $19.51$ & $0.91$ \\
$35\%$        & CPL-0  & $0.30$ & $0.91$ & $1.82$ & $0.44$ & $19.72$ & $0.91$ \\
\bottomrule
\end{tabular}}
\end{center}

\begin{center}
\resizebox{0.9\textwidth}{!}{%
\begin{tabular}{l l l l l l l l}
\toprule
SmartBay 1705 & Run & \multicolumn{2}{l}{Significant wave height} & \multicolumn{2}{l}{Peak period} & \multicolumn{2}{l}{Peak direction} \\
Sea partition &     & RMSE $(\si{\m})$ & Correlation & RMSE $(\si{\s})$ & Correlation & RMSE $(^\circ)$ & Correlation \\
\midrule
Total sea     & SWAN-0 & $0.16$ & $0.95$ & $2.85$ & $0.48$ & $28.40$ & $0.40$ \\
              & CPL-0  & $0.15$ & $0.95$ & $2.76$ & $0.50$ & $28.55$ & $0.44$ \\
Swell         & SWAN-0 & $0.13$ & $0.90$ & $3.38$ & $0.16$ & $32.40$ & $0.17$ \\
$66\%$        & CPL-0  & $0.12$ & $0.92$ & $3.24$ & $0.16$ & $32.52$ & $0.07$ \\
wind wave      & SWAN-0 & $0.19$ & $0.94$ & $1.52$ & $0.51$ & $19.27$ & $0.70$ \\
$34\%$        & CPL-0  & $0.18$ & $0.94$ & $1.40$ & $0.53$ & $18.66$ & $0.74$ \\
\bottomrule
\end{tabular}}
\caption{Mean wave statistics for the SmartBay wave buoy, for the 1701 and 1705 runs, divided into swell dominated sea state and wind waves dominated sea state. An excellent agreement is found for the significant wave height, which is even better when the wind waves dominate. There is a clear indication looking at the peak parameters that the wind waves are better captured than the swell inside Galway Bay.}
\label{smartbay}
\end{center}
\end{table}

\begin{figure}
  \centering \includegraphics[width=14cm,height=10cm]{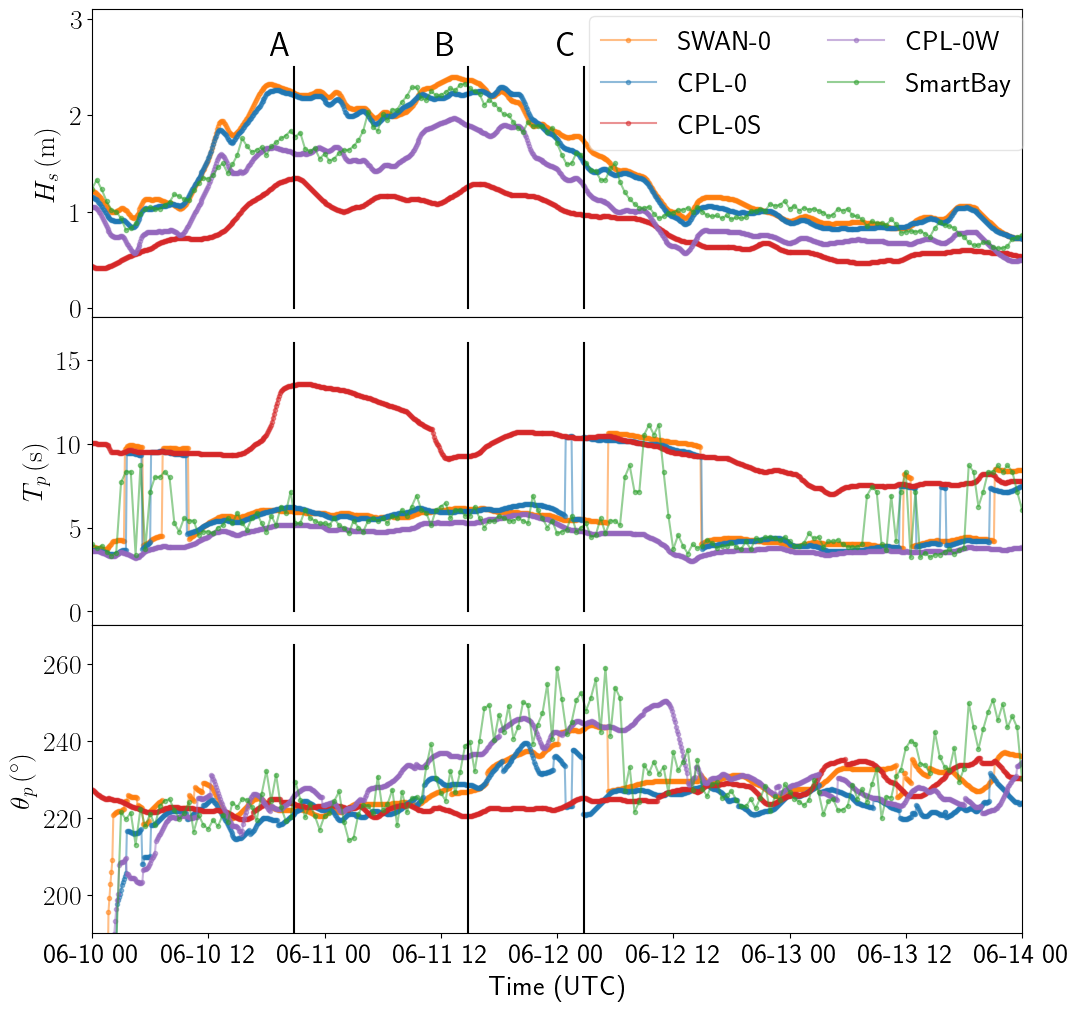}
  \caption{Time series of the significant wave height $H_s$, peak period $T_p$ and peak direction $\theta_p$ during a strong wave event at the SmartBay location. The two runs SWAN-0 and CPL-0 are compared against the measurements highlighting the weak impact overall of the coupling as far as the wave propagation is concerned. The runs CPL-0S and CPL-0W highlight the behavior of the swell component and wind wave component at the SmartBay location. They are shown to equally explain the wave heights meaning the sea state inside the bay is strongly mixed.
  The three slices A, B and C correspond to the time at which the wave spectrum is plotted in Figure \ref{smartbay1705spectra}.}
  \label{smartbay1705waves}
\end{figure}

Table \ref{smartbay} shows the statistics for the SmartBay wave buoy for the two runs $1701$ and $1705$. In both cases a good agreement is found for the significant wave height. The low absolute errors must be put in perspective with the low mean wave height value, $0.94\,\si{\m}$, for the run 1701 and $0.71\,\si{\m}$ for the run $1705$. It gives relative errors around $23\%$ for both runs.
Looking at the overall impact of the coupling, a slight improvement for all the statistics is observed compared to the standalone run, by less than $1\%$. No significant impact on either spectral component is observed. Both the swell and wind waves seem to be improved by the same marginal amount. Even during the strong wave event highlighted in Figure \ref{smartbay1705waves} the time series do not show any strong impact induced by the coupling.

As opposed to what is observed outside of the bay, the wind wave component gives a better agreement between the model and the observations. The correlation coefficient is slightly better for the wind waves as far as the significant wave height is concerned, but way better for the peak period and peak direction. The peak period and peak direction correspond well to the value of each wave partition, and there is a clear indication that the swell is poorly captured inside the bay. This is especially true for the wave direction. Errors are of the order of $30^\circ$ with correlation coefficients close to $0$. The agreement is significantly better for the wind waves with errors around $20^\circ$ and correlation coefficients between $0.7$ and $0.9$.
The swell is characterised by longer wavelengths and is more sensitive to the bathymetry than wind waves. However the quality of the bathymetry inside Galway Bay is excellent, since the raw surveys are conducted with spatial resolutions of a few meters. It is more likely that the propagation of the swell through the narrow entrance points of the bay is poorly captured by the model.

The significant wave height still includes both the swell and wind wave components regardless of which one dominates. The small changes in terms of statistics between the two wave components highlight that the sea state is in general always strongly mixed. The partition between swell dominated and wind waves dominated sea states is indeed close to equal, with $65\%$ for the swell and $35\%$ for the wind waves.
This is also highlighted by Figure \ref{smartbay1705waves}, that shows time series of the significant wave height, peak period and peak direction at the SmartBay location for the run $1705$ during a strong wave event. Only looking at the measurements the time peak period shows several jumps between low values around $4\,\mathrm{s}$ and higher values around $10\,\mathrm{s}$. The low values match well those of the run CPL-0W where the wave model is only forced with wind and only including wind waves, while the higher values match those of the run CPL-0S only including swell. The time series of the runs CPL-0W and CPL-0S for the significant wave height also indicate that the swell and wind wave components both equally contribute to explain the wave heights. It is also worthwhile noticing that the superposition of the two runs CPL-0W and CPL-0S overestimates the result given by CPL-0. This highlights the heavily non-linear behavior of the wave propagation.

\begin{figure}
  \centering \includegraphics[width=10cm]{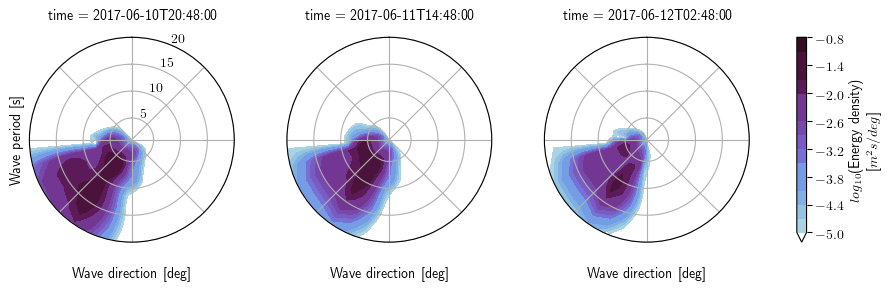}
  \caption{Wave spectra at the SmartBay location from run CPL-0, shown at the three instants A ($2017/06/10$ $20\mathrm{:}48\mathrm{:}00$), B ($2017/06/11$ $14\mathrm{:}48\mathrm{:}00$) and C ($2017/06/12$ $02\mathrm{:}48\mathrm{:}00$) in this order. In all three cases a main swell component coming from the South Sound and a wind wave component appear, as well as a weaker secondary swell component coming from the North Sound according to the direction.}
  \label{smartbay1705spectra}
\end{figure}

Figure \ref{smartbay1705spectra} shows the wave spectra at the SmartBay location from the run CPL-0 at the three marked instants A, B and C, highlighting once again the strong mixed sea state in the bay. A strong swell component appears in all three cases, with a pronounced southwest origin which would correspond to swell entering through the South Sound. An equally strong wind wave component appears, featuring shorter time periods. Finally a weaker secondary swell component still emerges, with an origin more west-southwest which would correspond to swell entering through the North Sound. The wind wave partition is well aligned with the main swell partition for the first case, but for the two following cases it is miss-aligned as observed previously in Figure \ref{smartbay1705waves}.

\subsection{Statistical impact of the sea state on the currents}

\begin{figure}
  \centering \includegraphics[width=14cm]{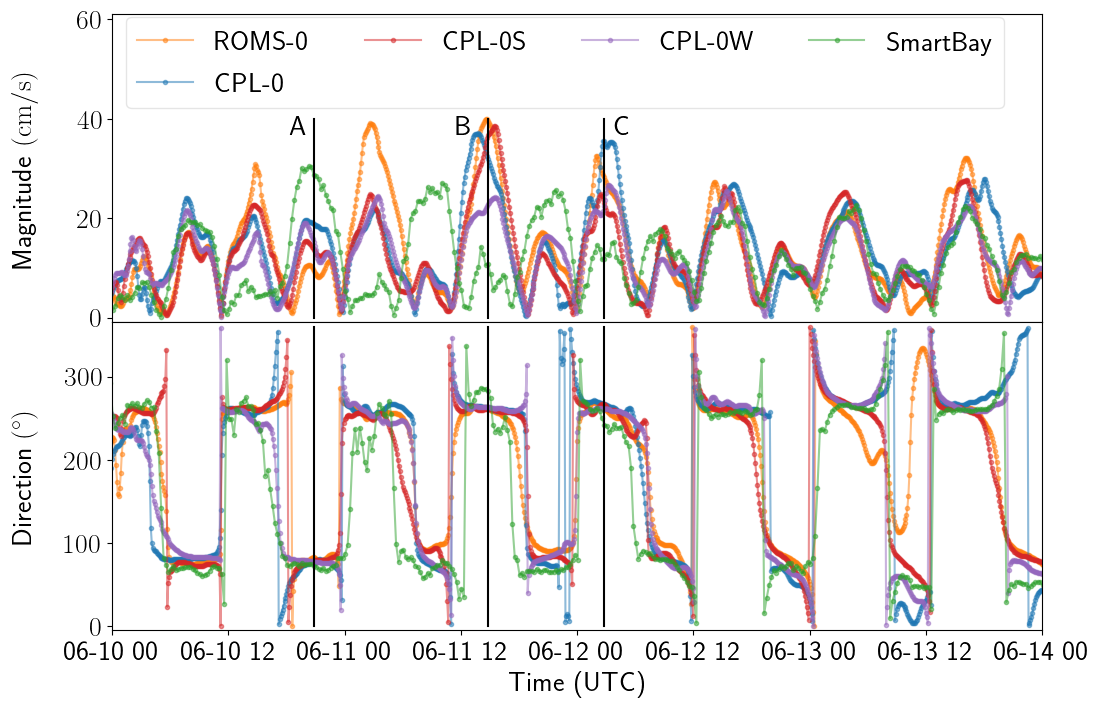}
  \caption{Time series of the current magnitude and direction at the SmartBay ADCP location, for the run 1705 during a storm event.
  The standalone run ROMS-0 is compared against the three coupled runs CPL-0, CPL-0S and CPL-0W, highlighting the effects of either swell or wind waves on the coupling.
  Little impact of the coupling is observed outside of the storm window. During the storm the coupling improves the accuracy of the model but it still fails to capture correctly the velocities.
  Overall there is a surprising good agreement between the three coupled runs. Discrepancies are only observed for the marked cases B and C.}
  \label{smartbay1705currents}
\end{figure}

The only station where measurements of the currents and the wave parameters are both available is the SmartBay station, during the run 1705. A storm is isolated for the study and highlighted in Figures \ref{smartbay1705waves} and \ref{smartbay1705currents}, with a maximum significant wave height of $2.3 \,\si{\m}$ inside the bay corresponding to a swell outside the bay reaching $6\,\si{\m}$. During this time period the winds are also strong, around $12\,\si{\m\per\s}$, and are mostly south, southwest. Winds certainly have an effect on the upper layer of the water column but since they are mostly constant for this period their effect is not studied. 

The global statistics shown in Table \ref{smartbay}, and the time series shown in Figure \ref{smartbay1705waves} indicate that the wave parameters at the SmartBay location are only marginally impacted by the currents, even during a strong storm event. No further analysis is necessary with this aspect of the interaction and instead we focus on the impact of the sea state on the currents.

\subsubsection{Impact of the significant wave height}

During the peak of the storm, between slices A and B, the currents are badly predicted by the model. The coupling seems to improve the agreement but a large error is still observed. The impact of the coupling during this event is also quite counter-intuitive. Compared to the standalone run ROMS-0 the coupling reduces the mass transport during the tidal flood (current direction of $260^\circ$) while it increases the mass transport during the tidal ebb (current direction of $80^\circ$). This effect actually goes against the waves, which are mostly propagating eastward. On the other hand, outside the time window of the storm, almost no impact of the coupling is observed.

In order to quantify this behavior the statistics for the SmartBay ADCP are recomputed discriminating against the wave height (Table \ref{statshs}). The subset labelled calm sea only includes data with wave heights lower than $1.2\,\si{\m}$, and the subset labelled storm is the complementary set.
Including the full set of data gives an acceptable error of $6.3\,\si{\cm\per\s}$ that is not improved by the coupling. The subset of calm sea gives a better error of $5.5\,\si{\cm\per\s}$ that is again not improved by the coupling. This is not surprising as during calm seas the impact of waves is minimal, and most of the data correspond to calm sea events.
The storm subset gives a higher error of $10.5\,\si{\cm\per\s}$ for the run ROMS-0. The error is reduced to $9.6\,\si{\cm\per\s}$ with the coupling. The correlation coefficient is also improved with the coupling. The error is still high but it indicates that the coupling improves the capabilities of the model under strong sea conditions, which is also when the model is struggles the most to accurately capture the circulation inside the bay.

\begin{table}
\begin{center}
\resizebox{0.5\textwidth}{!}{%
\begin{tabular}{l l l l}
\toprule
SmartBay 1705 & Run & \multicolumn{2}{l}{Current magnitude} \\
Sea condition &     & RMSE $(\si{\cm\per\s})$ & Correlation \\
\midrule
Total sea     & ROMS-0 & $6.3$ & $0.581$  \\
              & CPL-0  & $6.2$ & $0.579$  \\
Calm sea      & ROMS-0 & $5.5$ & $0.663$  \\
$89\%$        & CPL-0  & $5.6$ & $0.639$  \\
Storm         & ROMS-0 & $10.5$ & $0.181$  \\
$11\%$        & CPL-0  & $9.6$ & $0.258$  \\
\bottomrule
\end{tabular}}
\caption{Statistics for the depth-integrated current velocity magnitude for the two runs ROMS-0 and CPL-0, comparing a subset of the data only containing calm sea conditions with a significant wave height lower than $1.2\,\si{\m}$, and the complementary subset with stronger sea conditions. There is a clear indication that the currents are better predicted during a calm sea, and badly picked up during strong sea events. The coupling slightly improves the accuracy of the model during those stronger wave events.}
\label{statshs}
\end{center}
\end{table}

\subsubsection{Impact of the sea partition}

For most of the time period shown in Figure \ref{smartbay1705currents} the two coupled runs CPL-0S and CPL-0W surprisingly show similar results for the currents.
Between the two cases A and B the coupled runs CPL-0 and CPL-0W predict similar mean wave parameters (wind wave dominated), as shown by Figure \ref{smartbay1705waves}. The interaction terms appearing in the coupling mostly involve the mean wave parameters, namely the significant wave height, peak wavenumber and peak wave direction. One would therefore expect the velocities of CPL-0 and CPL-0W to be similar, which is not the case for case B.
Similarly for case C, the mean wave parameters of CPL-0 are similar to CPL-0S (swell dominated), but the responses in terms of mass transport of CPL-0 and CPL-0S are still quite different. This highlights that other parameters come into play, and probably a local analysis is not enough to explain the impact of waves on the currents.

Similarly to what was done previously to highlight the impact of the wave heights on the wave effect on currents, the global statistics for the current velocity are recomputed discriminating against the dominating sea partition (Table \ref{statspp}). The wave effects on the currents seem to be slightly better captured when the swell dominates the sea state, with an error of $5.9\,\si{\m\per\s}$ against an error of $6.7\,\si{\m\per\s}$ for when the wind wave dominates. The coupled run CPL-0, compared to the standalone run ROMS-0, only improves statistics when the wind waves dominate, and slightly deteriorates them when the swell dominates.
The statistics are much closer than when discriminating against the wave height, which would indicate that using the dominant sea partition as an indicator is not the best option. Because of the strong mixed sea state the dominant sea partition can fluctuate rapidly between swell and wind waves as well, making the statistics less reliable in the end.

\begin{table}
\begin{center}
\resizebox{0.5\textwidth}{!}{%
\begin{tabular}{l l l l}
\toprule
SmartBay 1705 & Run & \multicolumn{2}{l}{Current magnitude} \\
Sea partition &     & RMSE $(\si{\cm\per\s})$ & Correlation \\
\midrule
Total sea     & ROMS-0 & $6.3$ & $0.581$  \\
              & CPL-0  & $6.2$ & $0.579$  \\
Swell         & ROMS-0 & $5.5$ & $0.650$  \\
$20\%$        & CPL-0  & $5.9$ & $0.623$  \\
Wind wave     & ROMS-0 & $7.3$ & $0.499$  \\
$80\%$        & CPL-0  & $6.7$ & $0.512$  \\
\bottomrule
\end{tabular}}
\caption{Statistics for the depth-integrated current magnitude for the two runs ROMS-0 and CPL-0, comparing a subset of the data where wind waves dominate with a peak period lower than $6.8\,\si{\s}$, and the complementary subset where the swell dominates. There is a small indication that the model captures better the currents when the swell dominates and that the coupling improves slightly the solution when wind waves dominate, but the differences remain almost marginal.}
\label{statspp}
\end{center}
\end{table}

\section{Analysis of the coupling during Storm Hector \label{hector}}

In the previous section we compared the model results with measurements. We found that there is a good overall agreement for both the wave parameters and the current velocities. Discrepancies appear for the velocities during strong sea states. A mixed sea state is observed inside Galway Bay. Wind waves are accurately predicted but the model fails to capture correctly the swell component. The coupling is unable to improve the capabilities of the model.
A small to marginal impact of the interaction is observed on the global statistics. In this section we look more closely at the impact of the wave-induced forcing on the sea level and currents during Storm Hector. There is also a noticeable impact of currents on wave propagation but it is not as striking, and given that we have some doubts on the capabilities of the model to capture accurately the swell partitions inside Galway Bay, we prefer to focus on the wave effects on the currents.
Storm Hector passed through the western part of Ireland between $2018/06/13$ and $2018/06/15$, from south-west to north-east. At the peak of the storm, wave heights up to $3\,\si{\m}$ were observed inside Galway Bay.

\subsection{Wave effects on the surge}

The tidal analysis conducted in the previous section (Figure \ref{surge}, top plot) shows a $3\,\si{\cm}$ increase of the sea level during Storm Hector induced by the coupling, comparing the run CPL-0 to ROMS-0. Whether or not this contributes to an increase in accuracy is difficult to assess. For instance the bottom plot of Figure \ref{surge} shows that the coupling increases the error. However, we argued that most of the observed surge signal is contained in the boundary conditions, so the coupling could very well increase the capabilities of the model by capturing accurately the local wave effects.

\subsubsection{Averaged surge and wave effects}

\begin{figure}
  \centering \includegraphics[width=7cm]{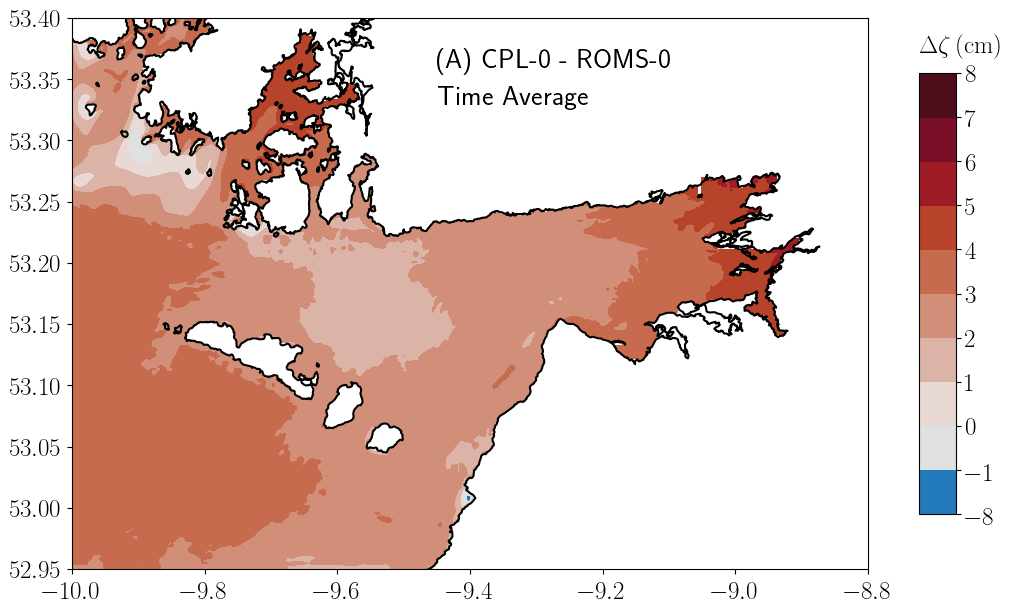} \includegraphics[width=7cm]{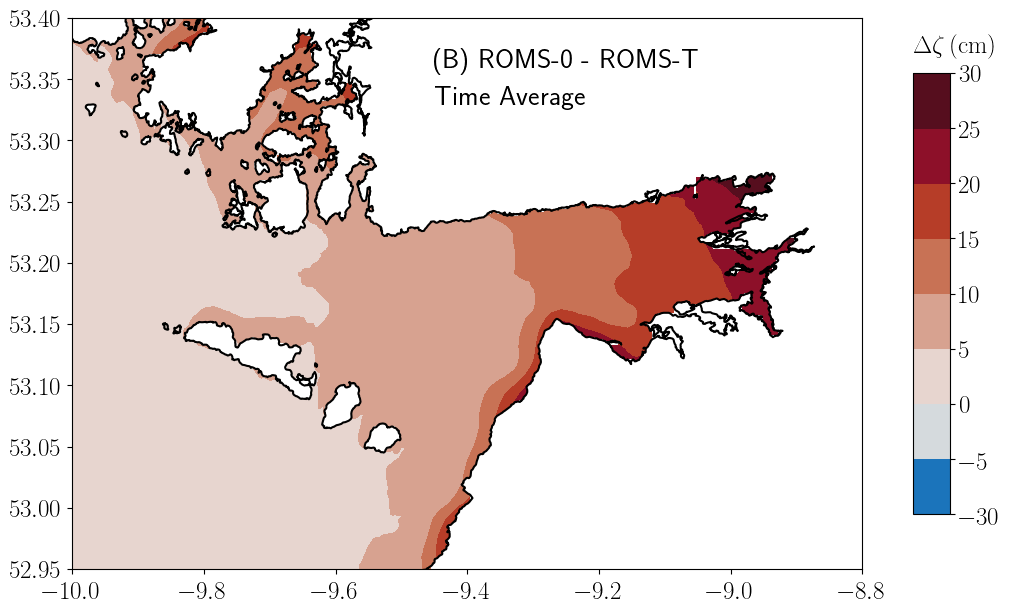}
  \centering \includegraphics[width=7cm]{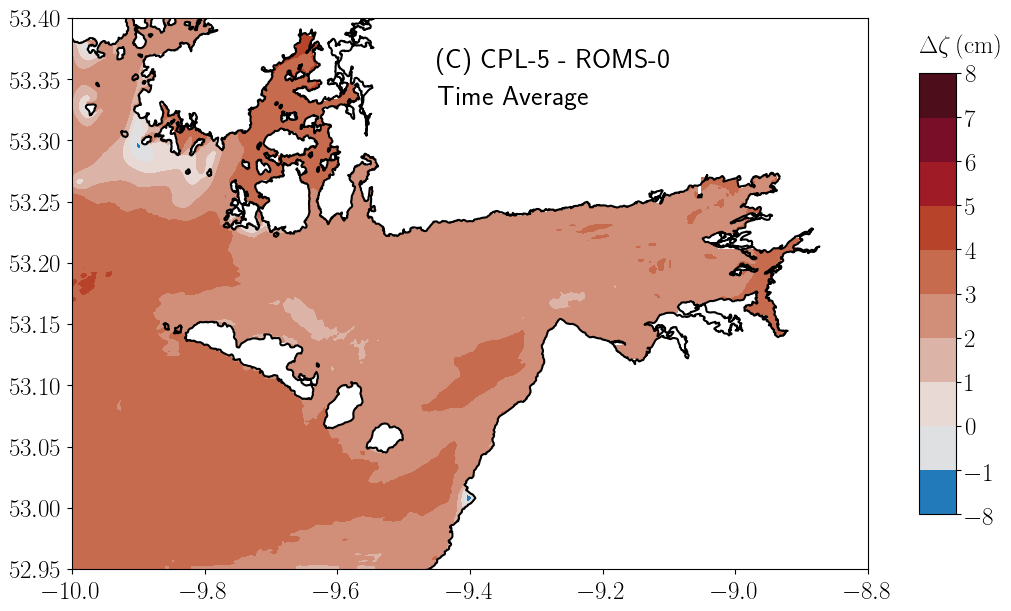} \includegraphics[width=7cm]{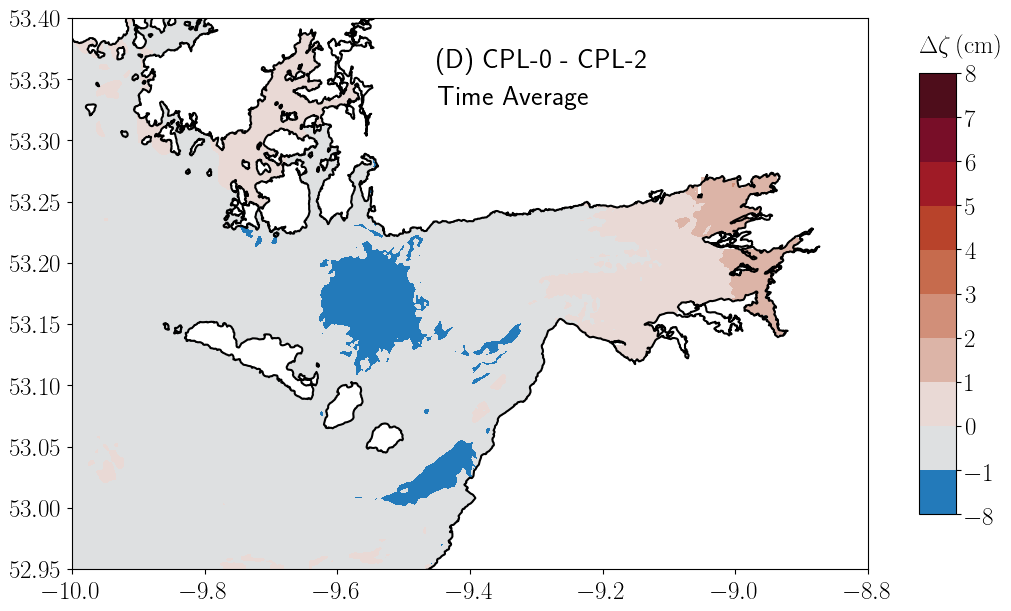}
  \caption{Difference in sea surface elevation between different runs during the peak of Storm Hector.
  The different runs highlight the individual impact of each process included in the model, summarized in Table \ref{runs}. To save space the runs CPL-3A, CPL-3B, CPL-3C and CPL-4 are not reported but no difference is observed compared to CPL-0.}
  \label{surge_runs}
\end{figure}

Assuming now that the model captures accurately the effects of the waves we have a closer look at the surge induced during Storm Hector. The small changes in sea level, of the order of $10\,\si{\cm}$ at best, are negligible compared to the tidal range. We therefore rely on the difference in sea level between different test runs to observe the wind or wave effects, shown in Figure \ref{surge_runs}. All the plots correspond to a time average over the peak of the storm, between $2018/06/14$ $02\mathrm{:}00\mathrm{:}00$ and $2018/06/14$ $06\mathrm{:}00\mathrm{:}00$. The wave-induced surge is quickly losing intensity, which is why we reduce the window for the time-average, focusing on the strong response at the peak.

Plot (A) shows CPL-0 minus ROMS-0, which includes a wide range of wave-induced processes. The difference in sea level is the strongest in the back of Galway Bay and Kilkieran Bay with a maximum increased sea level of $6\,\si{\cm}$ induced by the coupling.

Plot (B) features ROMS-0 minus ROMS-T and highlights the impact of the atmospheric forcing. It is not shown here but with a better tuned configuration of the model the atmospheric response is attributed solely to the wind impact. Looking at the scaling it can induce a surge up to $30\,\si{\cm}$. The wave-induced surge is almost four times smaller than the wind-induced surge, but it is not small enough to be considered negligible and justifies the inclusion of wave effects.

Plot (B) shows CPL-5 minus ROMS-0, which corresponds to the impact of the conservative wave-induced terms. It explains most of the impact of waves up to roughly $4.5\,\si{\cm}$. Oddly enough the same wave-induced increase in sea level is observed out of the bay, and also attributed to the conservative forcing terms. The vortex force formalism is composed of different forcing terms. The contribution of each of those terms will be studied below in more depth to identify which effect actually dominates.

Plot (D) corresponds to CPL-0 minus CPL-2 and highlights the impact of the wave-enhanced surface roughness. It explains a smaller portion of the surge up to $1.5\,\si{\cm}$, which is significant and should not be neglected.
The runs CPL-3A, CPL-3B, CPL-3C and CPL-4 are not reported here but no significant difference is observed. The wave breaking and bottom streaming contributions have a negligible impact on the sea surface elevation.

\subsubsection{Explaining the wave-induced surge}

The wave-induced surge inside Galway is attributed principally to the conservative terms, and to a smaller extent to the increase in surface stress caused by the wave-enhanced surface roughness (Figure \ref{surge_runs}, plots C and D).
The overall impact reaches $7\,\si{\cm}$, which is smaller but not negligible compared to the $30\,\si{\cm}$ induced by the winds (Figure \ref{surge_runs}, plots A and B), especially in a populated area like Galway, which is regularly suffering from winter floods during strong storms. We believe that the impact of waves can improve the accuracy of the forecast and the quality of the warning.

The wave-enhanced surface roughness is seen to increase the surface stress. It induces a stronger mass transport towards the coast, in the back of the bay. The momentum terms are introduced and commented later, but one can refer to Figure \ref{map_budget_cpl2}. The enhanced wind surface stress and increased mass transport is met by a stronger pressure gradient, which explains well a strong variation in sea level.

\begin{figure}
  \centering \includegraphics[width=4.8cm]{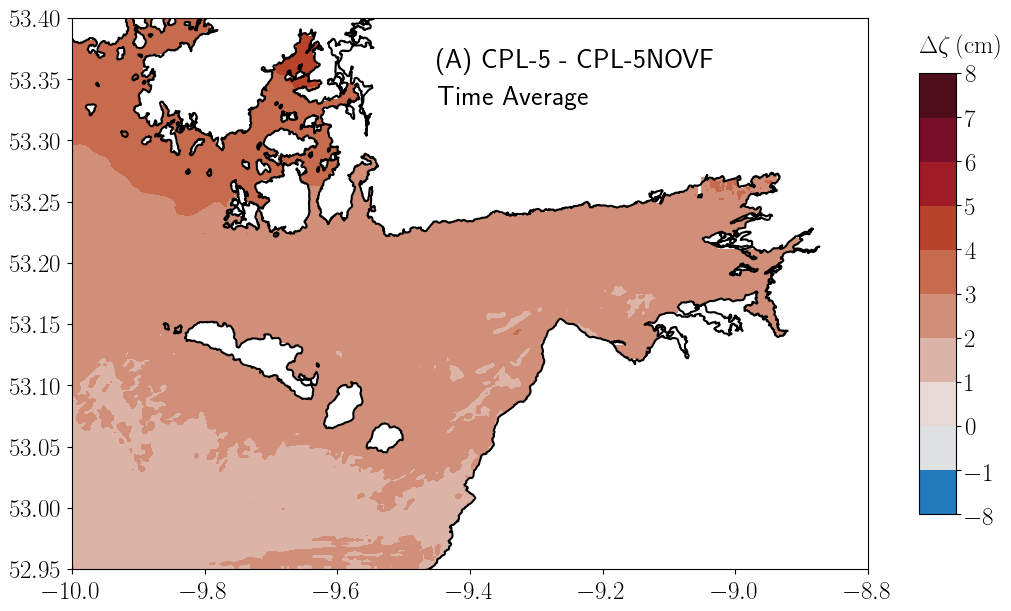} \includegraphics[width=4.8cm]{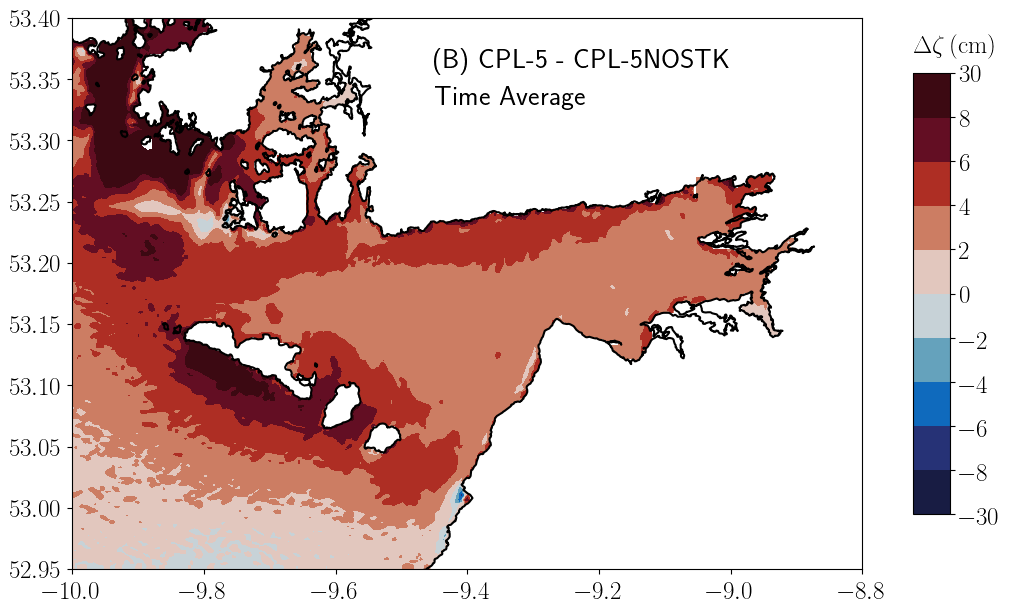}
  \includegraphics[width=4.8cm]{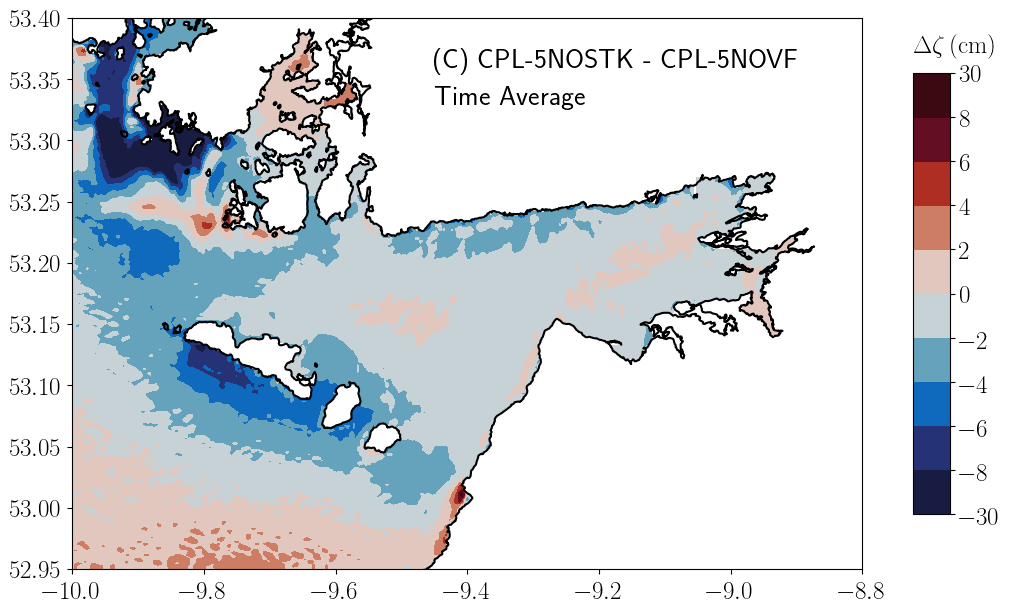}
  \centering \includegraphics[width=4.8cm]{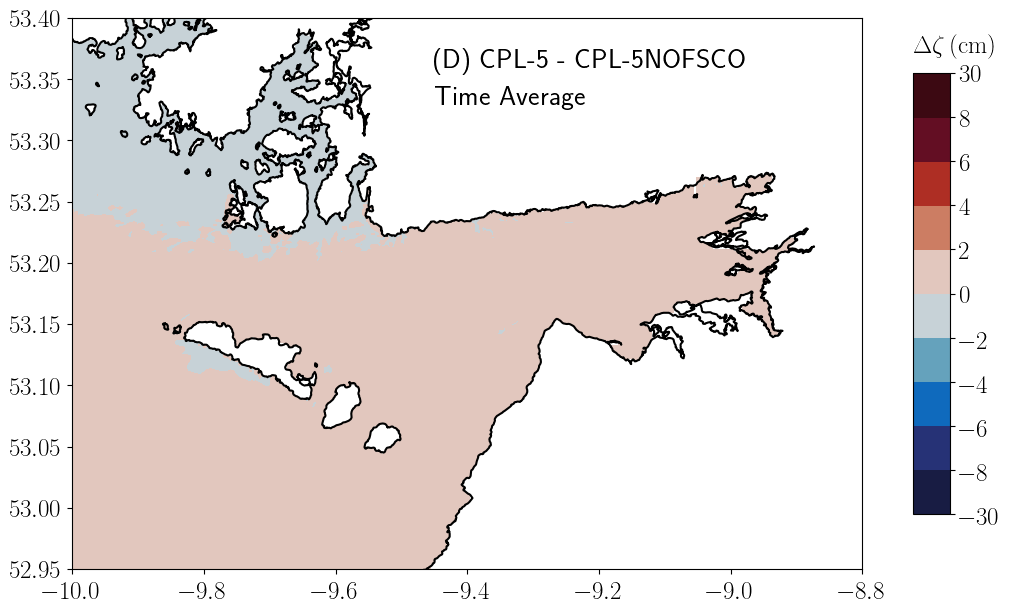} \includegraphics[width=4.8cm]{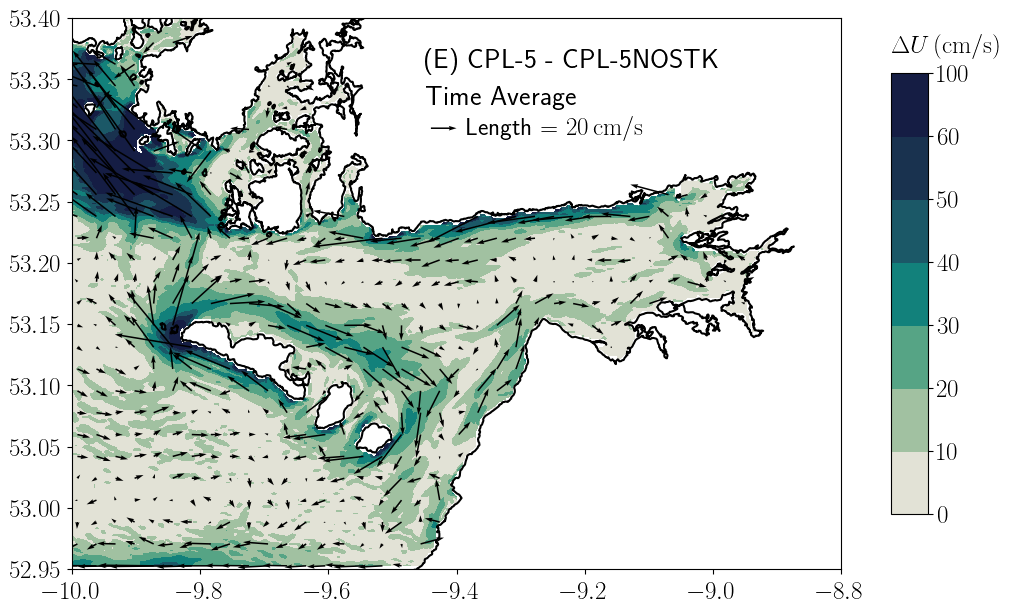}
  \includegraphics[width=4.8cm]{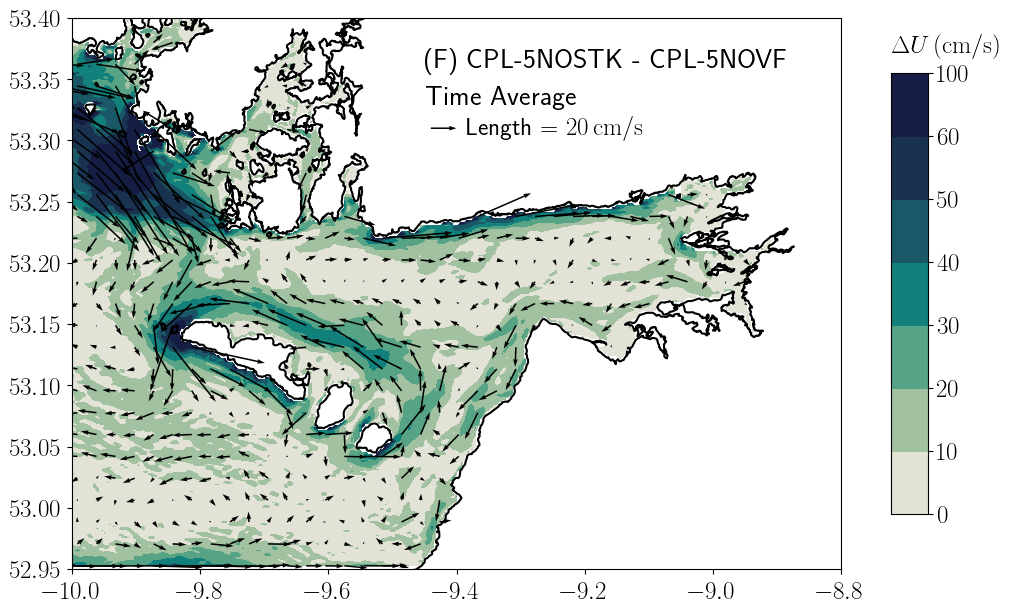}
  \caption{Difference in sea surface elevation and current velocity averaged over the peak of Storm Hector. Plot (A) shows CPL-5 minus CPL-5NOVF highlighting the combined impact of the vortex force and advection by the Stokes drift appearing in Eq. (\ref{sea_level_eq3},\ref{2d_momentum}). Plot (B) shows CPL-5 minus CPL-5NOSTK highlighting the specific impact of mass transport by the Stokes drift on the sea level in Eq. (\ref{sea_level_eq3}). Plot (E) features the impact on the currents. Plot (C) shows the specific impact of the momentum advection by the Stokes drift in Eqs. (\ref{2d_momentum}) on the sea level with CPL-5NOSTK minus CPL-5NOVF, neglecting the impact of the vortex force on the solution. Plot (F) features the same impact on the currents. Finally plot (D) shows the impact of the Stokes-Coriolis force on the sea surface elevation with CPL-5 minus CPL-5NOFSCO.}
  \label{surge_cpl5}
\end{figure}

Figure \ref{surge_cpl5} shows the contribution of the conservative wave-induced terms. Plot (A) features the surface elevation difference between the runs CPL-5 and CPL-5NOVF, where CPL-5NOVF has the same configuration as CPL-5 except that the vortex force $(\punderline{J},K)$ is removed from Eqs. (\ref{3d_momentum},\ref{2d_momentum}). The divergence of the Stokes drift appearing in Eq. (\ref{sea_level_eq3}) is also removed, as well as the advection by the Stokes drift appearing in Eq. (\ref{2d_momentum}). Those terms are all evaluated with the Stokes drift depth-integrated mass transport in \cwst{}. It is difficult to separate them any further because they are evaluated all at once using the Lagrangian velocity within a flux form formalism. Removing those terms provides a good agreement for the overall impact shown in Figure \ref{surge_runs}, plot (C). It will be argued later, when studying the momentum budget, that the direct impact of the vortex force is marginal, which means that only the mass and momentum transports are responsible for the wave-induced surge.

The small discrepancies are attributed to the Stokes-Coriolis force highlighted on plot (D), showing CPL-5 minus CPL-5NFSCO. The Stokes-Coriolis force induces a small reduction in sea surface elevation in the north-west part of the domain, and an increase in the south-east, consistent with the general direction of the Stokes-Coriolis force pointing south-east.

Plot (B) shows CPL-5 minus CPL-5NOSTK, which stands for CPL-5 only without the divergence of the Stokes drift appearing in Eq. (\ref{sea_level_eq3}). This plot highlights the specific effect of the mass transport by the Stokes drift on the sea level. It is strong in Kilkieran Bay and on the east side of the Aran Islands, up to $30\,\si{\cm}$. This is balanced by the effect of Stokes drift momentum advection in Eq. (\ref{2d_momentum}), highlighted in plot (C) with CPL-5NOSTK minus CPL-5NOVF. Unfortunately it was not possible to run a configuration where only the momentum advection by the Stokes drift was removed because the run was too unstable. It is understandable given the strong induced changes in the observed velocity by almost $1\,\si{\m\per\s}$.

The direct impact of the divergence of the Stokes drift on the sea surface elevation in Eq. (\ref{sea_level_eq3}) is well explained by an accumulation of water mass transported by the Stokes drift. The divergence is strongly negative near the shore where the incoming Stokes drift is not balanced by an outgoing Stokes drift because of the coastline. As a result it causes an increase in sea level, which is directly linked to the strength of the Stokes drift. This explains why the effect is stronger where the Stokes drift is also observed stronger (Figure \ref{stokes_coriolis_force}, plot C). The Stokes drift magnitude is mostly the result of a reduction in the water depth putting more weight in the surface levels where the drift is more consequent, especially in areas facing the open sea where both swell and wind waves can contribute. However, this is well balanced by the advection of momentum by the Stokes drift. Similarly to the water mass, momentum is also directly advected as shown in Eq. (\ref{2d_momentum}), with an opposite effect on the sea surface elevation which is not well understood.

The response in terms of currents is also well balanced, as shown by the plots (E) and (F). It only features long-shore induced currents. The increase in sea level on plot (B) is met with a clockwise long-shore circulation on plot (E), while the decrease on plot (C) is met with an anti-clockwise longshore circulation on plot (F). Long-shore circulations are strongly bounded by the Coriolis force, which means that plot (B) features an offshore current response deviated to its right, coherent with the nearshore increase in sea level. On the other hand plot (C) features an onshore current response, coherent with the increased advection caused by the Stokes drift. For plot (B) the increase in sea level is the cause of the current response, while in plot (C) it is the consequence of it. Both almost cancel each other perfectly, with only a small residual in the two bays.

To sum up, the Stokes-Coriolis force is observed to explain marginally the wave-induced response to the conservative forcing terms. A strong balance between the Stokes drift mass transport and Stokes drift momentum transport is observed, especially in Kilkieran Bay and on the west side of the Aran Islands, which explains most of the response. The Stokes drift mass transport appears as the leading mechanism, with the transport of momentum being the control mechanism. This is understandable as the transport of momentum does not directly impact the sea surface elevation. In both cases it is triggered by the coastline stopping the advection, and forcing a long-shore circulation. No fundamental argument is found as to why the increase is stronger in the back of Galway bay and in Kilkieran bay compared to the rest of the domain.

\subsection{Wave effects on the circulation inside Galway Bay}

The hydrodynamics is complex inside Galway Bay. As mentioned in \textcite{ren2017effect} the currents are strongly driven by the winds as well as the tides. Strong wave events like Storm Hector highlighted in this paper usually come with strong winds as well. A more complete analysis would also include the impact of wind on the circulation but for the sake of conciseness, we decided to focus only on the impact of waves.

\subsubsection{Impact on the depth-averaged transport}

\begin{figure}
  \centering \includegraphics[width=4.8cm]{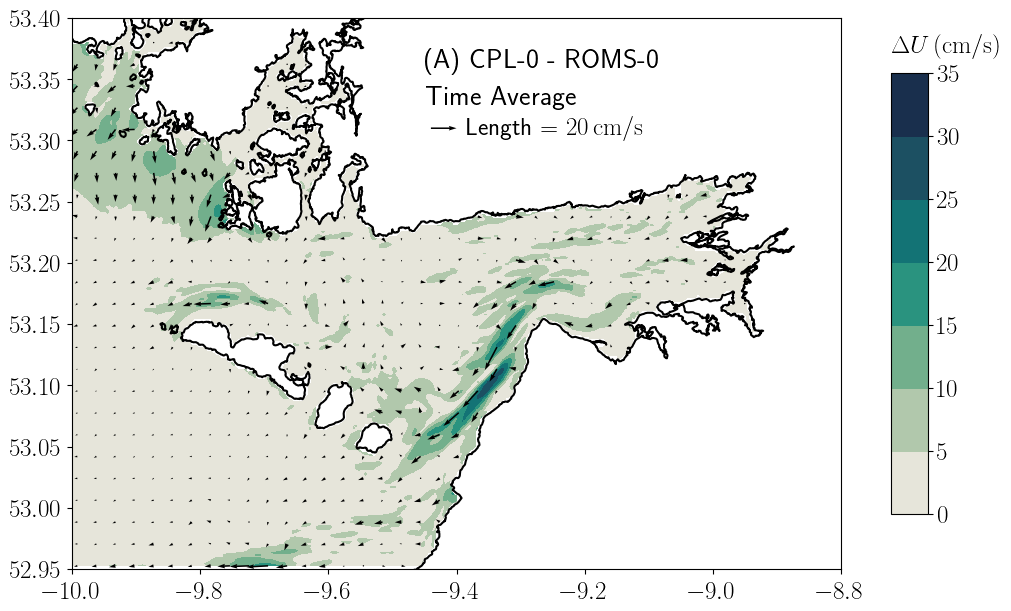} \includegraphics[width=4.8cm]{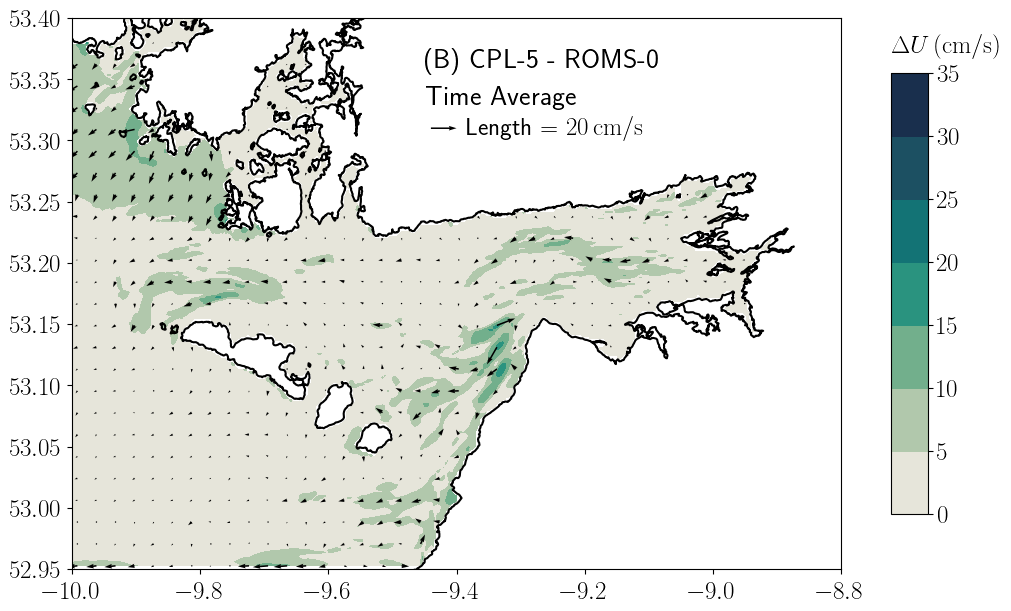} \includegraphics[width=4.8cm]{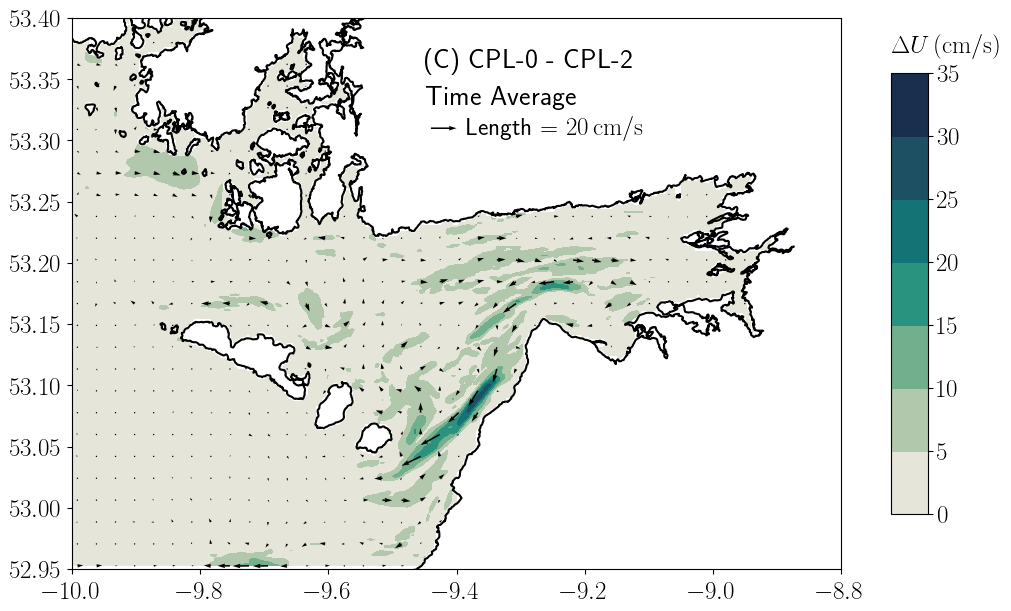}
  \caption{Depth-averaged difference of the current field averaged over the duration of Storm Hector. The quivers are computed as the difference between the two vector fields of the runs featured. The color map corresponds to the magnitude of this residual current field.
  Most of the wave effects shown on plot (A) are attributed to the wave-enhanced surface roughness, shown on plot (C). The conservative terms also contribute to a lesser extent, shown on plot (B).}
  \label{currents_runs}
\end{figure}

The specific impact of the different wave-induced effects on the depth-averaged velocities is studied. Figure \ref{currents_runs} shows the depth-averaged and time-averaged current field for different run configurations.

On plot (A) CPL-0 minus ROMS-0 shows the overall impact of all the wave-induced terms in the model. A strong response in terms of currents is observed. It is directed against the overall wave propagation and stays constant during the duration of the storm regardless of the tides. During a tidal flood the coupling reduces the inward flow, while during a tidal ebb it increases it, by at most $30\,\si{\cm\per\s}$ which is significant for the area. The maximal velocities observed in the model are around $80\,\si{\cm\per\s}$. This agrees well with the previous observations made during the validation (Figure \ref{smartbay1705currents}).
The wave effects are mostly attributed to the wave-enhanced surface roughness highlighted on plot (B) with CPL-2, and to a smaller extent to the conservative wave-induced terms with CPL-5 shown on plot (C). The sum of those two contributions indeed agrees quite well with the total observed wave effects.
A strong wave impact is also observed in the Kilkieran Bay, and even in the nearshore Connemara region although this part of the domain is not shown here. The wave-induced response is also seen to be directed against the wave propagation, and is induced by the conservative wave-induced forcing terms, highlighted by plot (C).

\subsubsection{Momentum balance inside Galway Bay}

\begin{figure}
  \centering \includegraphics[width=7cm]{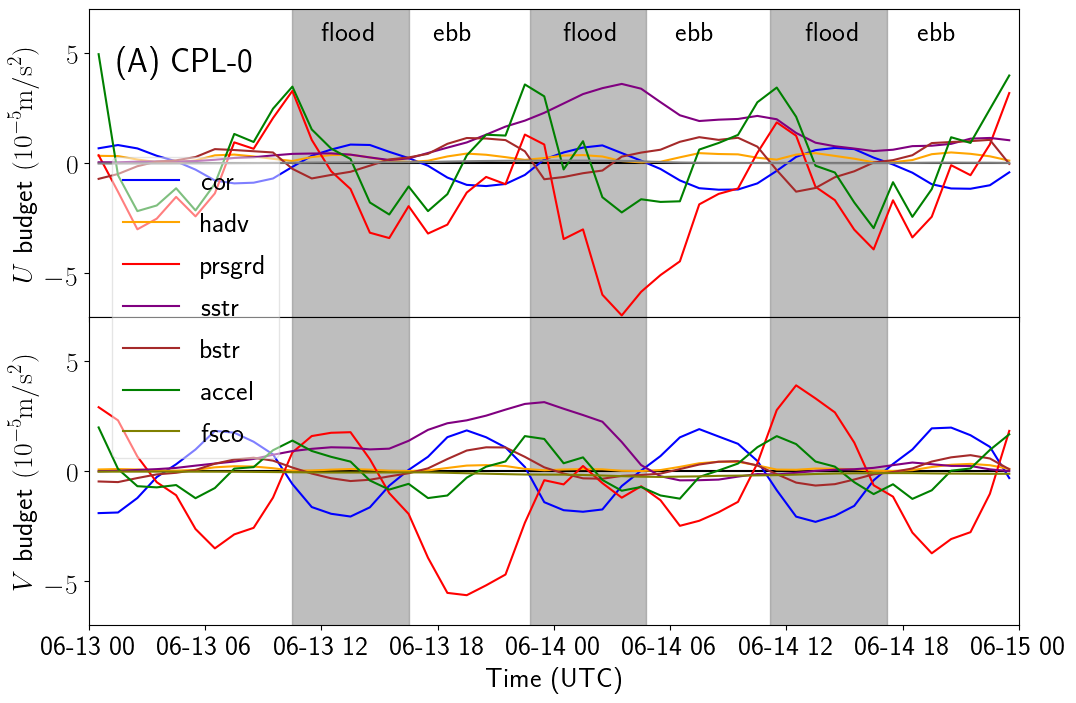} \includegraphics[width=7cm]{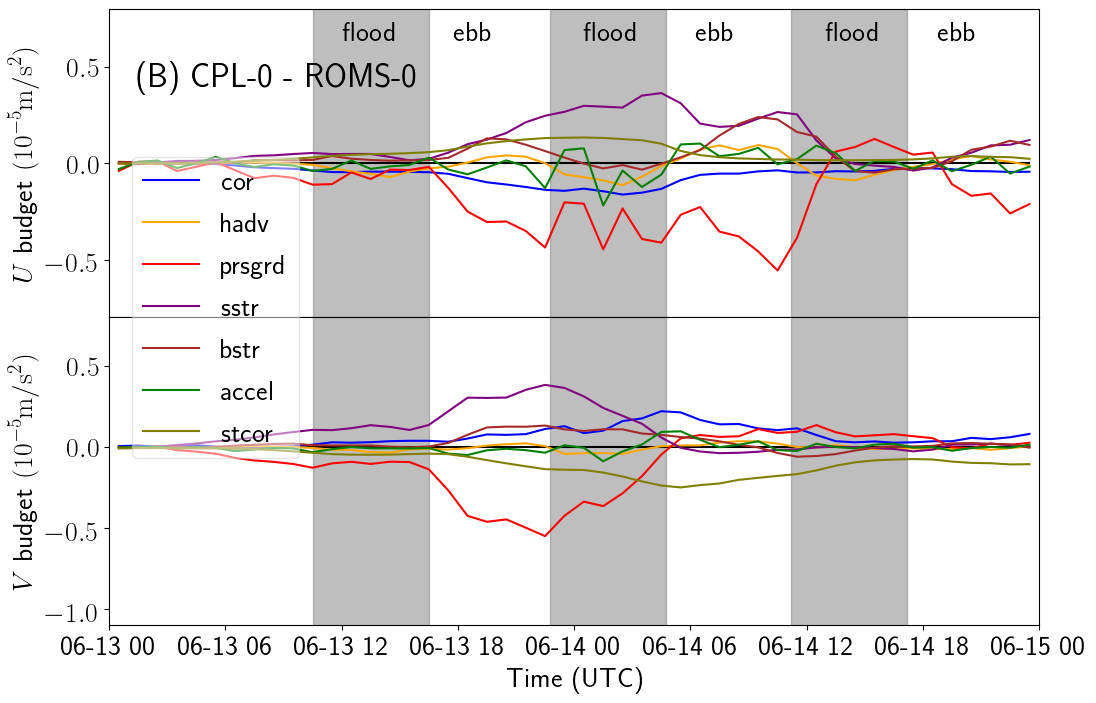}
  \centering \includegraphics[width=7cm]{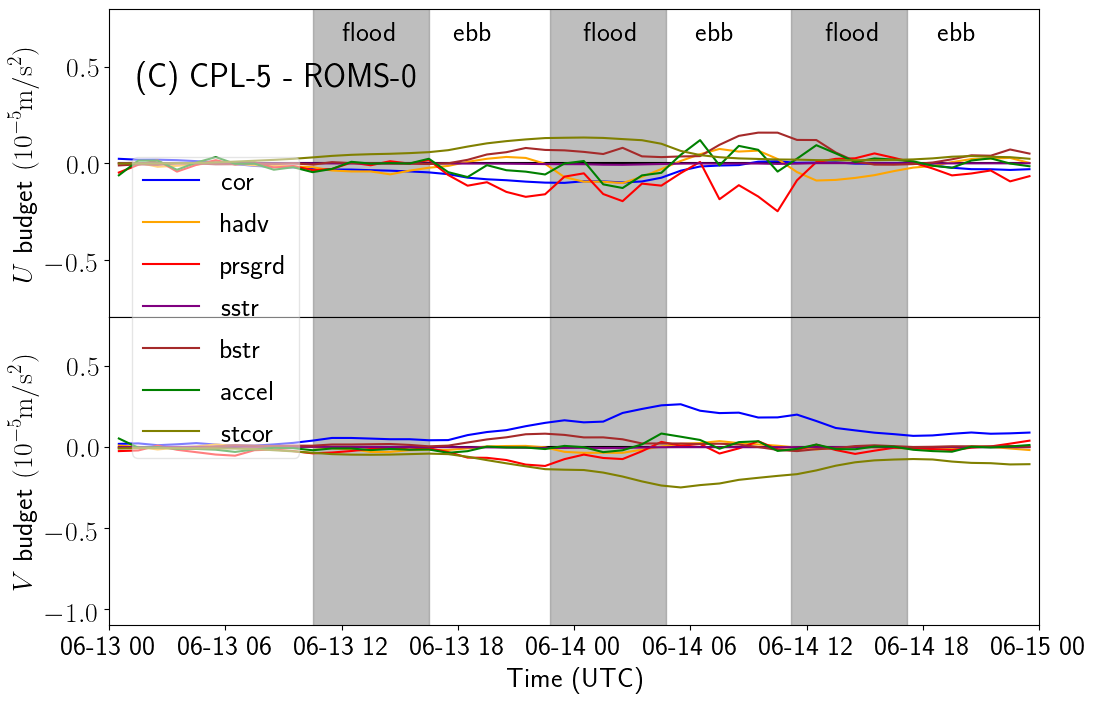} \includegraphics[width=7cm]{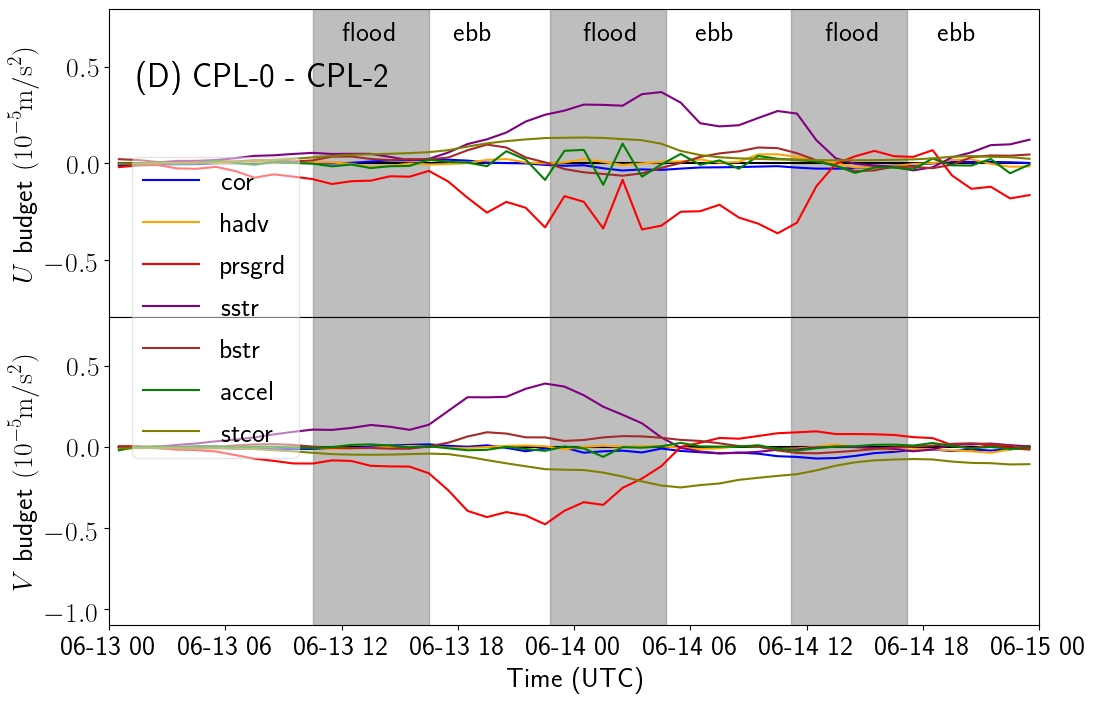}
  \caption{Time series of the depth-integrated budget terms in the Navier-Stokes equations during Storm Hector around $2018/06/14$, integrated over the area of Galway Bay. Only the dominant terms in the budget are retained. Plot (A) shows the terms introduced in Eq. (\ref{2d_momentum}). The difference between the runs CPL-0 and ROMS-0 is shown on plot (B), between CPL-5 and ROMS-0 on plot (C) and between CPL-0 and CPL-2 on plot (D).
  It highlights the change in budget caused by the wave effects.}
  \label{ts_budget}
\end{figure}

Time series of the dominant momentum terms in Eq. (\ref{2d_momentum}) are shown on plot (A) in Figure \ref{ts_budget}. The depth-integrated momentum terms are integrated horizontally in a squared box encompassing Galway Bay to obtain a general idea of which forces drive the overall circulation inside the bay. 

Outside the storm the acceleration terms follow closely the pressure gradient. During the storm the surface stress perturbs this balance but both the acceleration and the pressure gradient are still closely correlated. The pressure gradient is mostly dominated by the tidal motion through the gradient of the sea surface elevation. It is then correctly observed in quadrature with the sea level tidal motion.
The Coriolis force has a stable and important impact on the circulation. It appears in phase with the tidal motion described by the surface elevation and in quadrature with the pressure gradient and acceleration. This is expected as the Coriolis force is proportional to the current field itself. The amplitude of Coriolis forcing is roughly half the amplitude of the pressure gradient on the eastward direction, and of similar amplitude on the northward direction.
The Stokes-Coriolis force provides the largest direct wave-induced effects, but appears as a second order driving mechanism, which is expected for a wave-induced effect. All the other wave-induced forcing terms, such as the horizontal or vertical vortex force, the wave-breaking induced momentum or the bottom streaming are completely negligible.

The bottom stress is also noticeable. It is always delayed and opposed to the tidal forcing, which is also expected from a friction force. During the storm it seems that the bottom stress in the main eastward direction is enhanced, almost matching the Coriolis term. Outside the storm window the amplitude of the bottom stress is roughly half the amplitude of the Coriolis force.

The intensity of Storm Hector is well illustrated by the contribution of the surface stress. At the peak of the storm on $2018/06/14$ the surface stress is the most important forcing term, especially in the eastward direction where its amplitude accounts for more than twice the amplitude of the Coriolis term. It is balanced by an excess in the pressure gradient term in the opposite direction.

The impact of the coupling on the budget balance is shown in Figure \ref{ts_budget}. Plot (B) features the time series of the difference in budget terms between the runs CPL-0 and ROMS-0. Plot (C) shows these between CPL-5 and ROMS-0 and plot (D) between CPL-0 and CPL-2. When ROMS-0 is involved in the difference, its contribution for the Stokes-Coriolis force is null. Looking at the scaling, the wave effects roughly induce a $10\%$ change in the budget. The impact of the conservative terms highlighted on plot (C) is twice smaller than the impact of the wave-enhanced surface roughness shown on plot (D).
The surface stress is clearly seen to capture most of the wave effects on plot (B). It is induced by the wave-enhanced surface roughness and balanced by the pressure gradient as reproduced on plot (D). The Stokes-Coriolis force is the second wave effect, balanced by the Coriolis force as shown on plot (C). The horizontal advection which is enhanced by the Stokes drift is the last significant wave effect appearing in the budget.

Plot (C) is a little redundant but it shows indeed that the Coriolis force balances the Stokes-Coriolis force, and it highlights the contribution of the Stokes drift in the horizontal advection term. During the tidal flood the advection of currents by the Stokes drift causes a reduction of the eastward horizontal advection, and an increase during the tidal ebb.
During the two ebb periods covered by the time series the bottom stress in the eastward direction is seen to be enhanced by the waves on plot (B). It is mostly captured by the conservative terms, inducing an increase of the bottom stress during the tidal ebb and a reduction during the tidal flow.

A small semi-diurnal variability is also observed and induced by the wave-enhanced surface roughness, reducing consistently the bottom friction. Those effects are not explained here. The increased bottom stress is also balanced by the pressure gradient.

Two distinct main wave-induced responses are therefore observed. The first one appears from the conservative wave forcing terms and generates a balance between the Coriolis force and the Stokes-Coriolis force, while the second one is induced by the wave-enhanced surface roughness.

\subsubsection{Spatial distribution of the momentum balance}

The time series in Figure \ref{ts_budget} give an overall picture of balance inside the bay, but they do not accurately describe the local balances of the forces involved. In order to smooth out the variability spawning at short temporal scales the budget terms are averaged over the duration of Storm Hector, between $2018/06/13$ $23\mathrm{:}00\mathrm{:}00$ and $2018/06/14$ $10\mathrm{:}00\mathrm{:}00$.

\begin{figure}
  \centering \includegraphics[width=4.8cm]{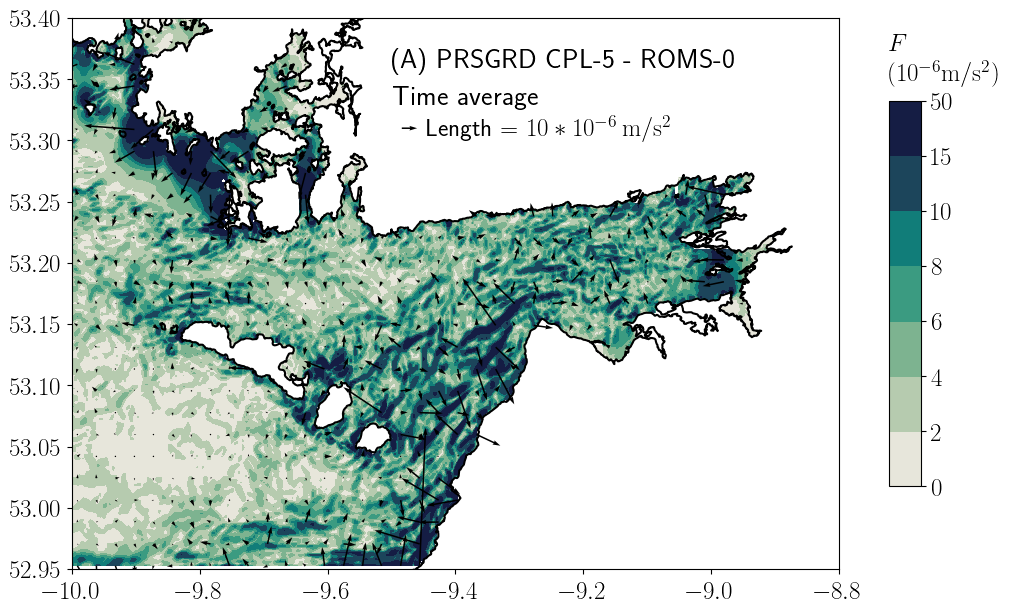} \includegraphics[width=4.8cm]{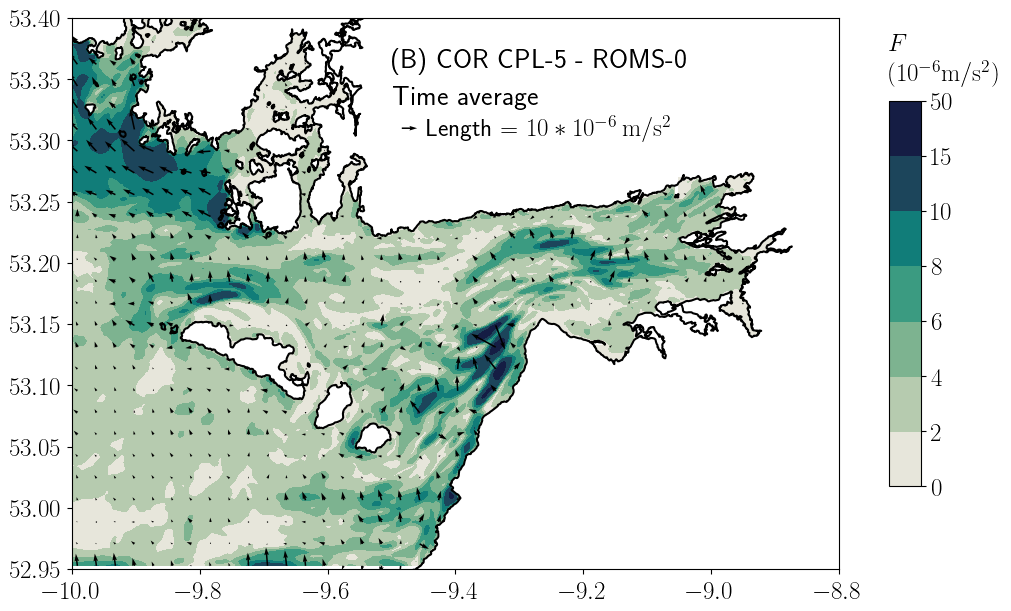} \includegraphics[width=4.8cm]{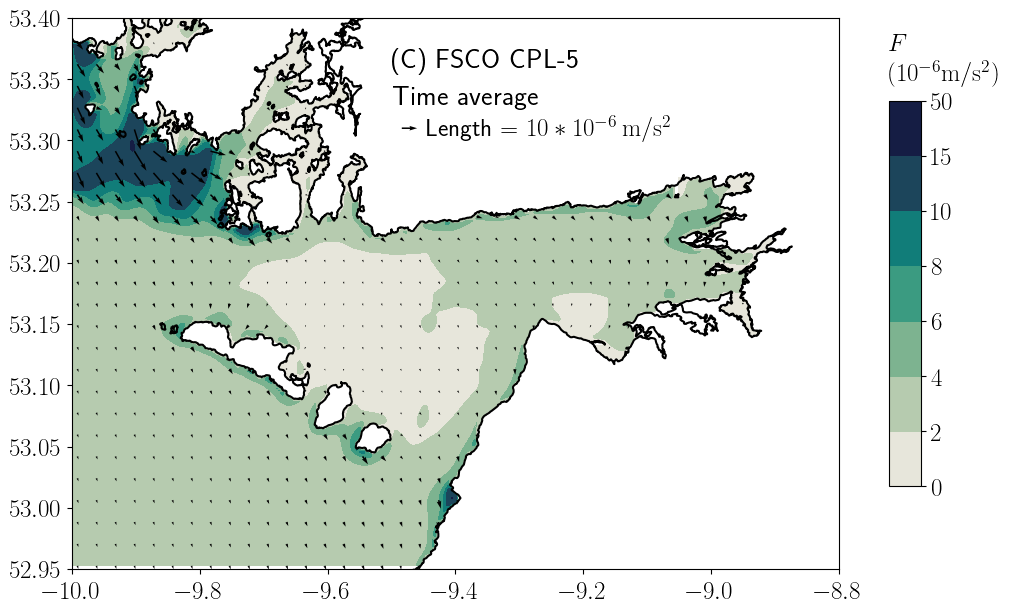}
  \caption{Difference in the significant budget terms, between the runs CPL-5 and ROMS-0, averaged over the duration of Storm Hector. Plot (A) features the pressure gradient, plot (B) the Coriolis force and plot (C) the Stokes-Coriolis force.
  The balance between the Coriolis and Stokes-Coriolis forces observed in Figure \ref{ts_budget} is not reproduced for the strong response along the coast of Clare. Instead a balance between the Coriolis force and the pressure gradient appears.}
  \label{map_budget_cpl5}
\end{figure}

Figure \ref{map_budget_cpl5} focuses on the response to the conservative wave-induced terms by showing the dominant forces involved in the balance. They consist in the Coriolis force on plot (B), which seems to be balanced by the pressure gradient shown on plot (A), and not the Stokes-Coriolis force shown on plot (C). On the contrary the Stokes-Coriolis force is well distributed spatially inside Galway Bay. This goes slightly against what Figure \ref{ts_budget} shows and will be investigated in a later paragraph. 

\begin{figure}
  \centering \includegraphics[width=4.8cm]{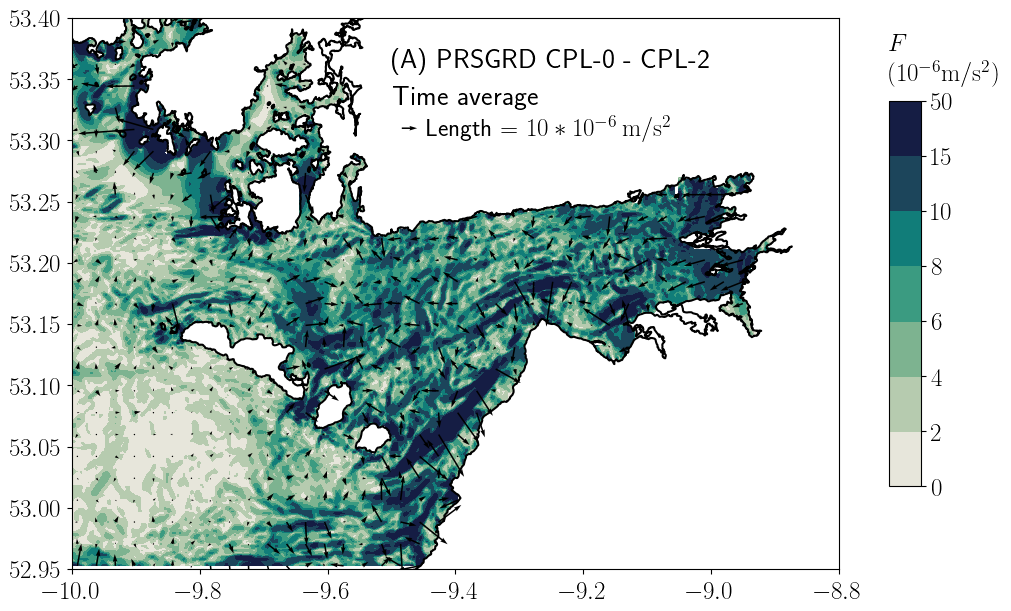} \includegraphics[width=4.8cm]{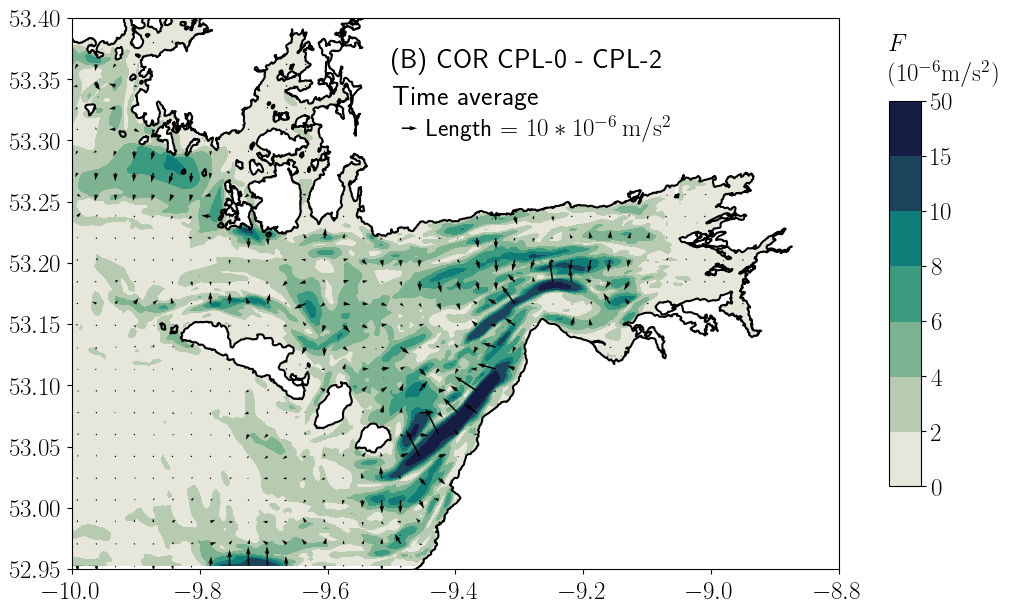} \includegraphics[width=4.8cm]{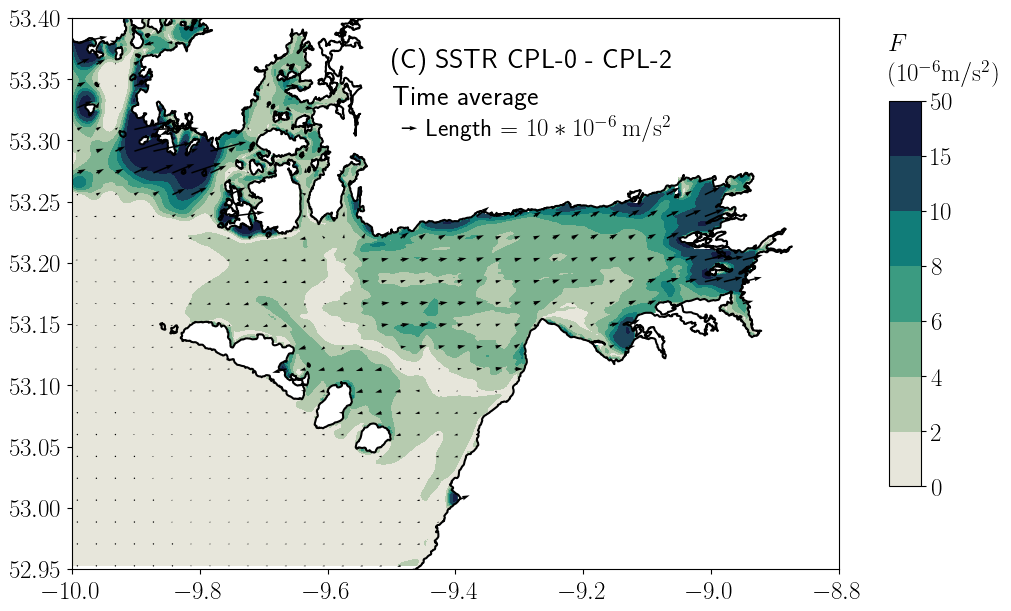}
  \caption{Difference in the significant budget terms, between the runs CPL-0 and CPL-2, averaged over the duration of Storm Hector. Plot (A) features the pressure gradient, plot (B) the Coriolis force and plot (C) the surface stress.
  Although the highlighted processes are different, the balance between the Coriolis force and the pressure gradient is similar to that observed in Figure \ref{map_budget_cpl5}.}
  \label{map_budget_cpl2}
\end{figure}

The main wave effect on the current is induced by the wave-enhanced surface roughness. Figure \ref{map_budget_cpl2} highlights the strongest budget terms that are modified by including this option. A strong balance between the Coriolis force and the pressure gradient is observed again, shown on plots (A) and (B). Although the effect and process highlighted are different, the conservative wave-induced terms and the wave-enhanced surface roughness both induce a similar response both in terms of current and budget balance. A significant difference is the increase in surface stress shown in plot (C). It is even stronger in the back of the bay where the water depth is lower. It is then well balanced by the pressure gradient specifically in the back of the bay.
Figure \ref{ts_budget}, plot (C), shows that the balance between the Stokes-Coriolis and Coriolis forces is not picked by the difference between the runs CPL-0 and CPL-2, and rightfully so. However a strong response in terms of Coriolis force is still observed here.

It is also worthwhile mentioning that the strong local Coriolis force difference is directly related to the strong response in terms of currents and agrees with its southward direction.
The figures show the wave-induced effects, but both the full pressure gradient and the full Coriolis force are tidal dependent. During a flood the tidal flow is going inside the bay and the balance is between a Coriolis force pointing east against a pressure gradient pointing west. It is the other way around during the ebb.
What those figures show is the wave effect on those forces, which is similar during both flood and ebb. During a flood the Coriolis force is reduced and the pressure gradient enhanced, and vice versa during the ebb.

\subsubsection{Current response to the conservative forces}

\begin{figure}
  \centering \includegraphics[width=6cm]{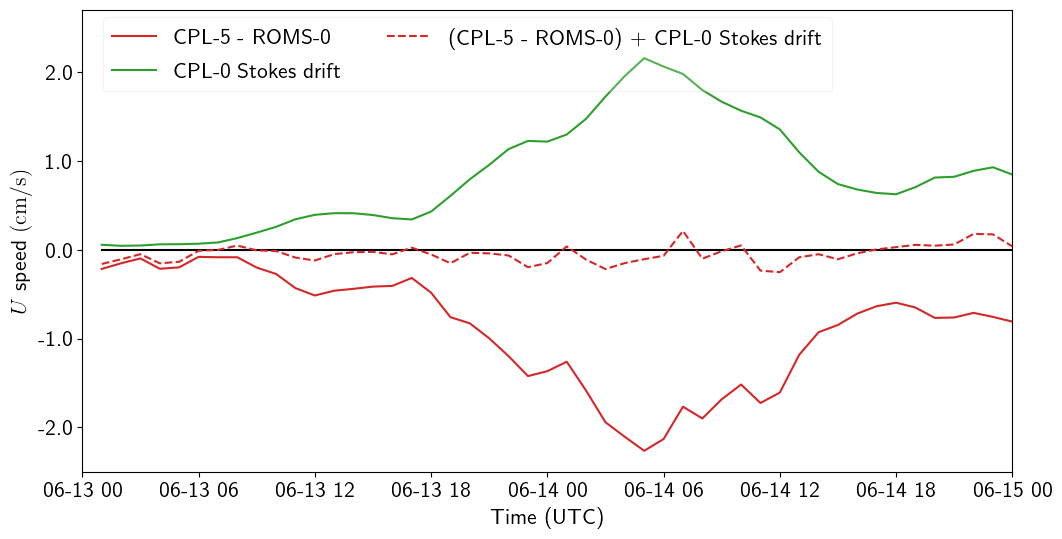} \includegraphics[width=6cm]{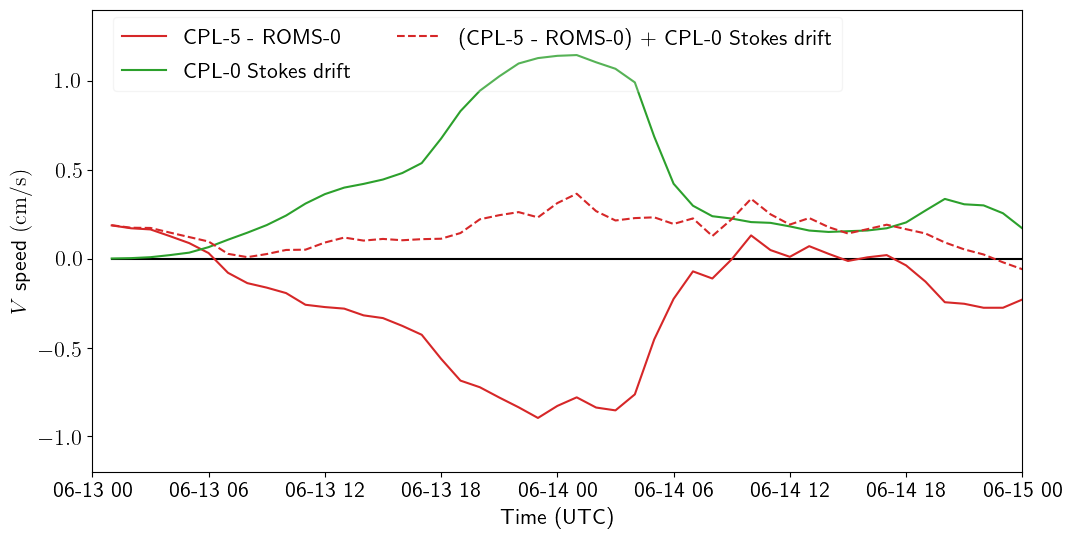}
  \caption{Time series of the depth-averaged and horizontally averaged currents and Stokes drift over Galway Bay, on the left is the eastward velocity $U$ and on the right the northward velocity $V$.
  The wave-induced currents balance well the Stokes drift on average.}
  \label{ubar_stokes}
\end{figure}

The Stokes-Coriolis force is the only significant wave-induced forcing appearing in the current velocity depth-averaged budget (Figure \ref{ts_budget}). It is small overall compared to the other forcing terms, but dominates the wave-induced response and is balanced by a modification in the Coriolis force. Given the formulations of those two forces, it only highlights a direct balance between the depth-averaged Stokes drift and the depth-averaged Eulerian velocity, which is shown in Figure \ref{ubar_stokes}, where only the response to the conservative terms with CPL-5 is considered. Overall and on average inside the bay, the Eulerian response to the wave forcing matches quite well the Stokes drift. The amplitude of the response is quite small, around $2\,\si{\cm\per\s}$ at best in the eastward direction and $1\,\si{\cm\per\s}$ in the northward direction, but it corresponds to a spatial average.

\begin{figure}
  \centering \includegraphics[width=4.5cm]{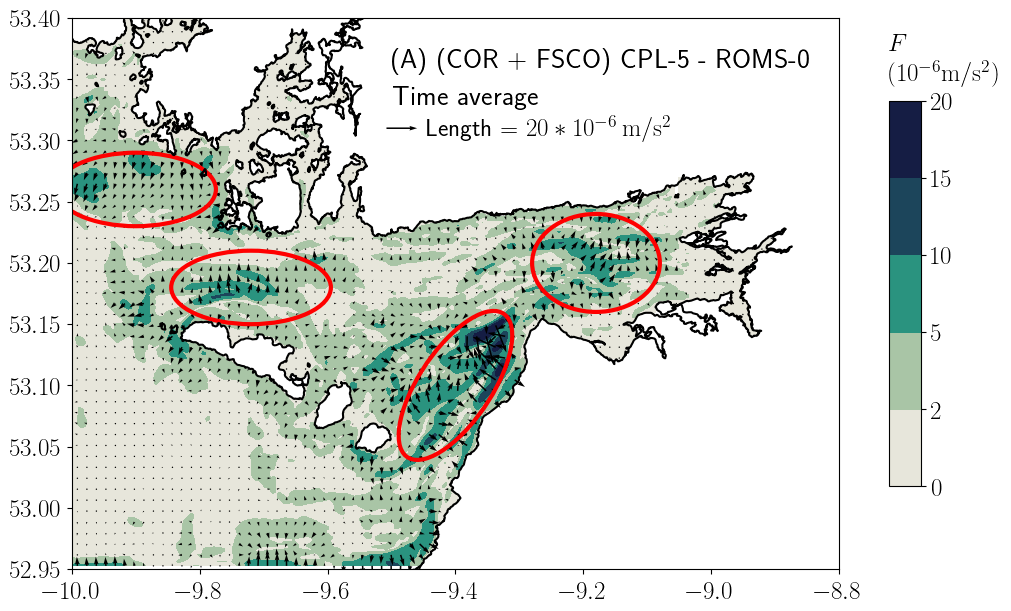} \includegraphics[width=4.5cm]{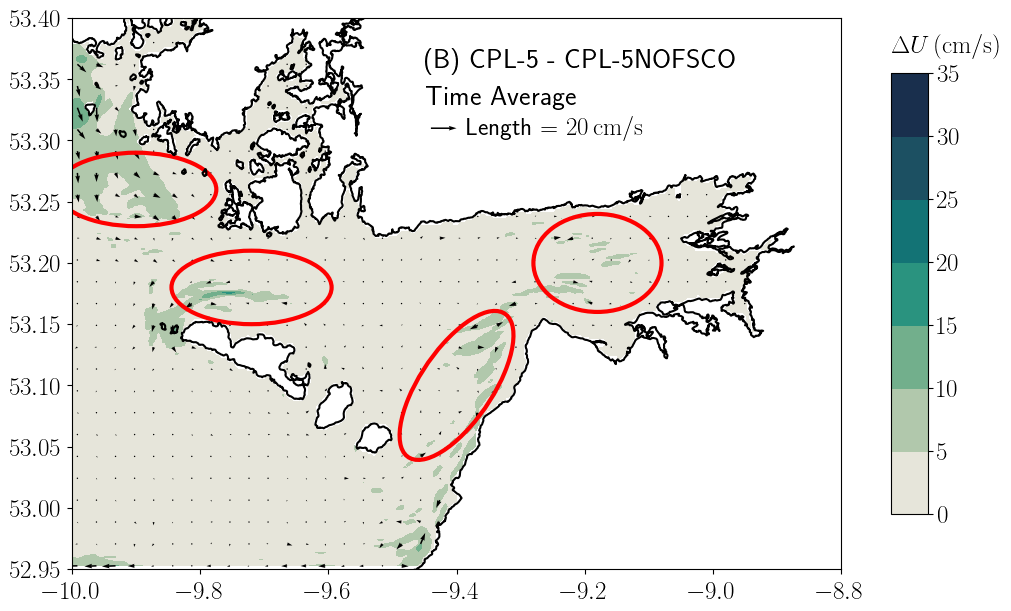}
  \includegraphics[width=4.5cm]{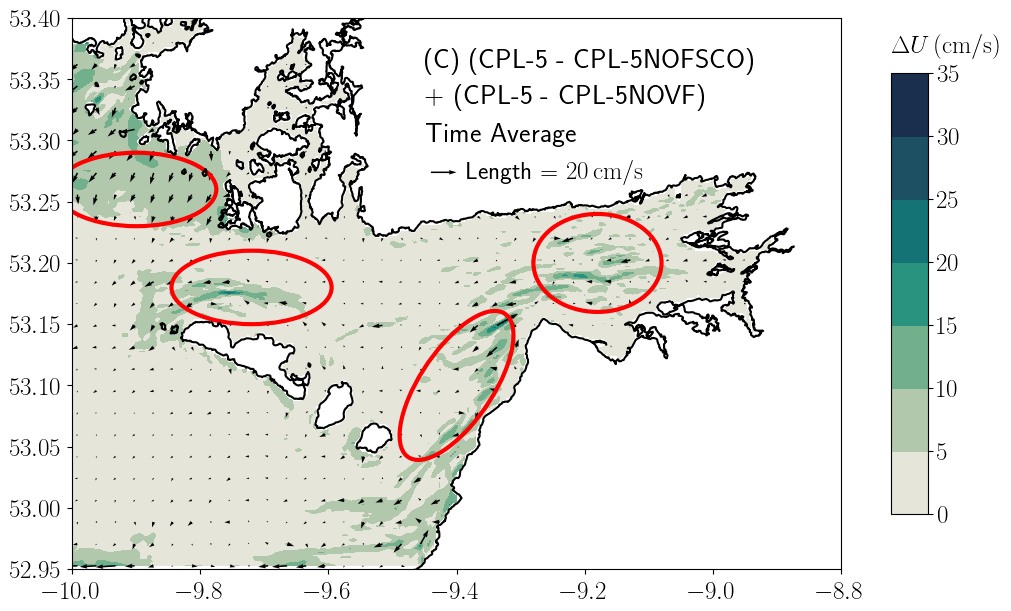}
  \caption{Plot (A) shows the resultant of the Coriolis force plus the Stokes-Coriolis forces, between the runs CPL-5 and ROMS-0, averaged over the duration of Storm Hector. It is well correlated with areas of strong responses observed in Figure \ref{currents_runs}, plot (B).
  Plot (B) shows the impact of the Stokes-Coriolis force only with CPL-5 minus CPL-5NOFSCO, while plot (C) shows the impact of the vortex force and Stokes drift mass transport with CPL-5 minus CPL-5NOVF. Both effects are able to individually capture the response to the conservative terms, but added together they explain most of it.}
  \label{stokes_coriolis_force}
\end{figure}

Looking at Figure \ref{map_budget_cpl5} it is not obvious whether a local equilibrium or at least partial equilibrium is reached between the Coriolis and Stokes-Coriolis forces. The Stokes-Coriolis force is seen quite weak in the center of Galway Bay behind the Aran Islands, and stronger in the inner bay and along the coast of Clare, which also corresponds to the areas where the Coriolis force is stronger. Plot (A) in Figure \ref{stokes_coriolis_force} shows the resultant of those two forces. The areas still left with a strong forcing are well correlated with areas where a response is observed (Figure \ref{currents_runs}, plot B), and equivalently where the wave-induced Coriolis force is seen strong (Figure \ref{map_budget_cpl5}, on plot B). They are highlighted in the figure by a red ellipse and correspond to Kilkieran Bay, the north side of Inishmore, along the coast of Clare and off to the west of Black Head, and inside Inner Galway Bay.
In general those two forces do not locally balance each other, indicating that the Eulerian response is a local feature, which balances the global input of momentum provided by the wave forcing.

The Stokes-Coriolis force appears then as a natural candidate to explain the wave-induced impact on the currents, but it is found unable to capture fully the response as shown on plot (B) in Figure \ref{stokes_coriolis_force}. However the full response is well reproduced when adding the impact of the mass and momentum transports by the Stokes drift. Plot (C) shows the sum of the individual responses from the Stokes-Coriolis force with CPL-5NOSFCO and from the vortex force plus the momentum transport induced by the Stokes drift with CPL-5NOVF. Those last two effects are paired due to the implementation done in \cwst{}, but given the study of the budget terms, the momentum transport is assumed to be the only relevant mechanism and the vortex force negligible.

\subsubsection{Current response to wave-enhanced surface roughness}

\begin{figure}
  \centering \includegraphics[width=4.8cm]{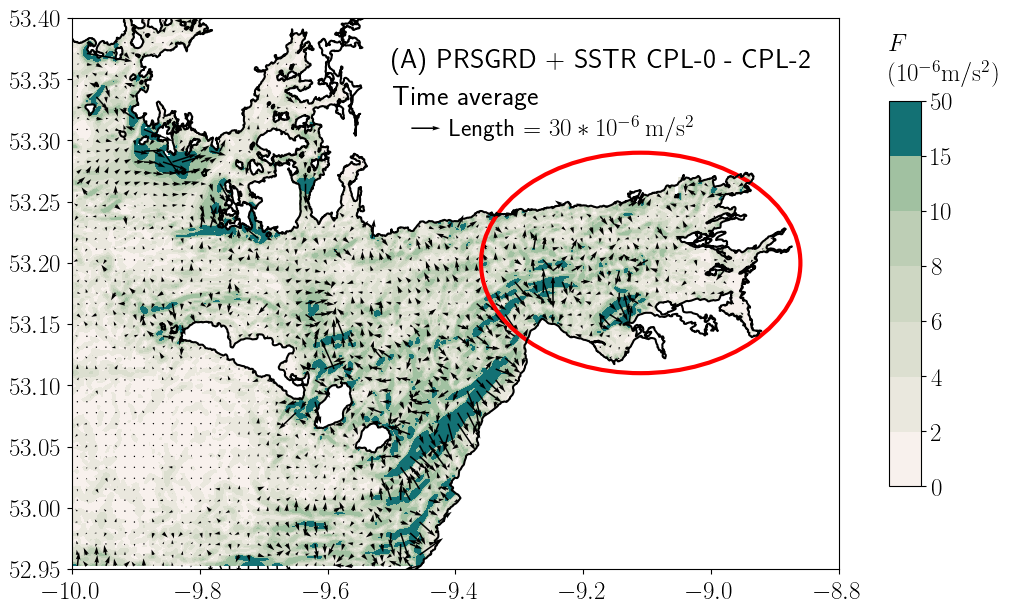} \includegraphics[width=4.8cm]{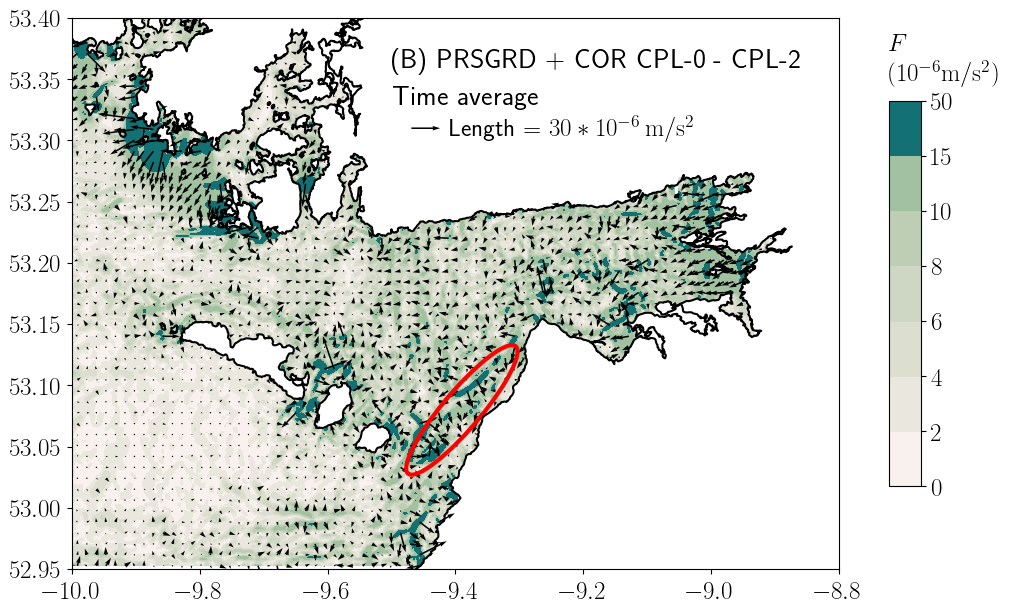} \includegraphics[width=4.8cm]{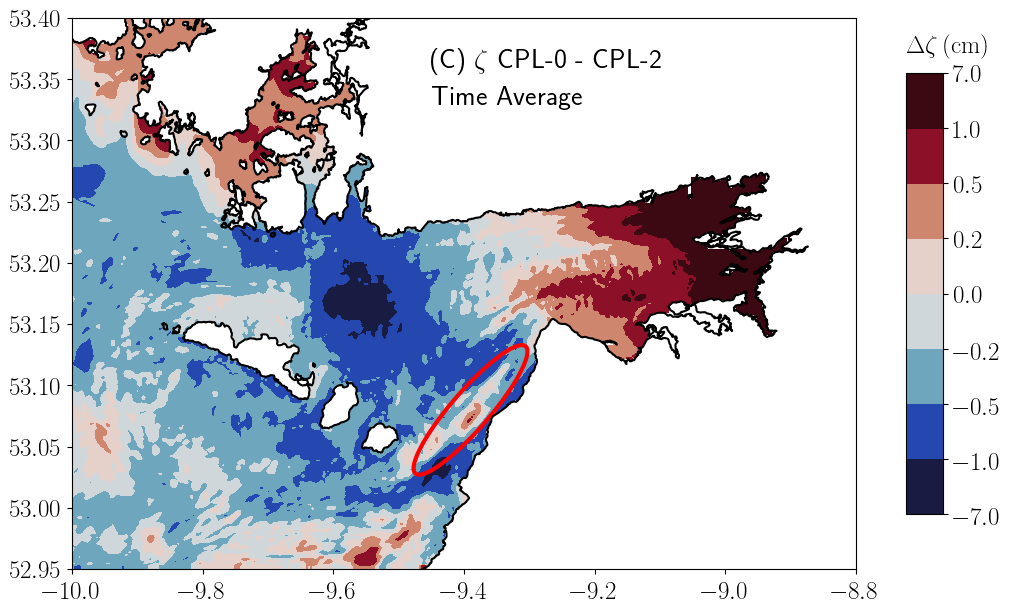}
  \caption{Plot (A) shows the resultant of the pressure gradient plus the surface stress computed for the run CPL-0 minus CPL-2. The contribution of those two budget terms inside the inner bay is well canceled. Plot (B) similarly shows the resultant of the pressure gradient plus the Coriolis force. The contribution of those terms along the coast of Clare is mostly canceled as well. Plot (C) zooms on the sea surface elevation induced by the wave-enhanced roughness, which is estimated to be strong enough to generate the observed response in pressure gradient.
  The red ellipses are highlighting the areas of interest in each plot.}
  \label{roughness}
\end{figure}

Although the response in current to the wave-enhanced surface roughness is similar to the response to the Stokes drift forcing (Figure \ref{currents_runs}), it doesn't involve the same processes. In terms of balance (Figure \ref{ts_budget}), the overall forces involved are the surface stress, which is enhanced by the waves, and the pressure gradient balancing the surface stress. Most of this balance occurs inside Inner Galway Bay (Figure \ref{map_budget_cpl2}, plots A and C), and explains well the wave-induced surge. It is highlighted in Figure \ref{roughness}, plot (A).

A strong balance between the Coriolis force and the pressure gradient appears along the coast of Clare on plot (B). However, it is mostly a consequence of the southwards response in current. The Coriolis force response pointing west is induced by the south response in current. It induces a stronger mass transport and a slight water level increase to the west of the current response. A pressure gradient is thus generated that is opposed to the gradient in surface elevation that is emerging. This is verified on plot (C), which shows the difference in sea surface elevation, where the scale has been modified to highlight this particular location. The gradient in surface elevation can be roughly estimated. A variation of $0.7\,\si{\cm}$ over $2\,\si{\km}$ gives a pressure gradient of $35\cdot10^{-6}\,\si{\m\per\square\s}$. This shows that the Coriolis force and pressure gradient are able to maintain the balance, but it does not explain the origin of this excess of momentum.

\begin{figure}
  \centering \includegraphics[width=4.8cm]{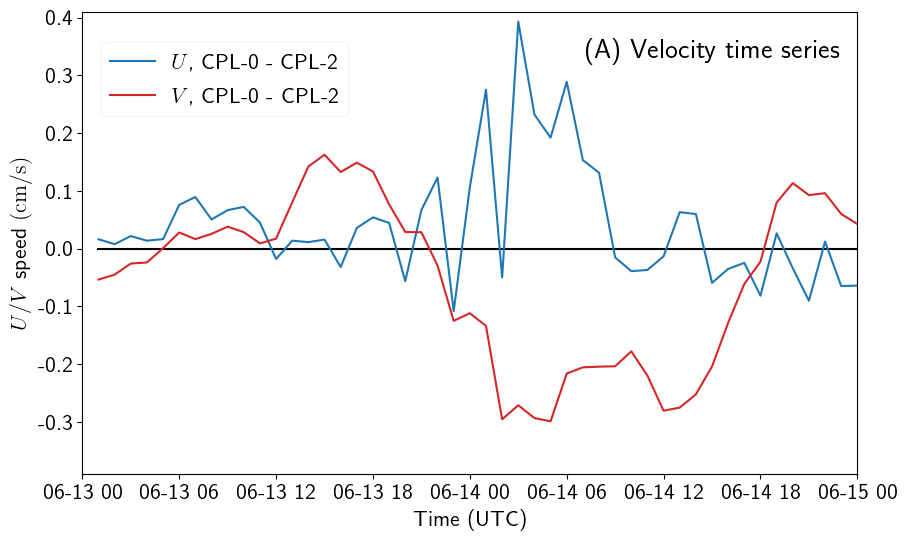} \includegraphics[width=4.8cm]{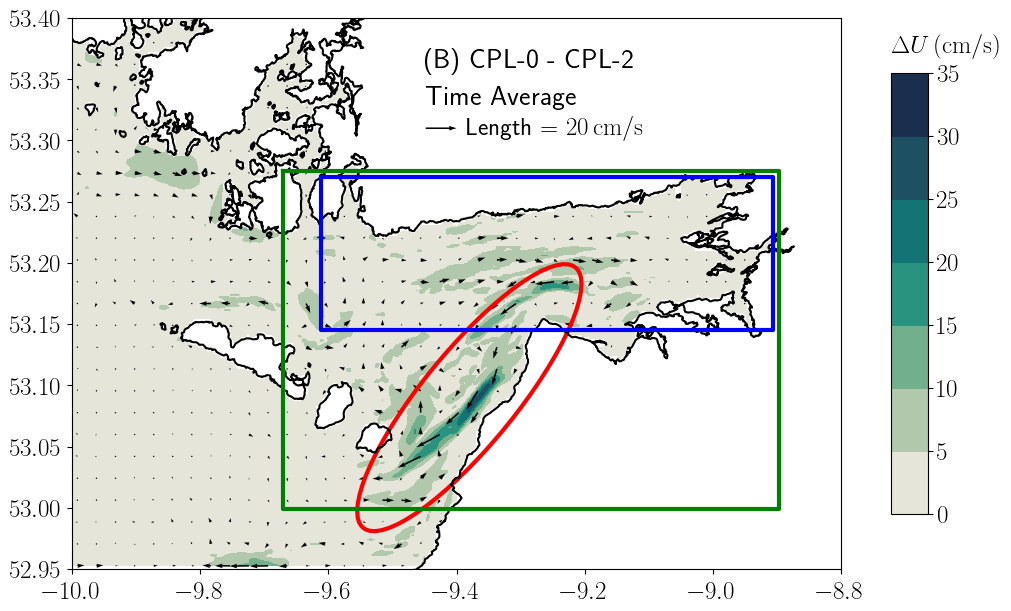} \includegraphics[width=4.8cm]{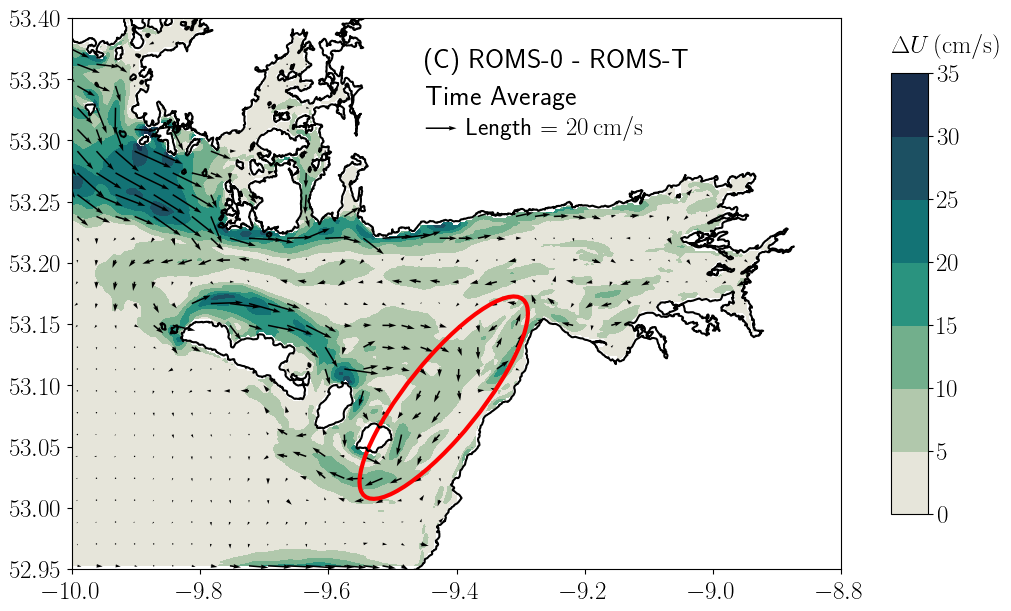}
  \caption{Plot (A) shows the time series of the depth-averaged and spatially averaged eastward and northward velocities inside the green box shown in plot (B). The southward velocities match well the eastward velocities, preserving the conservation of mass. Plot (B) shows the response in terms of currents to the wave-enhanced surface roughness. The eastward transport in the blue box is believed to be balanced by the southward transport in the red ellipse going around Black Head and then through the South Sound. Plot (C) shows the wind-induced circulation with ROMS-0 minus ROMS-T, already featuring a southward response in the red ellipse.}
  \label{roughness_momentum}
\end{figure}

Plot (A) in Figure \ref{roughness_momentum} shows the time series of the depth-averaged eastward and northward velocities averaged inside the green box shown in plot (B). The southward velocities, mostly defining the response along the coast of Clare inside the red ellipse in plot (B), match quite well the eastward velocities mostly strong in the blue box. The strong southward response can hence be explained through the conservation of mass of the enhanced wind-induced velocities due the wave-enhanced surface roughness. The excess of transport going inside Inner Galway Bay has to be compensated by an outward flow, which is contained on the south part of the bay. This is consistent with the global rotation forced by the Coriolis force inside the bay. Plot (C) shows the averaged circulation forced by the wind by showing ROMS-0 minus ROMS-T. It already induces a southward flow along the coast of Clare highlighted by the ellipse. Therefore the wave forcing increases this particular feature of wind-induced circulation.
Again, the main circulation is still forced by the tides, so the response just described above consists in a reduction of the inward flood tidal flow, or an increase of the outward ebb tidal flow. In both cases it is induced by a stronger wind-stress caused by the action of waves inside Galway Bay.

\section{Conclusion}

A coupled wave-ocean model is set up for Galway Bay using \cwst{}. The model domain is the same as currently used in an operational forecasting system based on \roms{} at the Marine Institute, Ireland. The impact of wave-ocean coupling was tested in this study. This coupling comes with a computational and financial cost. It is adamant to assess the impact of the coupling on the capabilities of the model.
We follow closely the configuration presented in \textcite{kumar2012implementation}. The wave-induced surface streaming is however not included. It makes our model highly unstable, despite several refinements of the surface layers and reduction of the time-step. Candidly tested during Storm Hector, it is observed to have a noticeable impact on the solution, but negligible compared to the wave-enhanced surface roughness for instance.

The model is validated using tidal gauges, ADCPs, a wave buoy, and covering three time periods of roughly four months each. It gives an overall good agreement. The errors in terms of velocities are of the same order, if not slightly less, than a previous model found in the literature for Galway Bay (\nptextcite{ren2017effect}). Excellent agreement on the tidal signal is found for the surface elevation. The swell outside Galway Bay agrees well with the measurements. It validates the boundary conditions. However a loss of accuracy is observed inside Galway Bay when it comes to the swell partition. This is especially true when looking at the peak wave direction and peak frequency associated to the swell systems. The locally generated wind waves are still very well captured by the model.
We suspect that the model fails to capture accurately the depth-induced processes as the swell passes through the North and South Sounds. We also show that the sea state inside Galway Bay is characterised by a mixed sea between one wind wave partition and two swell partitions, entering through the North and South Sounds.

The impact of the coupling is not seen to improve significantly the statistics at the stations where measurements are available. 
First, the coupling is only significant inside Galway Bay during strong sea conditions. This is necessary to have a strong wave forcing impacting the currents and sea level, but also necessary to have enough wave energy inside the bay that can be significantly impacted by the currents and sea level. The long-term statistics are smoothing out those rare events.
The model is significantly less accurate during those storm events, especially for the currents. The coupling is only slightly improving the agreement. An interesting follow-up would be to specifically study the influence of the wind conditions on the accuracy of the model. Strong sea states and storms are commonly accompanied with strong winds that are known to drive significantly the circulation in Galway Bay.
Second, focusing on Storm Hector, we show that although a small increase in sea surface elevation is observed everywhere in the domain by $2\,\si{\cm}$ on average. The strongest effects are observed in the back of Galway Bay, reaching $7\,\si{\cm}$. The wave effects are even more local in terms of currents. A weak Eulerian response is observed in the surface layers everywhere in the model, going against the wave propagation (not shown in this paper), but the strongest response in terms of currents appears along the coast of Clare. The response induces a reduction of the tidal flood and an increase of the tidal ebb by $30\,\si{\cm\per\s}$ at most, which is significant.

The response in sea level is divided between the direct impact of the Stokes drift inducing a strong mass transport directed in the back of the bay, and the wave-enhanced surface roughness increasing the wind-induced momentum and therefore mass transport also pushing water in the back of the bay. The Stokes drift accounts for most of the effect, around $4.5\,\si{\cm}$, while the increased wind transport by caused by the waves only accounts for $1.5\,\si{\cm}$.
In terms of currents the response is also divided between the conservative and non-conservative effects. The Stokes-Coriolis force and the Stokes momentum transport induce a noticeable Eulerian response in the bay. Although the wave forcing is well spread inside the bay, the response is well localised inducing changes in the currents up to $10\,\si{\cm\per\s}$ along the coast of Clare. When integrated inside the bay it is still found to balance the Stokes drift. The wave-enhanced surface roughness overall increases the wind-induced circulation in the bay. It dominates over the conservative wave effects by accounting up to $20\,\si{\cm\per\s}$ along the coast of Clare.

\section*{Acknowledgments}{The authors wish to acknowledge the Cullen Fellowship Programme, under the fellowship ``Coupled wave-ocean models'' funded by the Marine Institute. The authors also wish to acknowledge the DJEI/DES/SFI/HEA Irish Centre for High-End Computing (ICHEC) for the provision of computational facilities and support. Finally, the
authors are grateful to Teledyne RD Instruments (TRDI) for providing the ADCP and for their useful contributions.}


\section*{Credit authorship contribution statement}{C Calvino: Conceptualisation of this study, Methodology, Software, Data
curation, Writing - Original draft preparation.
T Dabrowski and F Dias: Overall supervision - Contributed to manuscript
preparation and approved the submitted version.}

\appendix
\section{Statistical parameters \label{statistics}}

The following statistical parameters are used for the validation, using the generic notation $X$ and $Y$ for the time series with $N$ records:
\begin{empheq}{align}
 & \text{mean :} && \mathrm{m}(X) = \frac{\sum_{i=1}^{N} {X_i}}{N} \,,\\
 & \text{standard deviation :} &&  \sigma(X) = \sqrt{\frac{\sum_{i=1}^{N} (X_i - m(X))^{2}}{N}} \,,\\
 & \text{root-mean-square error :} && \mathrm{rmse}(X,Y) = \sqrt{\frac{\sum_{i=1}^{N} (X_i - Y_i)^{2}}{N}} \,,\\
 & \text{Pearson correlation :} && \text{cor}(X,Y) = \frac{\sum_{i=1}^{N} (X_i - \mathrm{m}(X))(Y_i - \mathrm{m}(Y))}{\sqrt{\sum_{i=1}^{N} (X_i - \mathrm{m}(X))^2 \sum_{i=1}^{N} (Y_i - \mathrm{m}(Y))^2}} \,.
\end{empheq}

The formulas are adjusted in the case of circular variables to take into account their periodicity. Below the time series $X$ and $Y$ are assumed to be in radians between $-\pi$ and $+\pi$. For the standard deviation the formula given below is not in radians. The values range from $0$ to $\infty$ with $0$ describing a distribution with no variance at all:
\begin{empheq}{align}
 & \text{mean :} && \mathrm{m}_\mathrm{r}(X) = \mathrm{atan}\left(\mathrm{m}(\sin(X)),\mathrm{m}(\cos(X))\right) \,,\\
 & \text{standard deviation :} &&  \sigma_\mathrm{r}(X) = \sqrt{-2 \ln{\left(\sqrt{\mathrm{m}(\cos{X})^2 + \mathrm{m}(\sin{X})^2}\right)}} \,,\\
 & \text{root-mean-square error :} && \mathrm{rmse}_\mathrm{r}(X,Y) = \sqrt{\frac{\sum_{i=1}^{N} ((X_i - Y_i + \pi \mod{2\pi}) - \pi)^2}{N}} \,,
\end{empheq}
\begin{equation}
 \text{Pearson correlation :} \;\;
 \mathrm{cor}_\mathrm{r}(X,Y) = \frac{\sum_{i=1}^{N} \sin(X_i - \mathrm{m}_\mathrm{r}(X)) \sin(Y_i - \mathrm{m}_\mathrm{r}(Y))}{\sqrt{\sum_{i=1}^{N} \sin^2(X_i - \mathrm{m}_\mathrm{r}(X)) \sum_{i=1}^{N} \sin^2(Y_i - \mathrm{m}_\mathrm{r}(Y))}} \,.
\end{equation}

\printbibliography

\end{document}